%% file: main.tex
\documentclass[10pt,journal,compsoc,nofonttune]{IEEEtran}

\usepackage{amsmath,amsfonts,amssymb}
\usepackage{array}
\usepackage{textcomp}
\usepackage{stfloats}
\usepackage{url}
\usepackage{graphicx}
\usepackage{booktabs}
\usepackage{multirow}
\usepackage{cite}

\usepackage[acronyms,nonumberlist,nopostdot,nomain,nogroupskip,acronymlists={hidden}]{glossaries}
\newglossary[algh]{hidden}{acrh}{acnh}{Hidden Acronyms}
\glsdisablehyper

\usepackage{tikz}
\usetikzlibrary{arrows.meta, positioning, shapes, calc, patterns, shapes.geometric, fit, backgrounds}
\usepackage{pgfplots}
\usepgfplotslibrary{groupplots}
\pgfplotsset{compat=newest}
\usepackage{soul}
\usepackage{xspace}
\usepackage{ragged2e}
\usepackage{subcaption}
\usepackage{enumitem}
\setlist{nosep}

\usepackage{xcolor}
\usepackage{colortbl}

\newif\ifrevision
\revisionfalse
\newcommand{\blue}[1]{\ifrevision{\color{blue}\relax#1}\else#1\fi}

\newif\ifarxiv
\arxivtrue

\definecolor{inputcolor}{HTML}{4A90D9}
\definecolor{convcolor}{HTML}{5CB85C}
\definecolor{bncolor}{HTML}{F0AD4E}
\definecolor{relucolor}{HTML}{D9534F}
\definecolor{poolcolor}{HTML}{9B59B6}
\definecolor{fccolor}{HTML}{3498DB}
\definecolor{dropcolor}{HTML}{95A5A6}
\definecolor{softmaxcolor}{HTML}{E74C3C}
\definecolor{outputcolor}{HTML}{1ABC9C}
\definecolor{resblockcolor}{HTML}{ECF0F1}
\definecolor{arrowcolor}{HTML}{2C3E50}

\definecolor{lightgray204}{RGB}{204,204,204}  %
\definecolor{rawcolor}{RGB}{160,160,160}      %
\definecolor{avgcolor}{RGB}{230,159,0}        %
\definecolor{kalmancolor}{RGB}{86,180,233}    %
\definecolor{gtcolor}{RGB}{0,158,115}         %

\newif\ifexttikz
\exttikzfalse

\ifexttikz
  \usetikzlibrary{external}
\fi

\usepackage{microtype}
\usepackage{balance}

\input{acronyms.tex}
\newlength\fwidth
\newlength\fheight

\ifarxiv
\usepackage{tikzpagenodes,etoolbox}
\usetikzlibrary{calc}
\usepackage[contents={}]{background}
\AddEverypageHook{%
\ifnumequal{\thepage}{1}{%
    \tikz[remember picture,overlay]{%
        \node[
        minimum width=1.03\textwidth,
        text width=0.95\textwidth,
        font=\scriptsize
        ]
        at ($(current page header area) - (0,5pt)$)
        {%
        This work has been submitted to IEEE for possible publication.\\
        Copyright may be transferred without notice, after which this version may no longer be accessible.
    }}%
}{}%
}
\fi

\begin{document}
\title{Programmable GPU-Accelerated \blue{dApps}\\for Real-Time Edge Inference \blue{and} ISAC\\on NVIDIA Aerial Testbed}

\author{\IEEEauthorblockN{
Davide Villa,
Mauro Belgiovine,
Nicholas Hedberg,
Michele Polese,
Chris Dick,
Tommaso Melodia
}
\thanks{D. Villa, M. Belgiovine, N. Hedberg, and C. Dick are with NVIDIA Corporation, Santa Clara, CA. Email: \{dvilla, mbelgiovine, nhedberg, cdick\}@nvidia.com}
\thanks{D. Villa, M. Polese, and T. Melodia are with the Institute for Intelligent Networked Systems, Northeastern University, Boston, MA. Email: \{villa.d, m.polese, t.melodia\}@northeastern.edu}}

\makeatletter
\patchcmd{\@maketitle}
  {\addvspace{0.5\baselineskip}\egroup}
  {\addvspace{-1.5\baselineskip}\egroup}
  {}
  {}
\makeatother

\IEEEoverridecommandlockouts

\maketitle

\begin{abstract}

The transition of cellular networks to (i) software-based systems on commodity hardware and (ii) platforms for services beyond connectivity introduces critical system-level challenges. As sensing emerges as a key feature toward 6G standardization, supporting \gls{isac} with limited bandwidth and piggybacking on communication signals, while maintaining high reliability and performance, remains a fundamental challenge.
In this paper, we provide two key contributions. First, we present a programmable, open-source framework for processing \blue{PHY} signals through real-time, GPU-accelerated \gls{ai} applications on the edge \gls{ran} infrastructure. Building on the Open \gls{ran} dApp architecture, the framework interfaces with a GPU-accelerated gNB based on NVIDIA \gls{atb}, feeding \blue{PHY} data to custom AI logic with \blue{about $84~\mu\mathrm{s}$ of framework latency per slot and control loops with inference in about $123~\mu\mathrm{s}$ for lightweight TensorRT models}, multiple inference engines, and support for several \gls{ai} backends. We evaluate the framework on \blue{two} GPU platforms\blue{, with different GPU sharing modes and an increasing number of cells and dApps}.
Second, we demonstrate the framework capabilities through cuSense, an indoor localization dApp that consumes uplink DMRS channel estimates, removes static multipath components, and runs a neural network to infer the position of a moving person. Evaluated on a 3GPP-compliant 5G NR deployment, cuSense achieves a mean localization error of \blue{73}~cm\blue{ on runs unseen during training}, with \blue{79\%} of predictions falling within 1~meter, \blue{and runs live in $241~\mu\mathrm{s}$,}
without dedicated sensing hardware or modifications to the \blue{5G} signals. The framework is released as open source\blue{, together with the cuSense dataset}, providing a reference design for future AI-native RANs and \gls{isac} applications.

\end{abstract}


\glsresetall
\glsunset{nr}
\glsunset{5g}
\glsunset{dapp}
\glsunset{phy}

\section{Introduction}
\label{sec:intro}

\IEEEPARstart{T}{he} mobile cellular industry faces a reckoning: declining connectivity revenue and surging energy and operational costs 
make it harder to justify generational upgrades driven by mere connectivity gains~\cite{jiang2021road6g}. 
To this end, the evolution toward \gls{6g} networks encompasses software stacks on generic compute accelerators, to decrease costs and make the network more flexible, \emph{and} the identification of new revenue streams, beyond connectivity~\cite{giordani2020toward6g}. The \gls{ran} is transforming from connectivity infrastructure to an edge platform that can provide a variety of services, leveraging the availability of rich signals and real-time telemetry at the edge, and multiplexing additional compute tasks on a programmable, accelerated infrastructure. 
\gls{isac} represents an example in this space. Using signals available in communication stacks to enable sensing solutions has been a long-studied topic in the wireless literature~\cite{cui2021isac,liu2022isac}.
How to enable \gls{isac} at scale, on top of a commercial cellular network solution, with consistent and reliable performance, limited bandwidth, and specific reference signals, remains, however, an open question. 
The \gls{3gpp} is developing \gls{isac} study items for its Release~19~\cite{3gpp22137}, toward technical specifications for \gls{6g}. Similarly, several recent papers discuss waveform design, reference-signal structures, and sensing/detection algorithms for \gls{isac} in cellular systems, without, however, discussing system-level enablers~\cite{cui2022isac,liu2023isacmulti,wei2024multiple}. 

In this paper, we address critical \gls{isac} system challenges by demonstrating how to (i) evolve the network stack and edge \gls{ran} infrastructure to combine \gls{isac} signal processing with \gls{ai} for real-time, high-accuracy sensing, while (ii) leveraging signals already defined in 5G NR specifications, captured on a standards-compliant stack using commercial smartphones. By introducing real-time, open interfaces on an open-source protocol stack accelerated on \gls{gpu} and releasing a programmable \gls{isac} framework, we establish a practical pathway to 5G-based \gls{isac} and enable further development in this space. Next, we discuss the challenges addressed in this paper and summarize our main contributions.

\textbf{Challenge \#1: Lack of Real-Time Data Exposure.} 
\gls{ran} control and observability (e.g., with O-RAN interfaces)
remain limited to near- or non-real-time scales (i.e., above $10$~ms), and focused on the control plane. O-RAN programma-ble applications, xApps and rApps, are deployed in dedicated controllers that aggregate data from tens of \glspl{gnb}. They typically operate with timescales of tens of milliseconds to seconds and observe the \gls{ran} through aggregated \glspl{kpi}, per-device and per-cell statistics computed over many slots and subcarriers. This design is intentional, as sending granular control or user-plane data---for example, \acrshort{iq} samples or full \gls{csi} matrices---to a remote controller would incur prohibitive (Gbps-scale) data rates and raise security and privacy concerns. Today's programmable \glspl{ran} thus cannot leverage fine-grained \gls{phy}/\gls{mac} telemetry for per-slot tasks, such as fast link adaptation, beam management, or \gls{isac}.

\textbf{Challenge \#2: Real-time, High-Accuracy Estimates.}
Extracting accurate sensing estimates from communication signals raises both algorithmic and computational challenges. \gls{csi} is high-dimensional, noisy, and contains strong static multipath components that must be separated from the variations caused by a moving target. Thus, achieving sub-meter accuracy requires \gls{ai} models capable of learning complex spatial mappings, coupled with inference pipelines that execute in real time within slot-level timing constraints, a combination that demands high computational capacity.

\begin{figure}[t]
  \centering
  \includegraphics[width=\linewidth]{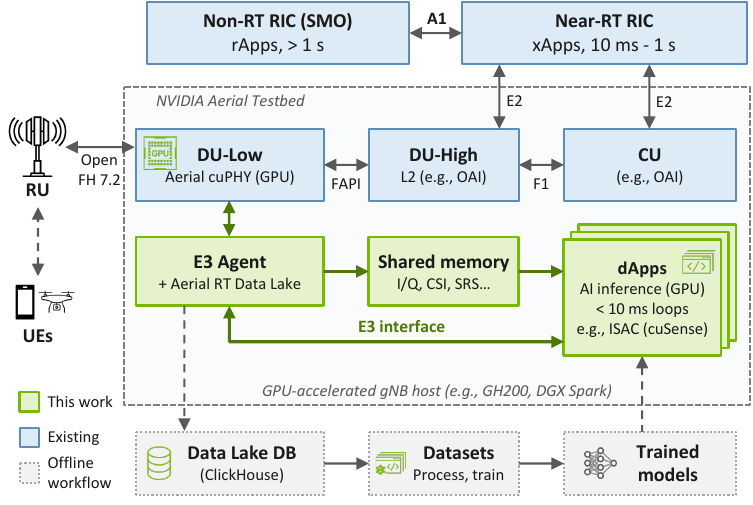}
  \caption{\blue{dApp framework in an O-RAN gNB on the NVIDIA Aerial Testbed.}}
    \label{fig:intro-diagram}
    \vspace{-.3cm}
\end{figure}

\textbf{This work: \blue{GPU-accelerated dApps for Real-Time ISAC} on the NVIDIA \gls{atb} 5G Stack.} In this paper, we address these challenges by designing, implementing, and evaluating \blue{a} \gls{gpu}-accelerated \blue{dApp framework for real-time \gls{isac}}, as illustrated in Figure~\ref{fig:intro-diagram}. 
To do so, we introduce the first dApp framework for the GPU-accelerated NVIDIA \gls{atb} 5G-NR stack. dApps are applications co-located with the \gls{gnb} that access \gls{phy}/\gls{mac} data and can directly influence \gls{ran} behavior on (sub-)millisecond timescales~\cite{D_Oro_2022,lacava2025dapps,ngrg-dapp-1,ngrg-dapp-2}. While CPU-based solutions are discussed in~\cite{lacava2025dapps}, in this paper, we design, develop, and open-source the first comprehensive dApp framework for the GPU-native NVIDIA \gls{atb} stack. 
This makes \gls{atb} into a platform where third-party dApps can access \blue{the \gls{phy} data produced on the \gls{gpu}} in real time and close tight control loops without compromising \gls{ran} performance, in a way that is aligned with the programmability and \gls{ai}-native objectives of the O-RAN and AI-RAN Alliances. \blue{As shown in Figure~\ref{fig:intro-diagram}, dApps run on the \gls{gnb} host and receive \gls{phy} data from an E3 Agent in the \gls{du}-Low through shared memory, closing control loops below $10$~ms, while the \glspl{ric} act on the \gls{gnb} at slower timescales through the E2 and A1 interfaces. The same data can also be stored by the \gls{adl} to build datasets and train the models deployed in the dApps.} \blue{We evaluate the framework on its own, independently of any specific application (Section~\ref{sec:evaluation}).}

To demonstrate the capabilities of this framework, we implement cuSense, an \gls{isac} dApp for person tracking. cuSense consumes real-time \gls{ul} \gls{csi} estimates from the \gls{gnb} and runs an \gls{ai} pipeline to detect channel perturbations caused by people moving and infer their two-dimensional position. We support this with an \gls{ota} data collection campaign and a labeling pipeline that combines \gls{5g} channel estimates with camera-derived ground truth, providing a concrete example of \gls{isac} over a \gls{3gpp}-compliant \gls{ran}. \blue{cuSense exercises the framework end to end, from data collection and training to live inference, and localizes a person who carries no device in one indoor environment.} %

This paper makes the following contributions:
\begin{itemize}[noitemsep, topsep=2pt, leftmargin=10pt]
  \item \textbf{GPU-accelerated dApp framework.} We design a real-time dApp framework for CUDA-accelerated \glspl{ran}\blue{, with a real-time export path and an E3 Agent inside the \gls{atb} L1 that expose \gls{phy} data, such as \gls{iq} samples, channel estimates, and \gls{srs},} via shared-memory buffers and an E3-based control/data-plane interface.
  \item \textbf{Reference dApp container design.} We realize this framework on NVIDIA \gls{atb} with a modular dApp container architecture comprising an in-\gls{gnb} E3 Agent, an out-of-\gls{gnb} E3 Manager, and a flexible dApp logic layer. We provide \blue{four inference engines (native TensorRT, Triton C~API, Triton gRPC, and embedded Python) supporting five \gls{ai} backends, and a standalone E3 Agent to develop and test dApps without a \gls{ran}.}
  \item \textbf{Experimental \gls{ota} evaluation\blue{ of the framework}.} \blue{We profile the framework on a live \gls{3gpp}-compliant \gls{ran}, where it takes about $84~\mu$s per slot and closes a control loop with inference in about $123~\mu$s at p50. We also evaluate how it scales with bandwidth, antennas, cells, and dApps, compare the \gls{gpu} sharing modes \gls{mig}, \gls{mps}, and time-slicing, and test its portability across the GH200 and the DGX Spark.}
  \item \textbf{cuSense \gls{isac} case study.} We implement cuSense, a \gls{csi}-based person-tracking dApp in a \gls{ul}-collaborative \gls{isac} setting\blue{ that validates the framework end to end}, and define comprehensive pipelines for data collection, inference, and testing. \blue{cuSense reaches a mean error of $73$~cm on unseen runs, runs live in $241~\mu$s with the native engine, and its preprocessing and \gls{cnn} run $4.8$ times faster on the \gls{gpu} than an optimized \gls{cpu} implementation on 16 cores. We also build a real-world \gls{ota} dataset with \gls{ul} \gls{5g} \gls{csi} records and camera-based ground truth.}
  \item \textbf{Open-source reference implementation.} We release the dApp framework and reference dApp container as open source\footnote{\url{https://github.com/NVIDIA/aerial-sample-apps}}, integrated with NVIDIA Aerial\footnote{\url{https://github.com/NVIDIA/aerial-cuda-accelerated-ran}}, providing a modular reference design and inference pipelines for dApps on a GPU-accelerated \gls{gnb}. \blue{The cuSense dataset is also publicly available\footnote{\label{fn:cusense-dataset}\url{https://huggingface.co/datasets/nvidia/aerial-isac-pusch-hest}}.}
\end{itemize}
The remainder of this paper is organized as follows. Section~\ref{sec:background} provides background and motivations for the dApp framework, which is introduced in Section~\ref{sec:design}. Section~\ref{sec:evaluation} evaluates the framework on \gls{atb}. Section~\ref{sec:cusense} introduces cuSense, and Section~\ref{sec:cusense-eval} its evaluation. Section~\ref{sec:related} discusses related work, and Section~\ref{sec:conclusion} concludes the paper.

\section{dApp Background and Motivation}
\label{sec:background}

This section reviews O-RAN and AI-native \glspl{ran}, introduces \glspl{dapp} and prior work, and summarizes NVIDIA \gls{atb} and \gls{adl} to motivate our design as illustrated in Figure~\ref{fig:intro-diagram}.

\textbf{AI-Native RAN Vision.}
Recent work in both standardization and industry is shifting towards an \gls{ai}-native \gls{ran} vision, where \glspl{gnb} are becoming programmable edge nodes that expose data and compute to \gls{ai} workloads. The O-RAN ALLIANCE promotes this transition through a disaggregated architecture with open interfaces and \glspl{ric} for data-driven control~\cite{polese2023understanding}, while initiatives such as the AI-RAN Alliance further refine how \gls{ai} workloads and services should be co-designed with the \gls{ran}~\cite{airan}.

\textbf{dApps: Real-Time Distributed Applications.}
\glspl{dapp} are lightweight applications, deployed directly on the \gls{gnb}, capable of accessing \gls{phy}/\gls{mac} telemetry and executing control actions at sub-$10$~ms timescales~\cite{D_Oro_2022}. The architecture in~\cite{lacava2025dapps} defines a new E3 interface to allow real-time interactions between \gls{ran} (e.g., \gls{cu} and \glspl{du}) and \glspl{dapp}, and an \gls{e3ap} to expose structured user-plane messages and control primitives, and demonstrates feasibility with use cases such as spectrum sharing and positioning on an \gls{oai}-based \gls{gnb}. The O-RAN nGRG research reports~\cite{ngrg-dapp-1,ngrg-dapp-2} further analyze dApp use cases, requirements, architecture, and interfaces.
\cite{neasamoni2025interforan} extends this line of work by embedding a GPU-accelerated dApp for \gls{ul} interference detection directly into the NVIDIA Aerial pipeline. 
In contrast, our work introduces a generic and comprehensive GPU-driven framework that exposes a shared-memory telemetry plane and structured E3 interface to support multiple data types and models, with dApps running outside the \gls{ran} process and decoupled from custom embedded kernels.

\begin{figure*}[htb]
  \centering
  \includegraphics[width=\textwidth]{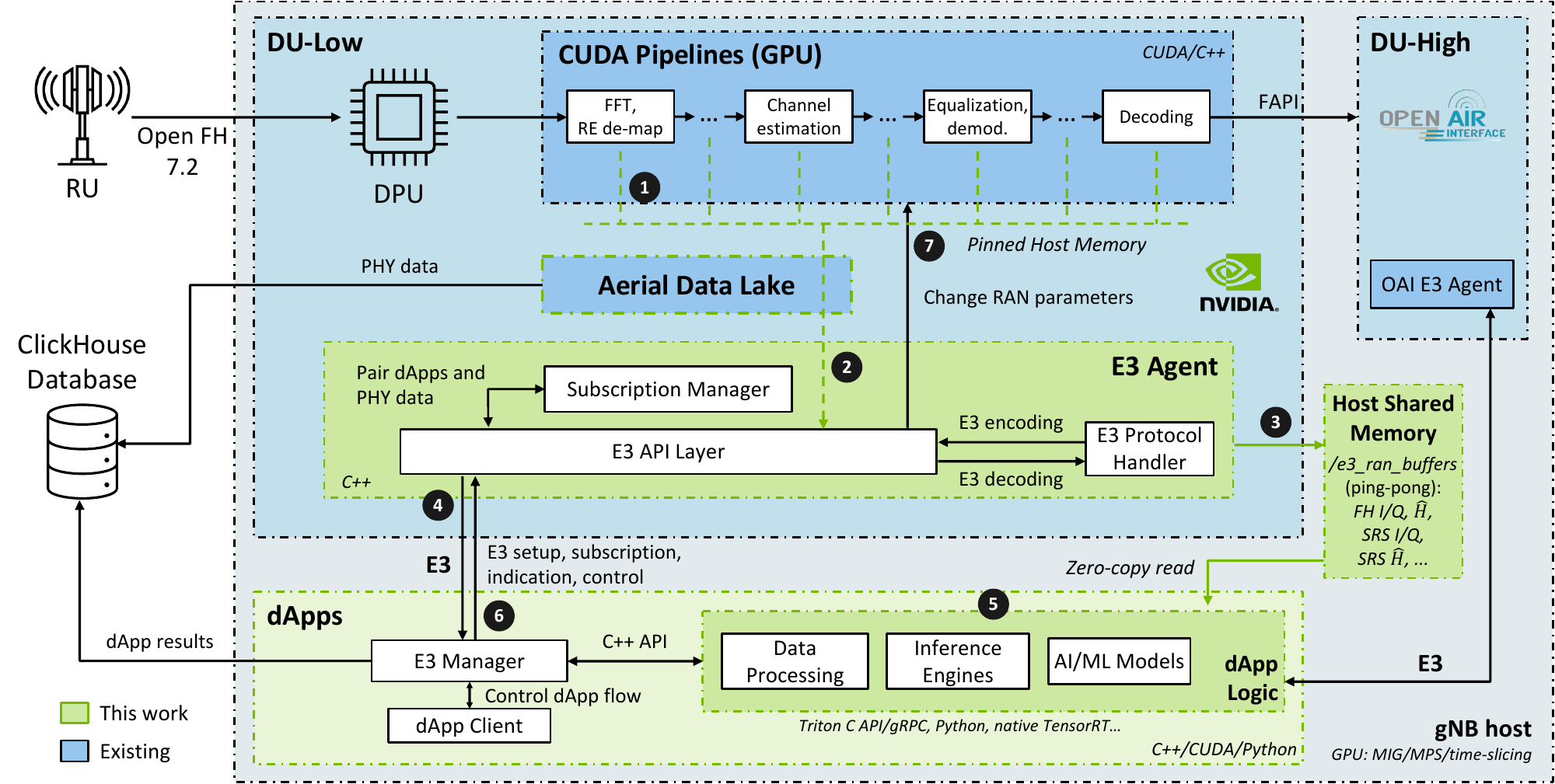}
  \caption{\blue{dApp framework integration in the NVIDIA Aerial Testbed gNB. Numbers mark the steps of the data path in Table~\ref{tab:datapath}.}}
  \vspace{-.15cm}
  \label{fig:high-level-diagram}
\end{figure*}

\textbf{NVIDIA Aerial Testbed.}
\label{subsec:atb-datalake}
NVIDIA \gls{atb} is a network stack that combines the NVIDIA Aerial in-line baseband accelerator\blue{, whose L1 runs on the \gls{gpu} through the cuPHY CUDA library,} with \gls{oai} L2 and 5G \gls{cn} open-source stack~\cite{kaltenberger2025driving} and commercial \glspl{ru}~\cite{atb}. It is designed as an \gls{ai}-enabled platform for \gls{ai}-\gls{ran} research, allowing developers to deploy and evaluate solutions on a commercial-grade, GPU-accelerated \gls{gnb}.
The \textbf{Aerial Data Lake}
is a data capture platform that records \gls{ota} signals and associated L2 metadata from \glspl{gnb} built on the Aerial stack, storing them in a database for offline post-processing~\cite{atb}. 

\vspace{-.3cm}
\subsection{Key Challenges and Requirements}
\label{subsec:challenges}

Enabling dApps on a GPU platform such as \gls{atb} raises several key challenges. They motivate the design of our \gls{gpu}-accelerated dApp framework, presented in Section~\ref{sec:design}.

\noindent \textbf{C1: Real-time data access.} dApps require low-latency access to selected data (e.g., \gls{iq} samples, channel estimates) without the overhead and latency of extra operations and remote \glspl{ric}.

\noindent \textbf{C2: Co-location with loose coupling.} dApps should be co-located with the \gls{gnb} to meet strict timing constraints, and also remain loosely coupled so they can be deployed and updated without modifying the production \gls{ran}.

\noindent \textbf{C3: GPU-native \gls{ai} tooling.} The framework should be able to take advantage of the existing \gls{gpu} infrastructure used by the \gls{ran} and support various \gls{ai} toolchains, providing a simple path to develop and deploy new models.

\noindent \textbf{C4: Isolation and scalability.} Third-party dApps must be isolated from the \gls{ran}, with controlled access to shared-memory regions and compute resources, and the framework should support multiple concurrent dApps without degrading \gls{ran} performance.

\noindent \textbf{C5: Alignment with standardization and existing frameworks.} Interfaces and message formats should align with existing frameworks and build on E3 abstractions and serve as a reference for ongoing O-RAN, AI-RAN, and \gls{3gpp} standardization efforts on AI-native \glspl{ran} and \gls{isac}.

\section{GPU-accelerated dApp Framework}
\label{sec:design}

This section describes our \gls{gpu}-accelerated dApp framework, including its main components and design choices.

\subsection{High-Level Architecture}

Figure~\ref{fig:high-level-diagram} shows the architecture of an \gls{atb} \gls{gnb}, extended to support our \gls{gpu}-accelerated dApp framework, with three main components: (i) \gls{du}-Low running the NVIDIA Aerial CUDA L1 \blue{pipelines}, \gls{adl}, and an E3 Agent component; (ii) \gls{du}-High and above running \gls{oai} open-source stack\blue{, connected to \gls{du}-Low through \gls{fapi}}; and (iii) one or more dApps.
\blue{The components introduced in this work are marked in green, together with the blocks they contain, while the others (e.g., Aerial CUDA pipelines, \gls{oai}, and ClickHouse) are existing components that we integrate or leverage.}
\blue{The numbered steps in Figure~\ref{fig:high-level-diagram} follow one \gls{ul} slot through the framework: \gls{phy} data are exported from \gls{gpu} memory to host shared memory (steps~1--3), signaled to the subscribed dApps with an E3 indication (step~4), and processed by the dApp logic (step~5), while an optional control message closes the loop back to the \gls{ran} (steps~6--7). The following subsections describe each part, and Section~\ref{sec:evaluation} quantifies each step.}
The entire framework runs on the same physical host, requiring partitioning of host resources, such as CPU core pinning and \gls{gpu} sharing mechanisms like NVIDIA \gls{mps} or \gls{mig}\blue{, or plain time-slicing}, to avoid degradation of \gls{ran} performance, which remains the highest priority.

\subsection{\blue{Real-Time Data Plane: ADL and Shared Memory}}
\label{subsec:rt-adl}

\begin{figure}[b]
  \centering
  \includegraphics[width=\linewidth]{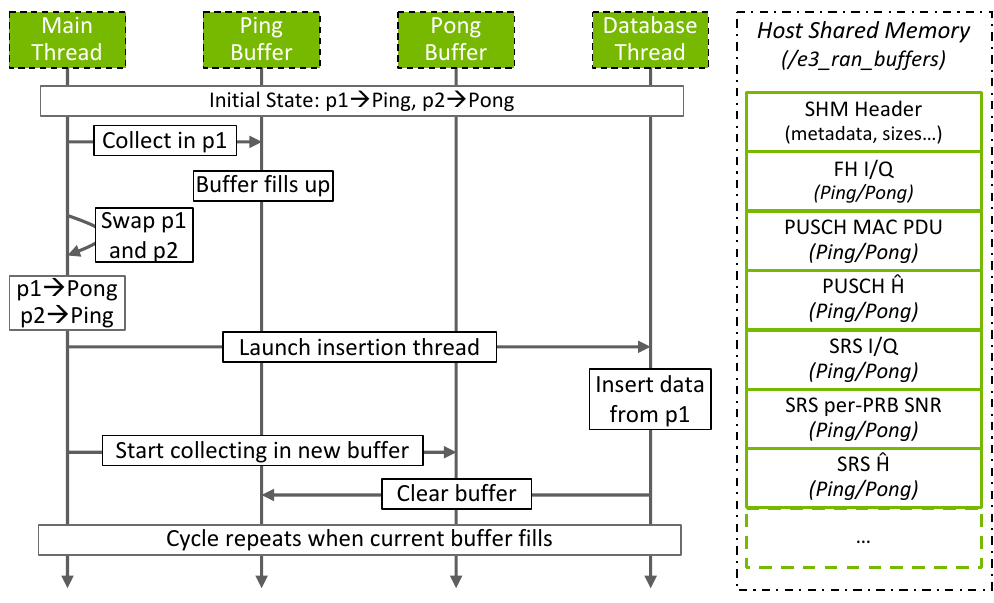}
  \caption{Aerial Data Lake ping-pong mechanism and \blue{host shared memory layout}.}
  \label{fig:adl-ping-pong}
\end{figure}

\textbf{Real-time Aerial Data Lake.}
To satisfy \textit{C1} (real-time data access), we extend \gls{adl}, discussed in Section~\ref{subsec:atb-datalake}, to a real-time version which uses a double buffering mechanism to capture and manage incoming data efficiently. As shown in Figure~\ref{fig:adl-ping-pong}, the main thread uses two pointers ($p1$ and $p2$) that point to two buffers ($ping$ and $pong$) and alternates writes between them. When one buffer is full, the thread swaps the pointers, starts collecting into the other buffer, and triggers a database insertion thread to write the completed buffer to the ClickHouse DB. The database can be configured at start time to use in-memory RAM tables or SSD storage. Since insertion is slower than capture, this ping-pong pattern allows continuous data collection without interruptions, as long as the buffers are large enough.

\textbf{E3 Agent Integration \blue{and Export Path}.}
We integrate an E3 Agent \blue{(Section~\ref{subsec:e3})} into the NVIDIA Aerial software, leveraging the same initial data path pipelines, triggering mechanisms (once per uplink \gls{tti}), and double-buffering abstraction of \gls{adl}.
\blue{The E3 Agent creates a POSIX shared-memory region in host memory that holds the ping-pong buffers with the structure shown in Figure~\ref{fig:adl-ping-pong},} enabling direct access by external dApps.
As illustrated in \blue{Figure~\ref{fig:high-level-diagram}}, when \gls{adl} and/or the E3 Agent are enabled, the \blue{cuPHY CUDA pipelines (e.g., \gls{pusch} and \gls{srs}) are} instructed to copy (\blue{step~1}) selected \gls{ul} data from device memory (GPU) into pinned host memory\blue{, i.e., page-locked CPU memory written directly by the \gls{gpu},} via an asynchronous CUDA \texttt{memcpy}. This allows the CPU to proceed with other work while the \gls{gpu} performs the transfers, minimizing the impact on the critical real-time L1 \gls{phy} path.
An atomic notification (\blue{step~2}) from the \blue{cuPHY} processing then triggers the \gls{adl} routines, which write (\blue{step~3}) the data from pinned host memory into the next \blue{row} of the appropriate \gls{shm} ping-pong buffer.
\blue{To keep this copy short, only the data streams requested by at least one active subscription are copied, \gls{iq} samples are limited to the \glspl{prb} of the configured \gls{ul} bandwidth, and the bulk copy can be split across helper threads pinned to dedicated cores, which we refer to as copy helpers. The cost of this copy grows with the amount of data copied per slot, as quantified in Section~\ref{sec:evaluation}. If a copy completes too late, when cuPHY may already reuse its pinned buffers, the slot is skipped, so that dApps never read partially written data.}
Although these \texttt{memcpy} routines are fast, an additional copy is currently required to decouple data ownership between \texttt{cuPHY} and \gls{adl}. We plan to optimize this path by removing this extra copy in future versions, e.g., by letting the \blue{cuPHY pipelines} copy directly into \gls{shm}.

\textbf{\blue{Exposed Data and Extensibility.}}
From the \blue{\gls{pusch} pipeline}, we export post-FFT, pre-equalization frequency-domain \gls{iq} samples and \gls{dmrs}-based \gls{hest}, \gls{mac} \glspl{pdu}, and a set of \gls{fapi} metadata\blue{. From the \gls{srs} pipeline, a second collector exports \gls{srs} \gls{iq} samples, \gls{srs} \gls{hest}, and per-\gls{prb} \gls{snr} at the \gls{srs} periodicity, with its own E3 indications. Both paths also expose per-\gls{ue} and per-cell measurements (e.g., \gls{snr} and noise variance) and an \gls{tai} timestamp for each slot. In multi-cell deployments, each cell writes its own buffer row, and a single indication per slot covers all cells.}
\blue{Each data source follows the same pattern, with its own collector, buffer regions, and stream identifiers, so that additional channels such as \gls{pucch} or \gls{prach} can be exposed without changes to the E3 procedures or to existing dApps. The full list of data streams is documented in the framework repository~\cite{aerial-sample-apps}.}

\textbf{Shared Memory.}
This pointer-based access pattern provides zero-copy data sharing between \gls{ran} and dApps, enabled by a copy path that decouples from critical GPU signal processing pipelines. Placing the buffers in host \gls{shm} also ensures portability across different deployment configurations. While NVIDIA provides direct GPU-to-GPU mechanisms, such as NVLink for peer-to-peer access and GPUDirect RDMA for inter-node communication, these require specific hardware topologies and do not support memory sharing across \gls{mig} partitions, which present isolated memory spaces even within the same physical GPU. By routing the data through pinned host memory and exposing it via \gls{shm}, our design provides a general access path for dApps regardless of whether they execute on a different \gls{mig} partition, a separate GPU, or the CPU. This also allows the same data to be accessed safely by multiple concurrent dApps while still meeting the low-latency requirements of \gls{isac} and other sensitive applications.

\textbf{\blue{Data Access and Buffer Lifetime.}}
\blue{\gls{adl} and the E3 Agent act as read-write producers, while dApps act as read-only consumers. \gls{gpu}-based dApps can further register the region with CUDA and read it in place from their kernels, avoiding an additional copy into dApp memory.}
Additionally, buffer sizes are configurable at startup, allowing for a trade-off between the amount of accessible history and memory usage\blue{, since each row remains valid until the other buffer of its ping-pong pair is filled. dApps can therefore read short histories in place, which avoids extra copies but is only safe as long as they consume the data before it is overwritten. Applications that need longer temporal contexts (e.g., micro-Doppler or gesture recognition over tens to hundreds of milliseconds) can instead accumulate per-slot inputs in their own memory and run inference once the window is complete, which decouples them from the buffer lifetime. cuSense supports both options (Section~\ref{subsec:cuSense-inference}).}
\gls{adl} and the E3 Agent share the same initial data paths and buffering mechanism, however, they can be enabled independently without changes to the \gls{ran}.
\blue{The same mechanism could also carry large payloads in the opposite direction, from dApps back to the \gls{ran} (e.g., beamforming weights), which we leave for future work.}

\vspace{-.3cm}
\subsection{E3 Agent and E3 Manager}
\label{subsec:e3}

The E3 Agent and E3 Manager implement the communication over the recently proposed E3 interface between a \gls{ran} node and a dApp, respectively, following the \gls{e3ap} procedures described in~\cite{lacava2025dapps} of setup, subscription, indication, and control (\textit{C5: Standardization}). 

\textbf{\blue{Indications and Control.}}
\blue{Once the data of a slot are in \gls{shm} (step~3), the E3 Agent} packs and sends the corresponding pointers and metadata to subscribed dApps via E3 Indication messages (\blue{step~4}). In particular, for small scalar values or metadata (e.g., slot index, cell ID), the indication carries the value directly. For larger data structures stored in \gls{shm}, the indication includes: (i) the data type (e.g., \gls{iq}, \gls{hest}); (ii) the buffer index (ping or pong); (iii) the \blue{row} within the buffer\blue{, i.e., the slot it was written to}; and (iv) \blue{per-\gls{ue} offsets and sizes within that row}. The dApp can then read the referenced memory while \gls{adl} and the E3 Agent prepare the next batch. \blue{In the opposite direction, dApps send E3 Control messages (step~6), which the E3 Agent decodes and hands over to the targeted \gls{ran} component, which can then apply them, e.g., by changing \gls{ran} parameters (step~7).}

\textbf{\blue{Transport and Encoding.}}
\blue{E3 messages are exchanged over \gls{zmq} sockets, either over \gls{tcp} or, for dApps co-located with the \gls{ran}, over \gls{ipc}. When several dApps are connected, each dApp can subscribe only to the messages addressed to it, so that the E3 Agent does not deliver the indications of other dApps. Messages are currently encoded in JSON, whose encoding and decoding time grows with the message size (Section~\ref{sec:evaluation}). Adopting a more compact binary encoding (e.g., ASN.1) is planned to reduce this cost and to align with ongoing standardization efforts.}

\begin{figure}[t]
  \centering
  \includegraphics[width=\linewidth]{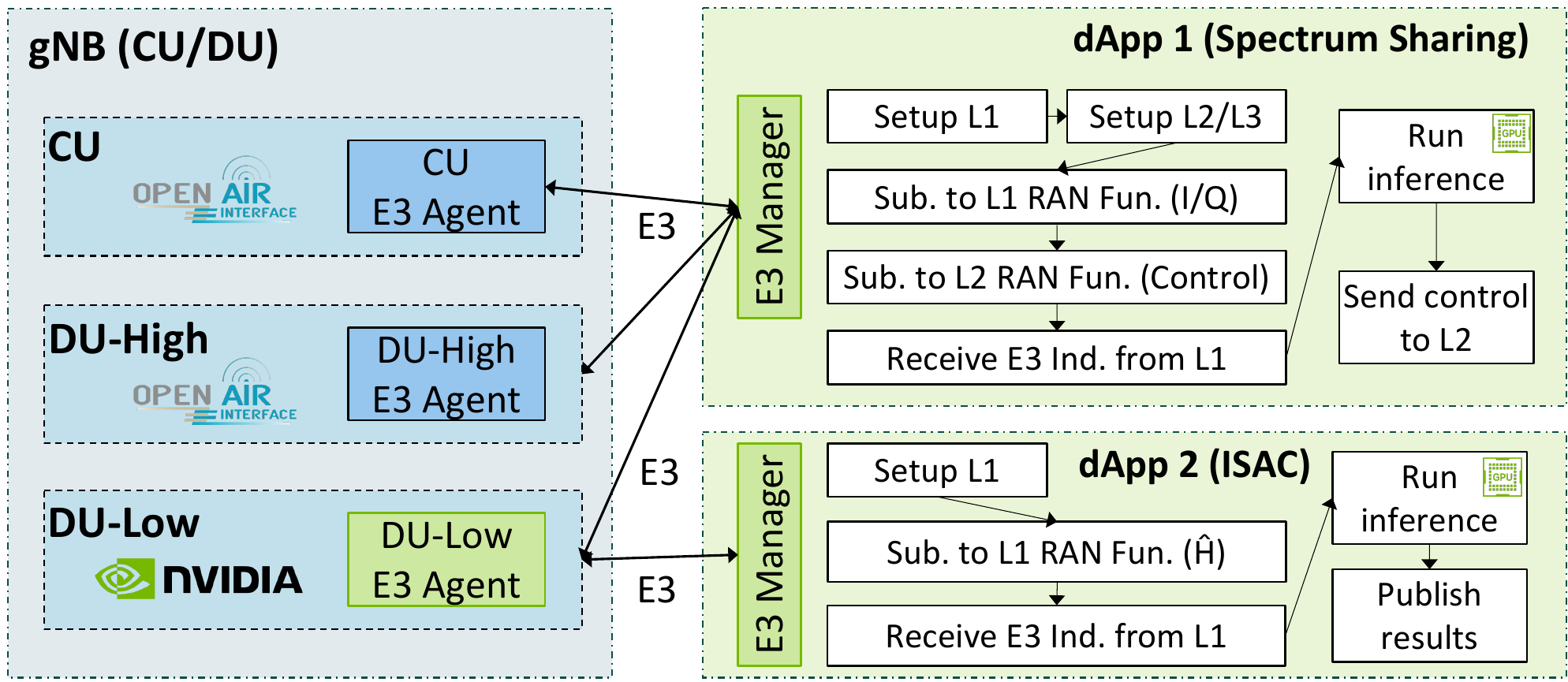}
  \caption{\blue{Support for multiple \gls{ran} nodes, E3 Agents, and dApps in the same gNB.}}
  \label{fig:multi-e3}
\end{figure}

\textbf{\blue{Multiple Agents and dApps.}}
As shown in Figure~\ref{fig:multi-e3}, our framework supports multiple E3 Agents running within a \gls{gnb}, one for each \gls{ran} \blue{node} (e.g., NVIDIA DU-Low, \gls{oai} DU-High)\blue{, each exposing the \gls{ran} functions of its node}.
An E3 Manager can independently register with multiple E3 Agents and create multiple subscriptions, for example, at different sampling rates or telemetry streams. This design enables system scalability (\textit{C4: Scalable}), allowing additional agents to be added without modifying existing dApps and for multiple dApps to subscribe to the same telemetry at different rates.
Figure~\ref{fig:multi-e3} shows two dApp examples: a spectrum sharing dApp that subscribes to L1 telemetry and sends control messages to L2 for scheduling decisions, and an \gls{isac} dApp like cuSense that only consumes channel estimates for inference without closing the control loop.

\subsection{dApp Reference Architecture}
\label{subsec:dapp-container}

\begin{figure}[t]
  \centering
  \includegraphics[width=\linewidth]{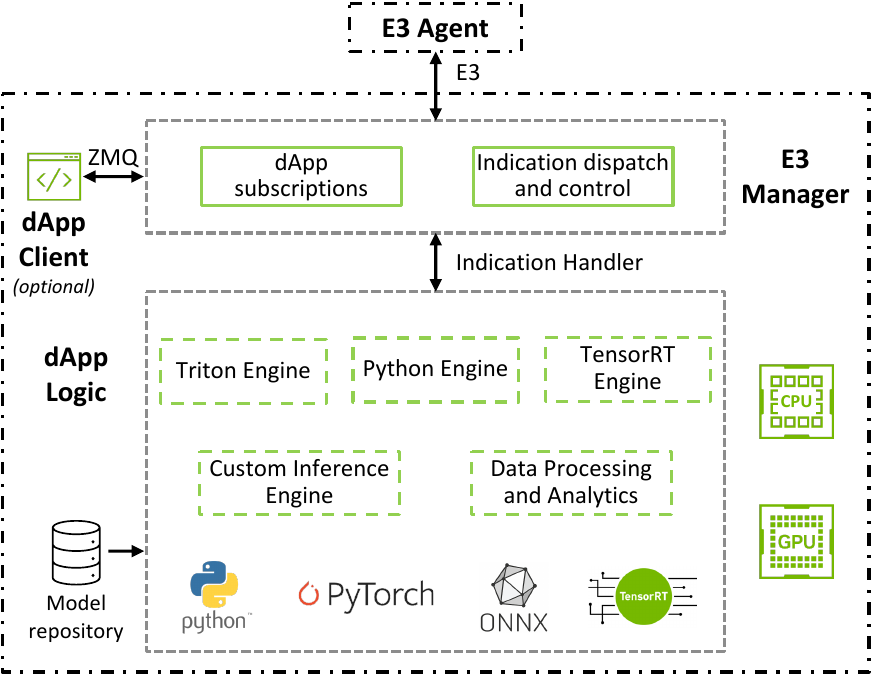}
  \caption{dApp container reference architecture with E3 Manager, dApp logic, and \blue{optional} dApp client.}
  \label{fig:dapp-container}
\end{figure}

To address \textit{C3 (GPU-native \gls{ai} tooling)} and \textit{C4 (isolation and scalability)}, we design a reference dApp container architecture with three main components, as shown in Figure~\ref{fig:dapp-container}: (i) an E3 Manager, the required dApp orchestrator; (ii) a dApp logic, the flexible application layer for data processing and inference; and (iii) \blue{an optional} dApp client, an application controller. This design serves as a concrete blueprint for deploying a \gls{gpu}-accelerated dApp within a single container, while allowing developers to modify the source code and internal layout to their specific needs. \blue{In Figure~\ref{fig:high-level-diagram}, the E3 Manager receives the indications (step~4) and sends the control messages (step~6), while the dApp logic executes step~5.} The three components separate concerns within the dApp and have the following roles.

\textbf{E3 Manager.}
The E3 Manager is a mandatory component that serves as the dApp orchestrator, interacting with the \gls{ran} nodes via the E3 interface. It manages all \gls{e3ap} communication with the E3 Agent (e.g., E3 setup and subscription procedures) and interactions with the dApp Client to implement the desired behavior, including application-specific \glspl{sm}. Additionally, it receives E3 Indications and dispatches them\blue{, in order of arrival,} to a registered application handler that executes the dApp-specific application logic. The E3 Manager does not contain any application-specific logic, and its procedures can be re-used by any dApp implementation. Its role is to decouple the E3 protocol from whatever processing the dApp performs.

\textbf{dApp logic.}
The dApp logic is the application-specific component registered with the E3 Manager. It receives an indication context for each E3 Indication and implements the desired processing. This processing is not limited to \gls{ai}/\gls{ml} inference, as a dApp may perform direct data processing such as statistical analysis or rule-based decisions on the received telemetry, or it may invoke an inference engine to run \gls{ai} models on the data, as shown in Figure~\ref{fig:dapp-container}. This separation allows developers to build arbitrarily complex or lightweight applications without modifying the E3 Manager or the underlying protocol stack.
As reference implementations, we provide \blue{four} inference engine configurations: (i)~a Triton C~API engine, where NVIDIA Triton Inference Server~\cite{triton} runs in-process \blue{with low overhead}; (ii)~a Triton gRPC engine, where Triton runs as a separate process with standard management and monitoring capabilities; (iii)~an embedded Python engine using \textit{pybind11}, bypassing Triton for rapid prototyping\blue{; and (iv)~a native \gls{trt} engine, also without Triton, which takes its inputs directly from the registered \gls{shm} and executes the model as a CUDA graph for the lowest latency}. \blue{The first three} engines support multiple \gls{ai} backends including Python/NumPy running on \gls{cpu}, and PyTorch, LibTorch, \gls{onnx} Runtime, and \gls{trt} running on \gls{gpu}.
Triton is an open-source inference serving software platform developed by NVIDIA and optimized for \gls{gpu}-accelerated inference. It exposes a model repository with support for multiple backends running on \gls{cpu} or \gls{gpu}, and provides \gls{grpc}/\gls{http} and C \glspl{api} for model life-cycle management, scheduling, and inference execution. It also supports combining different backends or running multiple models in parallel or in sequence, providing a flexible playground for benchmarking and experimentation. Additional engines or custom processing pipelines can be integrated following the same pattern. The performance characteristics of these configurations are evaluated in Section~\ref{sec:evaluation}.

\textbf{dApp client.}
The dApp client is a user-level component that allows a user to manage the dApp lifecycle at runtime by communicating with the E3 Manager. It provides commands to: (i) set up and release E3 Agent connections; (ii) create and delete subscriptions with parameters such as target agent, telemetry IDs, periodicity, duration, and inference model; and (iii) query the current status of agents and active subscriptions. This enables external control of any dApp implementation without modifying its internal components.

This dApp design enables flexibility, as shown in Section~\ref{sec:evaluation}, and simplifies the modular deployment of different processing and inference pipelines, as demonstrated by the cuSense \gls{isac} use case in Section~\ref{sec:cusense}. The source code is released as open source\blue{~\cite{aerial-sample-apps}}, with documentation, guides, and inline code comments, providing developers with reference points for developing and deploying their own dApp use cases\blue{, while some of the features described in this section will become available in upcoming releases}.

\begin{figure}[b]
  \centering
  \includegraphics[width=\linewidth]{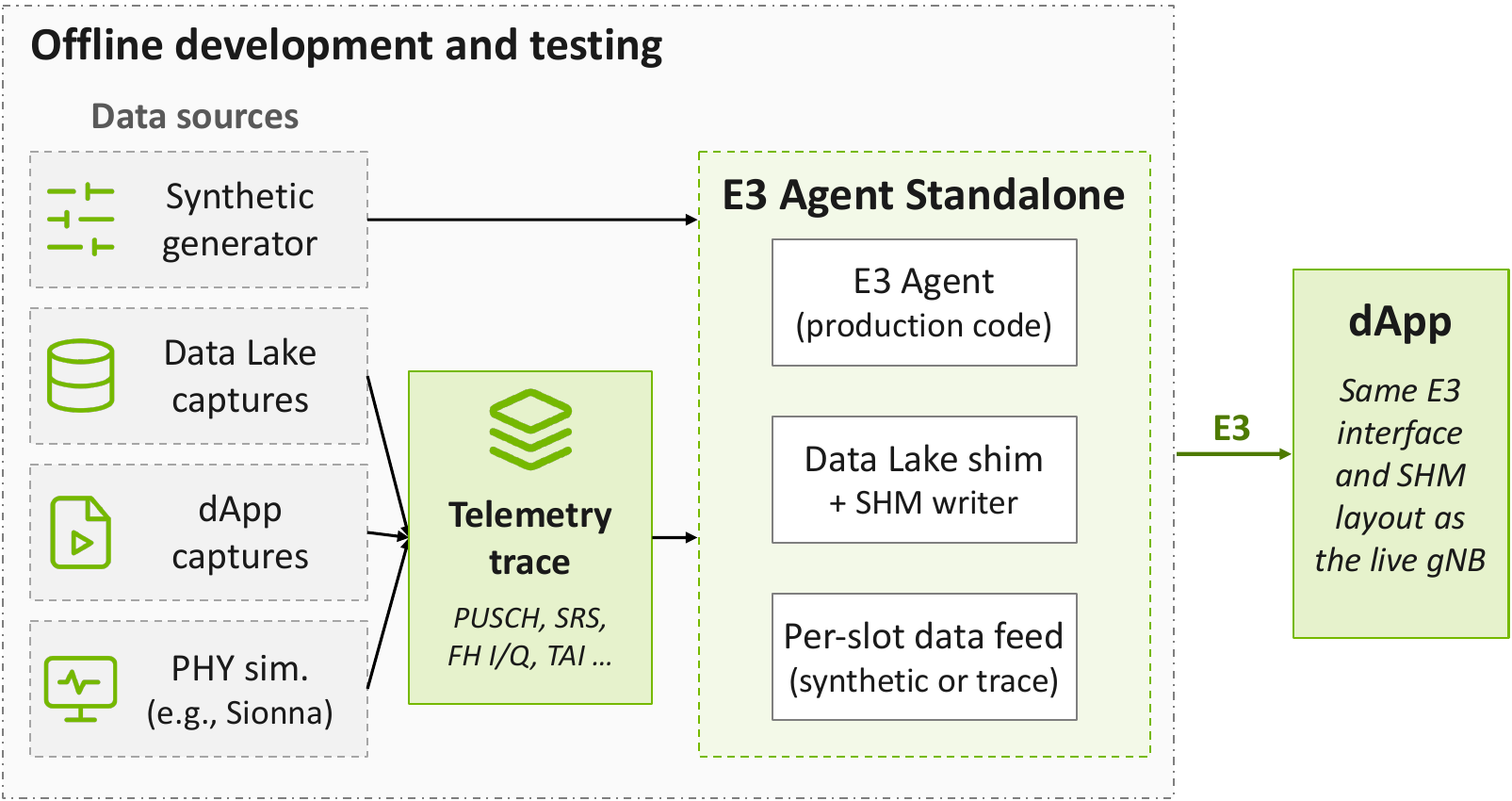}
  \caption{\blue{E3 Agent Standalone: dApps are developed and tested on synthetic or replayed telemetry through the same E3 interface used with the live gNB.}}
  \label{fig:standalone}
\end{figure}

\subsection{\blue{Development and Testing Without a RAN}}
\label{subsec:standalone}
\blue{Developing and debugging dApps on a live \gls{gnb} can be time-consuming and hard to reproduce, since every test requires the full \gls{ran} stack and radio hardware, and radio conditions and traffic can change between runs. To address this, we provide the E3 Agent Standalone tool, shown in Figure~\ref{fig:standalone}, which allows dApps to be developed and tested offline, without Aerial, \gls{oai}, or radio hardware. It builds the production E3 Agent code without modifications and replaces the \gls{adl} with a lightweight shim that exposes the same buffer descriptors to the agent. A \gls{shm} writer fills the ping-pong rows with the layout used by the live \gls{gnb} (Section~\ref{subsec:rt-adl}), and a per-slot data feed publishes the data of each \gls{ul} slot following the configured \gls{tdd} pattern. The \gls{shm} layout is checked against the \gls{adl} definitions at build time, so that the two cannot diverge. As a result, a dApp connects to the standalone agent exactly as to a live \gls{gnb}, with the same E3 procedures, subscriptions, indication schema, and processing code, and moves to a live deployment unchanged.}

\blue{The per-slot data feed operates in two modes. In synthetic mode, it generates telemetry for one or more cells from a configuration file. In replay mode, it reads a self-describing telemetry trace, extensible to new data types, that currently carries \gls{pusch}, \gls{srs}, and \gls{fh} \gls{iq} data, together with per-\gls{ue} metrics and \gls{tai} timestamps, for one or more cells, and paces the slots according to the recorded \gls{tai} timing. Traces can come from different data sources, such as \gls{adl} captures stored in ClickHouse, dApp captures, and \gls{phy} or ray-tracing simulators (e.g., Sionna~\cite{hoydis2022sionna}), which provide data with known ground truth, and other sources can be added by exporting to the same trace format. This enables deterministic replay for debugging and regression testing, the comparison of different dApps on identical inputs, and a direct transition from offline development to live deployment. The E3 Agent Standalone tool is released as part of NVIDIA Aerial~\cite{aerial-acar}.}

\vspace{-.15cm}
\section{Profiling dApp Framework Performance}
\label{sec:evaluation}

We now evaluate the performance of the proposed framework, integrated with \blue{NVIDIA Aerial, focusing on the options available to dApp developers: (i) the latency of each step of the data path and of the full control loop; (ii) how this latency scales with the number of cells, concurrent dApps, and payload size; (iii) the inference engines of Section~\ref{subsec:dapp-container} under different \gls{gpu} sharing modes; and (iv) portability across two \gls{gpu} platforms, NVIDIA GH200 and DGX Spark. All experiments run on the live \gls{atb} \gls{gnb} with a connected \gls{ue}, using a simple reference dApp in this section and the cuSense \gls{isac} dApp in Section~\ref{sec:cusense}.}

\subsection{Aerial Testbed Experimental Setup}
\label{subsec:setup}

\textbf{System configuration.} We perform our integration and evaluation on a standard \gls{atb} node as described in the documentation~\cite{atb}. The baseline setup consists of: (i) a single NVIDIA GH200 \gls{gh} server for the \gls{ran} workloads, featuring a 72-core Grace CPU, an NVIDIA H100 Tensor Core \gls{gpu}, two BlueField-3 \glspl{dpu}, and two ConnectX-7 \glspl{nic}; (ii) a VIAVI Qulsar QG-2 device as grandmaster synchronization clock; (iii) an NVIDIA Spectrum-2 SN3750-SX as \gls{fh} switch; (iv) a \blue{4T4R WNC} \gls{cbrs} \gls{fr1} \gls{ru} centered at \blue{$3.625$}~GHz with $100$~MHz of bandwidth; and (v) a Pixel 8 acting as \gls{ue}.
The complete \gls{gnb} stack, comprising NVIDIA Aerial L1, \gls{oai} L2 and above, the dApp framework, and \gls{oai} \gls{cn}, runs within the GH200 node.
Additionally, we validate the framework on an NVIDIA DGX Spark, a compact desktop system powered by the NVIDIA GB10 Superchip. It features a 20-core Arm CPU (10 Cortex-X925 high-performance cores up to 4~GHz and 10 Cortex-A725 efficiency cores), an integrated Blackwell \gls{gpu}, 128~GB of unified LPDDR5x memory, and a ConnectX-7 \gls{nic}. \blue{It runs the same NVIDIA Aerial L1, \gls{oai} L2, and dApp framework, while the \gls{oai} \gls{cn} runs on a separate host to leave more CPU cores to the \gls{ran} and the dApps.}
The framework has also been validated with other L2+ stacks such as \gls{odc}\blue{ and OCUDU. All components are Release builds of the public repositories~\cite{aerial-acar,aerial-sample-apps}.}

\blue{\textbf{GPU sharing modes and CPU allocation.} On the GH200, we evaluate three ways of sharing the \gls{gpu} between L1 and the dApps: (i) \gls{mig}, where the \gls{gpu} is partitioned into a \textit{4g.48gb} instance dedicated to L1 and a \textit{3g.48gb} instance dedicated to the dApps, isolating their memory and compute resources; (ii) \gls{mps}, where the dApps join the \gls{mps} server that L1 already uses for its own pipelines, so that their kernels run concurrently with L1 on the full \gls{gpu}; and (iii) time-slicing, where the dApps keep their own CUDA context and the \gls{gpu} alternates between contexts. DGX Spark does not support \gls{mig}, so the dApps share the \gls{gpu} through \gls{mps}, with the dApp CUDA context created before L1 starts, or through time-slicing.
On both platforms, L1, L2, and the operating system run on separate sets of isolated cores, and the E3 Agent, the \gls{adl} export threads, and the copy helper threads (Section~\ref{subsec:rt-adl}) run on further dedicated cores, which are configurable. On the GH200, the E3 Manager threads of the dApp are pinned to their own core, while on DGX Spark the whole dApp container is pinned to a single high-performance core, and the E3 Agent and the \gls{adl} threads run on efficiency cores.}

\begin{table*}[t]
    \centering
    \caption{\blue{Data path steps of Figure~\ref{fig:high-level-diagram} and their latency for one \gls{iq} slot (734~KB) processed by PRB Power with the native \gls{trt} engine on the GH200 with \gls{mps}, three copy helpers, and \gls{zmq} over \gls{ipc}. The p99 including step~1 is an upper bound.}}
    \label{tab:datapath}
    \scriptsize
    \setlength{\tabcolsep}{4pt}
    \def\sub#1{\hspace{0.9em}\textcolor{black!60}{#1}}
    \def\sv#1{\textcolor{black!60}{#1}\hspace{0.6em}}
    \begin{tabular*}{\textwidth}{@{\extracolsep{\fill}}l l l l l r r@{}}
        \toprule
        \textbf{Phase} & \textbf{Step} & \textbf{What happens} & \textbf{Scales with} & \textbf{Lever} & \textbf{p50 [$\mu$s]} & \textbf{p99 [$\mu$s]} \\
        \midrule
        Export & 1. \gls{gpu} $\rightarrow$ pinned host & cuPHY copies the \gls{ul} buffers to the host & bytes, cells & sharing mode & 26.0 & 36.3 \\
         & 2. Notify \gls{adl} & cuPHY wakes the \gls{adl} export thread & -- & dedicated core & 2.0 & 3.3 \\
         & 3. Copy to host \gls{shm} & & bytes, streams, cells & selective copy, copy helpers & 13.4 & 14.9 \\
         & \sub{3a. Metadata} & \textcolor{black!60}{slot, cell, and per-\gls{ue} metadata} & & & \sv{0.6} & \sv{0.9} \\
         & \sub{3b. Bulk copy} & \textcolor{black!60}{subscribed streams into the ping-pong row} & & & \sv{12.4} & \sv{13.8} \\
        \addlinespace[2pt]\arrayrulecolor{black!15}\midrule[0.3pt]\arrayrulecolor{black}\addlinespace[2pt]
        Indication & 4. E3 indication & & message size, dApps & \gls{ipc}, per-dApp filter & 23.8 & 27.4 \\
         & \sub{4a. Encode} & \textcolor{black!60}{E3 Agent builds the indication} & & \textcolor{black!60}{binary encoding (planned)} & \sv{4.5} & \sv{5.8} \\
         & \sub{4b. Send} & \textcolor{black!60}{\gls{zmq} publish} & & & \sv{2.6} & \sv{3.7} \\
         & \sub{4c. Deliver} & \textcolor{black!60}{\gls{zmq} transfer to the dApp} & & & \sv{13.4} & \sv{16.0} \\
         & \sub{4d. Decode} & \textcolor{black!60}{E3 Manager parses the indication} & & & \sv{2.9} & \sv{3.6} \\
        \addlinespace[2pt]\arrayrulecolor{black!15}\midrule[0.3pt]\arrayrulecolor{black}\addlinespace[2pt]
        Compute & 5. dApp processing & handler: \gls{shm} read, inference, output & model, engine & engine, sharing mode & 38.7 & 43.7 \\
        \addlinespace[2pt]\arrayrulecolor{black!15}\midrule[0.3pt]\arrayrulecolor{black}\addlinespace[2pt]
        Control & 6. E3 control & E3 Manager sends the control to the agent & dApps & \gls{ipc} & 15.3 & 17.4 \\
         & 7. Apply control & handed over to the targeted \gls{ran} component & -- & -- & 3.1 & 4.1 \\
        \addlinespace[2pt]\midrule\addlinespace[2pt]
        End to end & \multicolumn{4}{l}{One-way E3 path (steps~2--4)} & 39.2 & 43.7 \\
         & \multicolumn{4}{l}{Closed loop without processing (steps~2--4, 6--7)} & 57.8 & 63.4 \\
         & \multicolumn{4}{l}{Closed loop (steps~2--7)} & 96.5 & 104.4 \\
         & \multicolumn{4}{l}{Closed loop including step~1 (steps~1--7)} & 122.5 & $\leq$140.7 \\
        \bottomrule
    \end{tabular*}
\end{table*}
\textbf{Test configuration\blue{ and methodology}.} All benchmarking experiments use a single connected \gls{ue} generating $100$~Mbps of \gls{ul} \gls{udp} target traffic. The \gls{tdd} pattern is \textit{DDDDDDSUUU}, giving three \gls{ul} slots per half-frame (slots 7, 8, 9 and 17, 18, 19)\blue{, i.e., $600$ \gls{ul} slots and indications per second, arriving in bursts of three consecutive slots $0.5$~ms apart}. \blue{Since the dApps run on the same host as the \gls{gnb}, they connect to the E3 Agent with \gls{zmq} over \gls{ipc} instead of the default \gls{tcp} (Section~\ref{subsec:e3}). Each configuration is measured in a separate $60$~s run with a single active subscription, whose data streams (\gls{iq} samples, \gls{hest}, or all streams) are stated with each result, and the first $2$~s are discarded as warmup. The first inference after loading a model is slower, from $2.5$~ms with \gls{trt} to about $280$~ms with backends compiled at run time, and is avoided by preloading the models.}
\blue{We timestamp each step of the data path (Table~\ref{tab:datapath}) with lightweight instrumentation in the \gls{adl}, the E3 Agent, and the E3 Manager. All components run on the same host and use the same clock, and records are kept in memory and written out once per second from a non-isolated core, so that the instrumentation does not interfere with the measured path. Step~1, a \gls{gpu} copy, is timed separately with CUDA events in cuPHY. The dApp processing (step~5) is measured around the dApp handler, from the dispatch of the decoded indication to its return, including \gls{shm} access, input preparation, inference, and output processing. To measure the transport alone, the dApp skips its processing, and to close the loop the E3 Manager returns one echo control message per indication, which the E3 Agent decodes and hands over without changing \gls{ran} parameters. We report the median (p50) and the 99th percentile (p99) of each measurement.}

\textbf{Reference dApp\blue{: PRB Power}.} Our benchmark model PRB Power accepts as input a tensor of \gls{iq} samples with dimensions $[4, 14, 273, 12, 2]$, corresponding to $4$ antenna ports, $14$ \gls{ofdm} symbols, $273$ \glspl{prb}, $12$ subcarriers per \gls{prb}, and $2$ values (the $I$ and $Q$ components) in \gls{fp16} format. This corresponds to one full \gls{ul} slot of $100$~MHz bandwidth data with numerology~$\mu=1$.
The model computes the average power per \gls{prb} using
\begin{equation}
P_{\text{PRB}}(p) = \frac{1}{N_a N_s N_{sc}} \sum_{a=1}^{N_a} \sum_{s=1}^{N_s} \sum_{k=1}^{N_{sc}} \left( I^2_{a,s,p,k} + Q^2_{a,s,p,k} \right),
\end{equation}
where $P_{\text{PRB}}(p)$ is the average power for \gls{prb} $p \in \{1, ..., 273\}$, $N_a~=~4$ the number of antennas, $N_s~=~14$ the number of \gls{ofdm} symbols, $N_{sc}~=~12$ the number of subcarriers per \gls{prb}, and $I_{a,s,p,k}$ and $Q_{a,s,p,k}$ are the \gls{iq} components. The output is a $273$-element \gls{fp32} vector containing the power measurements for each \gls{prb}. \blue{PRB Power is a \gls{phy} telemetry application representative of use cases such as interference detection, spectrum monitoring, and \gls{isac}, and it is implemented for all engines and backends of Section~\ref{subsec:dapp-container}.}

\subsection{\blue{Data Path and Control-Loop Latency}}
\label{subsec:control-loop}

\blue{Table~\ref{tab:datapath} breaks down the data path of Figure~\ref{fig:high-level-diagram} into its steps for the best configuration we measured: one \gls{iq} slot of 734~KB, processed by PRB Power with the native \gls{trt} engine on the GH200 with \gls{mps}. From the copy of the slot out of \gls{gpu} memory to the delivery of the control message, the full loop takes $122.5~\mu$s at p50 and at most $141~\mu$s at p99 and $149~\mu$s at p99.9, about a quarter of a $500~\mu$s slot. The framework accounts for $84~\mu$s of this time and the dApp processing for $38.7~\mu$s. Therefore, the framework leaves nearly the full $0.5$~ms between the \gls{ul} slots of a burst to the dApp for its own logic, while heavier processing can run on a subset of the slots or batch several of them (Section~\ref{subsec:engine-comparison}). Figure~\ref{fig:datapath-bars} shows how this breakdown changes with the subscribed data streams and the \gls{gpu} sharing mode. Across all runs of this section, no slot, indication, or control message was lost.}

\blue{\textbf{Export (steps 1--3).} Step~1 takes $26~\mu$s without \gls{mig} and $40~\mu$s with it, regardless of the subscription, since cuPHY copies all \gls{pusch} outputs (about 1~MB per slot) to pinned host memory. The $15~\mu$s added by \gls{mig} is consistent with the L1 instance having only $4$ of the $8$ copy engines of the \gls{gpu}, and copying only the subscribed outputs in this step is left for future work. The notification (step~2) takes $2~\mu$s. Step~3 copies only the subscribed streams, so its cost grows with the bytes per slot: $8$, $15$, and $22~\mu$s for \gls{hest} (314~KB), \gls{iq} samples (734~KB), and all \gls{pusch} streams (1.07~MB), respectively, i.e., about $20~\mu$s/MB with three copy helpers. The same 1.07~MB copy takes $72~\mu$s on a single thread and $38~\mu$s with one helper, which makes the number of helpers the main lever for this step. On the \gls{srs} path, step~1 takes $16~\mu$s. The \gls{srs} collector copies its $324$~KB on a single thread in $22~\mu$s.}

\blue{\textbf{Indication and control (steps 4, 6, and 7).} E3 indications carry pointers and metadata, while the payload of the subscribed streams stays in \gls{shm}, so their cost does not grow with the payload size: across the configurations of Figure~\ref{fig:datapath-bars}, the indication (step~4) takes $20$--$26~\mu$s, the control (step~6) $12$--$16~\mu$s, and its hand-over to the \gls{ran} (step~7) $3~\mu$s. As shown in Table~\ref{tab:datapath}, the \gls{zmq} delivery to the dApp (step~4c) dominates with $13~\mu$s, while encoding and decoding the 256-byte JSON indication of the \gls{iq} subscription take $4.5$ and $2.9~\mu$s. Encoding and decoding costs grow with the amount of metadata, which matters for metadata-heavy subscriptions and many cells (Section~\ref{subsec:scaling}), and a planned binary encoding (e.g., ASN.1) will reduce them. \gls{zmq} over \gls{tcp}, which can also serve dApps on remote hosts, adds about $15~\mu$s to the closed loop compared to \gls{ipc}.}

\blue{\textbf{Compute (step 5).} With the native \gls{trt} engine, PRB Power processes one \gls{iq} slot in $39~\mu$s with \gls{mps} and $43~\mu$s with \gls{mig}. This is the largest single step of the loop and the only one that depends on the dApp logic, i.e., on the processing and model of its use case. We compare engines and models in Section~\ref{subsec:engine-comparison}.}

\begin{figure}[t]
    \centering
    \setlength\fwidth{0.67\linewidth}
    \setlength\fheight{0.52\linewidth}
    \input{figures_tex/datapath_bars}
    \caption{\blue{Per-step p50 latency of the data path for different subscriptions and \gls{gpu} sharing modes, on the GH200. Labels give the closed loop including step~1 as p50/p99 in $\mu$s; step~1 (hatched) is measured separately and its p99 is added.}}
    \label{fig:datapath-bars}
\end{figure}

\subsection{\blue{Scalability}}
\label{subsec:scaling}

\blue{Figure~\ref{fig:scaling-bars} shows how the data path scales with the payload of a cell, the number of cells, and the number of dApps, without dApp processing, so that only the framework is measured.}

\blue{\textbf{Bandwidth and antennas.} From $40$~MHz with $2$~Rx antennas to $100$~MHz with $4$~Rx antennas, i.e., from $213$~KB to $1.07$~MB of \gls{pusch} streams per slot, the closed loop grows only from $69$ to $91~\mu$s at p50. The increase comes from the two copies, step~1 ($19$ to $27~\mu$s) and step~3 ($9$ to $22~\mu$s), whose size follows the configured \gls{ul} bandwidth and the number of antennas, while the E3 messages are unchanged. At the same bandwidth, the numerology changes the slot duration and its payload in the same proportion, so the two copies scale with the slot duration, while the E3 messages remain one per slot. Massive \gls{mimo} cells mainly increase the same two copies: we estimate that a 32T32R cell, with about $10$~MB per slot, would cost about as much as the $10$ cells of Figure~\ref{fig:scaling-bars}, which makes selective copies and a direct copy from the \gls{gpu} into \gls{shm} the main optimizations.}

\blue{\textbf{Cells.} We connect up to $10$ cells of $100$~MHz with $4$~Rx antennas to one E3 Agent, using an \gls{ru} emulator with the Aerial test vectors of a fully loaded cell. Since one indication per slot covers all cells, with one \gls{shm} row per cell, only the per-cell work grows: step~1 by about $19~\mu$s and step~3 by about $12~\mu$s per cell, and the indication by about $5~\mu$s per cell of metadata, while the control is unchanged. With $10$ cells, the loop takes $399~\mu$s at p50, $78\%$ of which are the two copies. Copy helpers are the main lever at this scale: with $3$, $6$, and $9$ helpers, steps~2--7 take $336$, $247$, and $211~\mu$s. Copying only the cells and streams a dApp subscribes to would reduce both copies further and is planned for future work. The indication grows with the metadata\blue{, which also includes per-\gls{ue} metrics and thus grows with the number of scheduled \glspl{ue}}, as its JSON encoding and decoding take about $5$ and $10~\mu$s per KB: typical subscriptions produce a few hundred bytes for one cell and about $3$~KB for $10$ cells, but subscribing to all $86$ data streams of $10$ cells yields $42$~KB and a one-way latency of $772~\mu$s, so metadata-heavy subscriptions at scale benefit most from the planned binary encoding.}

\blue{\textbf{dApps.} Since one copy serves all dApps subscribed to the same streams, the export does not change with the number of dApps. The E3 Agent, however, encodes one indication per dApp and receives and applies all controls on a single socket and thread, so the last dApp served waits about $3~\mu$s per dApp ahead of it for its indication and about $8~\mu$s for its control. In this worst case, where all $10$ dApps send a control for every slot at the same time, the loop of the last dApp takes $190~\mu$s at p50, while dApps that only monitor receive their indication within $39~\mu$s plus $3~\mu$s per dApp ahead. A separate control thread would remove most of the control wait.} %

\begin{figure}[t]
    \centering
    \setlength\fwidth{0.72\linewidth}
    \setlength\fheight{0.95\linewidth}
    \input{figures_tex/scaling_bars}
    \caption{\blue{Per-step p50 latency of the data path without dApp processing (step~5) for various bandwidths and Rx antennas and for an increasing number of cells and dApps, on the GH200 without \gls{mig}. Copy helpers in parentheses. Labels give the closed loop including step~1 as p50/p99 in $\mu$s.}}
    \label{fig:scaling-bars}
\end{figure}

\subsection{\blue{Inference Engines and GPU Sharing}}
\label{subsec:engine-comparison}

\blue{In this section, we benchmark how the choice of inference engine, backend, and \gls{gpu} sharing mode affects the dApp processing (step~5) for the same dApp logic, represented by PRB Power. The results on the GH200 are shown in Figure~\ref{fig:backend-results}. PRB Power runs with five backends, namely Python/NumPy on the \gls{cpu}, and Torch, LibTorch, \gls{onnx} Runtime, and \gls{trt} on the \gls{gpu}, on each of the four engines of Section~\ref{subsec:dapp-container}, and with the three \gls{gpu} sharing modes of Section~\ref{subsec:setup}.}

\begin{figure*}[t]
    \centering
    \setlength\fwidth{0.29\textwidth}
    \setlength\fheight{0.17\textwidth}
    \input{figures_tex/engines_bars}
    \caption{\blue{Processing latency (step~5) of PRB Power on the GH200 per engine and backend with (a) \gls{mig}, (b) \gls{mps}, and (c) time-slicing. Bars and labels give the p50, lines the p99. The Python engine has no LibTorch, the native one runs \gls{trt} only.}}
    \label{fig:backend-results}
\end{figure*}

\blue{\textbf{MIG and MPS.} The native \gls{trt} engine processes one \gls{iq} slot in $41~\mu$s with \gls{mig} and $36~\mu$s with \gls{mps} (Figure~\ref{fig:backend-results}(a) and (b)), about $3\times$ faster than \gls{trt} in Triton through the C~API ($133~\mu$s), since it reads its input in place from the registered \gls{shm} and runs without a serving layer. Triton gRPC adds about $125~\mu$s to the C~API for serialization and inter-process communication, while the Python engine is faster than both Triton engines with Torch. Across backends, \gls{onnx} Runtime and LibTorch follow \gls{trt}, Torch takes about $380~\mu$s, and the \gls{cpu} baseline with NumPy about $580~\mu$s, $14\times$ the native engine. \gls{mps} matches \gls{mig} within about $10\%$ at p50 for every combination, and avoids the $15~\mu$s that \gls{mig} adds to step~1.}

\blue{\textbf{Time-slicing.} Without \gls{mig} or \gls{mps}, the dApp runs in its own CUDA context, and the \gls{gpu} alternates between this context and the one of L1. In this case, the engine and backend no longer determine the processing time, as shown in Figure~\ref{fig:backend-results}(c). All \gls{gpu} combinations take about $500~\mu$s at p50 and $720$--$860~\mu$s at p99, and the native engine needs $502~\mu$s instead of $36~\mu$s with \gls{mps}. The dApp kernels themselves take the same time as with \gls{mig} or \gls{mps}, but they need to wait for the \gls{gpu}. Since the E3 Agent sends each indication when cuPHY completes that slot, the dApp processes slot~$N$ while cuPHY processes slot~$N\!+\!1$, so that the three \gls{ul} slots of each burst (slots~7, 8, and~9 of the \textit{DDDDDDSUUU} pattern) take $703$, $502$, and $323~\mu$s, respectively, and only the last one finds the \gls{gpu} free of \gls{ul} work. With \gls{mig} and \gls{mps}, all three slots take the same time. The \gls{cpu} baseline is unaffected ($594~\mu$s) and becomes as fast as the \gls{gpu} backends, while at p99 all of them exceed the $0.5$~ms budget of Section~\ref{subsec:control-loop}.}

\blue{\textbf{Choosing an engine and a sharing mode.} The native engine is the fastest one but also the least flexible, as it runs a single prebuilt \gls{trt} model. On the other hand, Triton adds model management, several backends, ensembles to chain models, and monitoring, and the Python engine offers the shortest path from a prototype to a running dApp. Among the sharing modes, \gls{mig} isolates the memory and compute of the dApp from L1, at the cost of a fixed partition and $15~\mu$s more in step~1. \gls{mps} gives the lowest latency but no guaranteed isolation, as the dApp joins the \gls{mps} server of L1, although L1 reported no errors during subscriptions in our runs. Time-slicing is the default without \gls{mig} or \gls{mps}, but adds about $0.5$~ms, which makes it less suitable for latency-sensitive dApps. Other \gls{gpu} sharing mechanisms, such as CUDA green contexts that partition the streaming multiprocessors among workloads, will be evaluated in future work.}

\subsection{\blue{Cross-Platform Comparison on DGX Spark}}
\label{subsec:spark}

\begin{figure}[t]
    \centering
    \input{figures_tex/spark_path}
    \caption{\blue{Per-step p50 latency of the data path on the GH200 and DGX Spark with one copy helper, without processing and with the native \gls{trt} engine and \gls{mps}. Labels give the closed loop including step~1 as p50/p99 in $\mu$s.}}
    \label{fig:spark-path}
\end{figure}

\begin{figure}[t]
    \centering
    \input{figures_tex/spark_step5}
    \caption{\blue{Processing latency (step~5) of PRB Power on the GH200 and DGX Spark per engine with (a) \gls{mps} and (b) time-slicing. Bars and labels give the p50, lines the p99.}}
    \label{fig:spark-engines}
\end{figure}

\blue{We now compare the performance of the framework on the DGX Spark with the GH200. Figure~\ref{fig:spark-path} compares the data path of the two platforms with one copy helper each, the most that the efficiency cores of DGX Spark allow. Without processing, steps~2--7 take $65~\mu$s on both platforms. The copy to \gls{shm} is consistently faster on DGX Spark ($20$ vs.\ $25~\mu$s), while the E3 messages take slightly longer, since the E3 Agent runs on one of the slower efficiency cores (Section~\ref{subsec:setup}). The main difference is step~1, which takes $70$ instead of $25~\mu$s, consistent with the single copy engine of the GB10 against the $8$ of the H100. The closed loop including step~1 therefore grows from $90$ to $135~\mu$s without step~5, and from $128$ to $187~\mu$s including the processing with the native \gls{trt} engine.}

\blue{Figure~\ref{fig:spark-engines} compares step~5 for a subset of engines. With \gls{mps} (Figure~\ref{fig:spark-engines}(a)), DGX Spark is faster with the Python engine ($153$ vs.\ $226~\mu$s), whose input handling runs on the faster Cortex-X925 core, shows similar latency with the two Triton engines with \gls{trt}, and is slower with the native engine ($58$ vs.\ $36~\mu$s) and \gls{onnx} Runtime ($256$ vs.\ $153~\mu$s), possibly because the GB10 has $48$ streaming multiprocessors against the $132$ of the H100. The \gls{cpu} baseline with NumPy is about $1.5\times$ slower, as the whole dApp container runs on a single core on DGX Spark, while on the GH200 the backend threads can use several housekeeping cores.

With time-slicing (Figure~\ref{fig:spark-engines}(b)), both platforms behave as described in Section~\ref{subsec:engine-comparison}, with all \gls{gpu} engines at $430$--$630~\mu$s at p50, but DGX Spark shows larger tails, up to $1.3$~ms at p99 for the Triton engines. Overall, the differences follow the hardware, as fewer streaming multiprocessors and a single copy engine slow down the \gls{gpu} steps, while the faster \gls{cpu} cores speed up the \gls{cpu} work of the Python engine.}

\section{cuSense: Integrating ISAC in a dApp}
\label{sec:cusense}

Building upon the dApp framework presented in Section~\ref{sec:design} and evaluated in Section~\ref{sec:evaluation}, we now introduce cuSense, an \gls{isac} dApp for indoor person localization using uplink \gls{csi}. This case study validates our framework's ability to support intensive \gls{ai} workloads while showcasing the potential of \gls{isac} applications in \gls{5g} networks. \blue{Beyond its localization results, it exercises the framework end to end, from data collection and training to live inference on the \gls{ran}.}

\subsection{Overview and Goals}
\label{subsec:cusense-overview}

\begin{figure}[htbp]
  \centering
  \includegraphics[width=\linewidth]{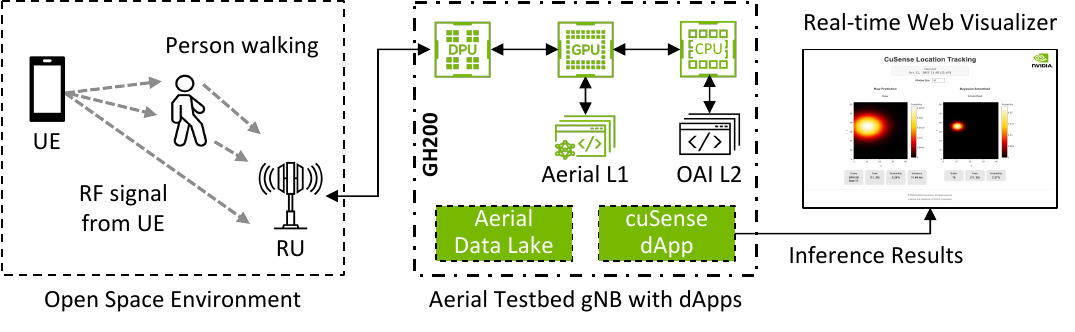}
  \caption{Overview of the cuSense UL DMRS-based ISAC dApp for person localization.}
  \label{fig:cusense-overview}
\end{figure}

cuSense targets indoor location estimation in a single-cell \gls{cbrs} deployment using \gls{csi} derived from \gls{ul} \gls{pusch} transmissions. We use \gls{dmrs} as it is natively present in every scheduled \gls{ul} transmission and its channel estimates are already computed by the \gls{phy} pipeline, requiring no additional signal overhead. The \gls{ue} transmits these signals and the E3 Agent exposes the resulting channel estimates to the dApp, in what is known as \gls{ul}-collaborative \gls{isac}~\cite{isacolab}. \blue{The framework also exposes \gls{srs} data (Section~\ref{subsec:rt-adl}), which a dApp can use in a similar way.} Following the reference dApp container architecture of Section~\ref{subsec:dapp-container}, the cuSense dApp processes this \gls{dmrs}-derived \gls{csi} through a lightweight signal-processing pipeline\blue{, served in Triton or natively with TensorRT (Section~\ref{subsec:cuSense-inference}),} and runs a \gls{nn} model to capture the variations in the channel given by the moving object, as shown in Figure~\ref{fig:cusense-overview}. The goal is to estimate the 2D position $(x_t,y_t)$ of a person walking in the space through a probability map over the area of interest.
The system addresses the following key challenges in \gls{ul}-\gls{csi}-based sensing: (i) extracting real-time channel perturbations from static multipath; (ii) learning spatial mapping from high-dimensional \gls{csi} data; and (iii) achieving real-time inference.

Finally, cuSense is designed as a reference \gls{isac} dApp to showcase our framework's ability to support real-time, slot-level inference, and provide a \blue{public} dataset for future \gls{isac} experiments. In the current prototype, we focus on single-user localization with a static background assumption and environment-specific calibration. %
The same design \blue{could} be extended \blue{to other tasks, such as occupancy monitoring and \gls{uav} or ground vehicle detection, without changes to the \gls{phy} processing}.

\begin{figure}[htbp]
  \centering
  \includegraphics[width=\linewidth]{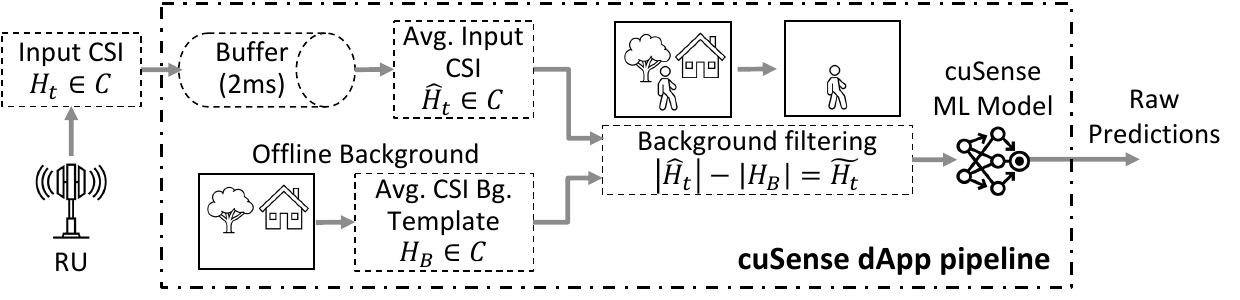}
  \caption{cuSense dApp processing pipeline overview.}
  \label{fig:cusense-pipeline}
\end{figure}

\subsection{Uplink CSI Processing Pipeline}
\label{subsec:cusense-pipeline}

cuSense processes raw \gls{ul} \gls{dmrs} \gls{csi} measurements to estimate spatial locations of targets within an indoor environment through a three-stage pipeline: (i) background environment characterization (computed offline); (ii) temporal noise reduction and feature normalization (online in the cuSense dApp pipeline); and (iii) \gls{nn}-based location estimation (online in the \blue{dApp}), as shown in Figure~\ref{fig:cusense-pipeline}.

\textbf{Stage 1: Background Environment Characterization.}
The objective of this first stage is to establish a baseline characterization of the wireless channel in the absence of dynamic targets. This background template captures an averaged \textit{snapshot} of multi-path reflections from static objects in a deployment location, such as walls, furniture, and other stationary objects, which can then be removed from live measurements to isolate target-induced perturbations.

Let $H_i[a,k,s] \in \mathbb{C}$ denote the complex \gls{csi} for measurement $i$ at antenna $a$, subcarrier $k$, and \gls{dmrs} symbol $s$, obtained after frequency-domain interpolation over all active subcarriers and before time-domain interpolation to all \gls{ofdm} symbols. To handle variable-length subcarrier allocations with zero-padding, we define a validity set for each $(a,k,s)$ position:
\begin{equation}
  \mathcal{V}_{a,k,s} = \{i : |H_i[a,k,s]| > \tau\}
\end{equation}
where $\tau = 10^{-10}$ is a tolerance power threshold, and $\mathcal{V}_{a,k,s}$ contains the indices of non-zero measurements. This ensures that statistics are computed exclusively from active subcarriers, avoiding contamination from zero-padded frequency bins that may vary across different channel allocations.

We compute a complex-valued background template by averaging over the $N_{a,k,s} = |\mathcal{V}_{a,k,s}|$ valid background measurements:
\begin{equation}
  H_{\mathrm{B}}[a,k,s] = \frac{1}{N_{a,k,s}}\sum_{i \in \mathcal{V}_{a,k,s}} H_i[a,k,s].
\end{equation}
The output of this stage is a complex background template $H_{\mathrm{B}}$ over active subcarriers, which is reused at inference time to remove static multi-path components.

\textbf{Stage 2: Temporal Noise Reduction and Feature Normalization.}
This stage applies temporal averaging to mitigate measurement noise, subtracts the static background to isolate dynamic components, and normalizes features to enable stable gradient-based optimization.
For each time $t$ in a target run, we construct a causal temporal window
\begin{equation}
  W(t) = \{ H_i \mid t - \Delta t \le t_i \le t \},
\end{equation}
containing all \gls{csi} samples whose timestamp $t_i$ falls within a window of length $\Delta t$ ending at $t$. Based on empirical hyper-parameter tuning, we set $\Delta t = 2$~ms, which provided the best trade-off between noise reduction and temporal resolution in our experiments. \blue{This window holds the three \gls{ul} slots of a \gls{tdd} period, i.e., one prediction every $5$~ms with the \textit{DDDDDDSUUU} pattern.}
We then average each $W(t)$ coherently, element-wise in the complex domain:
\begin{equation}
  \hat{H}_{\mathrm{t}}[a,k,s]  = \frac{1}{|W(t)|} \sum_{i\in W(t)} H_i[a,k,s],
\end{equation}
excluding zero-valued samples on inactive subcarriers. \blue{Since the \gls{pusch} allocation can start at a different \gls{prb} in each slot, we place the \gls{csi} of each slot at its \gls{prb} offset before averaging.} Temporal averaging reduces noise variance by approximately a factor of $1/|W(t)|$ under an \gls{awgn} assumption, improving the effective \gls{snr} while preserving the slowly varying channel characteristics associated with target motion.

To isolate dynamic components from static environment reflections, we subtract the background template in the magnitude domain from the averaged measurements,
\begin{equation}
  \tilde{H}_{\mathrm{t}}[a,k,s] =
  \bigl| \hat{H}_{\mathrm{t}}[a,k,s] \bigr|
  - \bigl| H_{\mathrm{B}}[a,k,s] \bigr|,
\end{equation}
and then average across the $S$ \gls{dmrs} symbols in the \gls{ul} slot\blue{, which produces} a real-valued tensor $\tilde{H}_{\mathrm{t}}^{\mathrm{D}} \in \mathbb{R}^{A \times K_{\mathrm{v}}}$, where $A$ is the total number of receive antennas and $K_{\mathrm{v}}$ the number of active (or valid) subcarriers. This step further reduces noise through intra-slot averaging under the assumption that the channel remains approximately constant within a slot.

Finally, we apply global $z$-score normalization,
\begin{equation}
  X_{\mathrm{t}}[a,k] =
  \frac{\tilde{H}_{\mathrm{t}}^{\mathrm{D}}[a,k] - \mu_{\mathrm{global}}}{\sigma_{\mathrm{global}} + \epsilon},
\end{equation}
where $\mu_{\mathrm{global}}$ and $\sigma_{\mathrm{global}}$ are computed over all available samples, antennas, and valid subcarriers, and are then applied identically to all splits, and $\epsilon = 10^{-8}$ is a small constant to prevent division by zero. This produces standardized input features $X_{\mathrm{t}}[a,k]$ for our \gls{nn} model with approximately zero mean and unit variance, preserving relative power relationships between samples, which facilitates stable training and inference.
In a deployed system, Stage~1 runs offline once per environment, while Stage~2 executes online inside the cuSense dApp on the \gls{csi} delivered by the E3 Agent via \gls{shm}.

\begin{figure*}[htbp]
\centering
\input{figures_tex/cusense_resnet}
\caption{Proposed \gls{nn} architecture of cuSense dApp for CSI-based location estimation.}
\vspace{-.15cm}
\label{fig:proposed-architecture}
\end{figure*}
\textbf{Stage 3: Neural Network Architecture and Training.}
In the last stage, we use a ResNet-inspired \gls{cnn} optimized for frequency-domain, multi-antenna \gls{csi} magnitude data to predict a 2D probability grid representing the likelihood of target presence at each location within the environment. In order to account for the relatively shallow architecture and to keep the parameter count low and allow fast inference, we consider only \textit{plain} blocks without residual connections for our architecture design (see \cite{He_2016_CVPR} for the original model).
The normalized features $X_{\mathrm{t}} \in \mathbb{R}^{A\times K_{\mathrm{v}}}$ are fed to our proposed architecture, as shown in Figure~\ref{fig:proposed-architecture}. The network consists of: (i) an initial 1D convolution and max-pooling layer; (ii) three sequential blocks with progressive channel expansion ($64\rightarrow128\rightarrow256\rightarrow512$); (iii) an adaptive global average pooling layer, to allow inputs of different shapes in case not all RBs are utilized in the \gls{ul} transmission; (iv) a three-layer \gls{mlp} stack that outputs an $H\times W$ (grid height and width) spatial probability map; and (v) a final softmax to ensure $\sum_{i,j} P_\mathrm{t}(i,j) = 1$ for each $(i,j)$ grid cell probability value $P_\mathrm{t}$.
Training uses the \gls{kl} divergence between the predicted distribution $P_{\text{X}}$ and a smoothed target distribution $P_{\text{Y}}$ as loss:
\begin{equation}
  \mathcal{L}_{\mathrm{KL}} =
  \sum_{i=0}^{H-1} \sum_{j=0}^{W-1}
  P_{\text{Y}}(i,j)
  \log \frac{P_{\text{Y}}(i,j)}{P_{\text{X}}(i,j)}.
\end{equation}
This formulation encourages learning the entire spatial distribution rather than only peak locations, improving robustness to measurement noise. The target distribution is obtained by convolving a one-hot probability grid\blue{, with 1 at the target cell and 0 elsewhere,} with a 2D Gaussian kernel $G_\sigma$ (label smoothing):
\begin{equation}
P_{\text{Y}} =
\frac{(G_\sigma * P_{\text{one-hot}})(i,j)}
     {\sum_{i,j} (G_\sigma * P_{\text{one-hot}})(i,j)},
\end{equation}
with $\sigma = 8.0$ in our experiments. Spatial smoothing encodes the intuition that nearby locations should receive similar probabilities, reflecting uncertainty in ground-truth annotations and measurement noise.
We train the model using Adam optimizer, mini-batches of size $B = 256$, and a learning rate $l_r = 10^{-3}$ for 100 epochs.
At inference time, cuSense runs Stage~2 pre-processing and the \gls{nn} inside the dApp container\blue{, with one of the serving options of Section~\ref{subsec:cuSense-inference},} receiving \gls{csi} for each \gls{ul} slot $t$, and outputting a 2D probability map $P_{\mathrm{t}}(i,j)$ over the $H\times W$ grid.
From this distribution, we can derive a maximum-likelihood estimate 
$(\hat{i}_t,\hat{j}_t) = \arg\max_{i,j} P_t(i,j)$ to obtain the grid coordinates of highest target location probability, which can be then mapped to the relative world coordinates.
In order to reduce output prediction errors, we average the output probability for the last 10 predictions and apply Kalman filtering \blue{to the estimated 2D location}.

\section{cuSense Experimental Evaluation}
\label{sec:cusense-eval}

To assess the feasibility and performance of cuSense, we perform an \gls{ota} data-collection campaign, develop labeling pipelines to obtain ground truth and build datasets, and train and test the cuSense \gls{nn} under realistic conditions.

\subsection{Data Collection and Labeling Methods}

\begin{figure}[tbp]
  \centering
  \includegraphics[width=\linewidth]{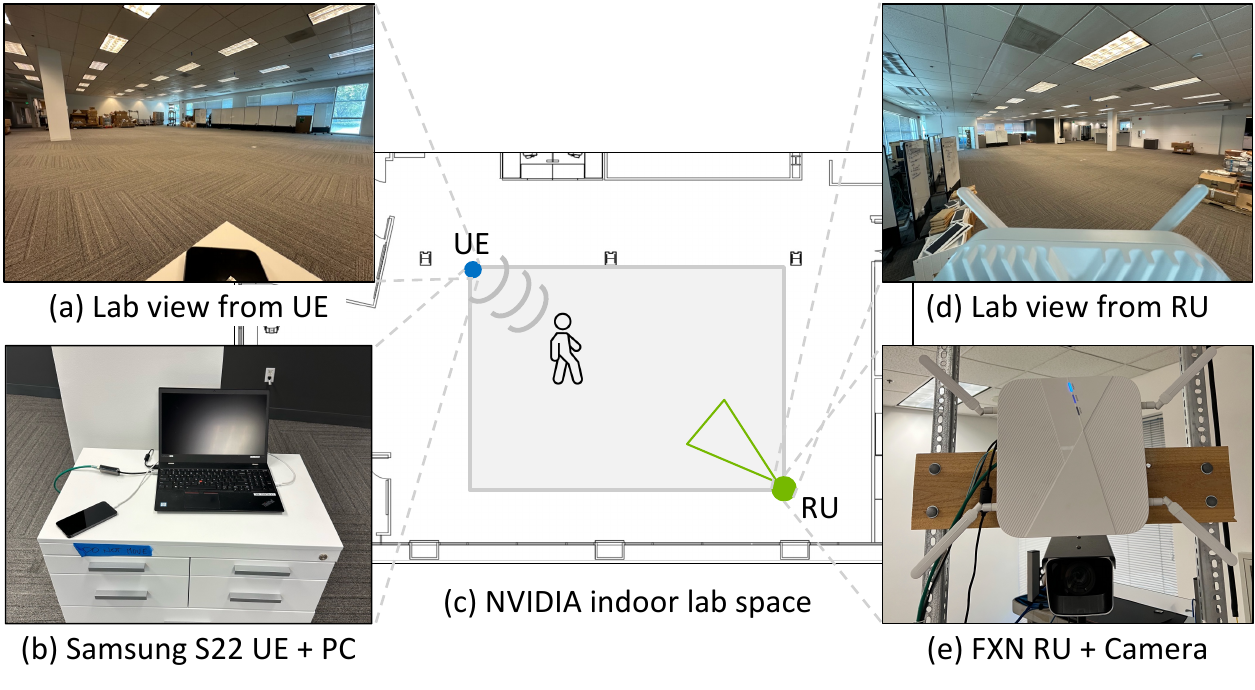}
  \caption{cuSense experimental environment.}
  \label{fig:cusense-setup}
\end{figure}

\textbf{Measurement Campaigns.}
\label{subsec:cusense-data}
\blue{The cuSense data campaigns use the GH200 node and software stack described in Section~\ref{subsec:setup}, with a 4T4R Foxconn \gls{cbrs} \gls{ru} and a Samsung Galaxy S22 \gls{ue}, as illustrated in Figure~\ref{fig:cusense-setup}.} The \gls{ru} is mounted on a custom stand~\cite{kelkar5g} facing an open indoor space in NVIDIA Lab. The \gls{ue}, together with a laptop used for remote access and traffic generation, is placed at the opposite end of a rectangular target area of approximately $6.78 \times 10.06$~m.
We collect multiple runs under two conditions: (i) \emph{background runs}, with no moving target present, to characterize the static channel; and (ii) \emph{target runs}, in which a person walks along various trajectories (e.g., lawnmower, spiral, random) at normal walking speed. Each run follows the same procedure: (i) start camera recording at $30$ or $60$~\gls{fps} (for target runs); (ii) start the \gls{atb} \gls{gnb} with \gls{adl} enabled; (iii) connect the \gls{ue}; (iv) start uplink UDP traffic through \texttt{iperf} with a target rate of $100$~Mbps; (v) perform the walking trajectory within the target area (for target runs); and (vi) stop the run after approximately $2$--$3$ minutes.
\blue{Across 18 runs, nine target runs and nine background runs, we collect about $1.19$ million \gls{csi} records together with about $64{,}000$ video frames, which we use in the next step for labeling and dataset construction. The full campaign, with the channel estimates, their per-slot metadata, and the camera-derived positions, is publicly available as a dataset\footref{fn:cusense-dataset}.}

\begin{figure}[bp]
  \centering
  \includegraphics[width=\linewidth]{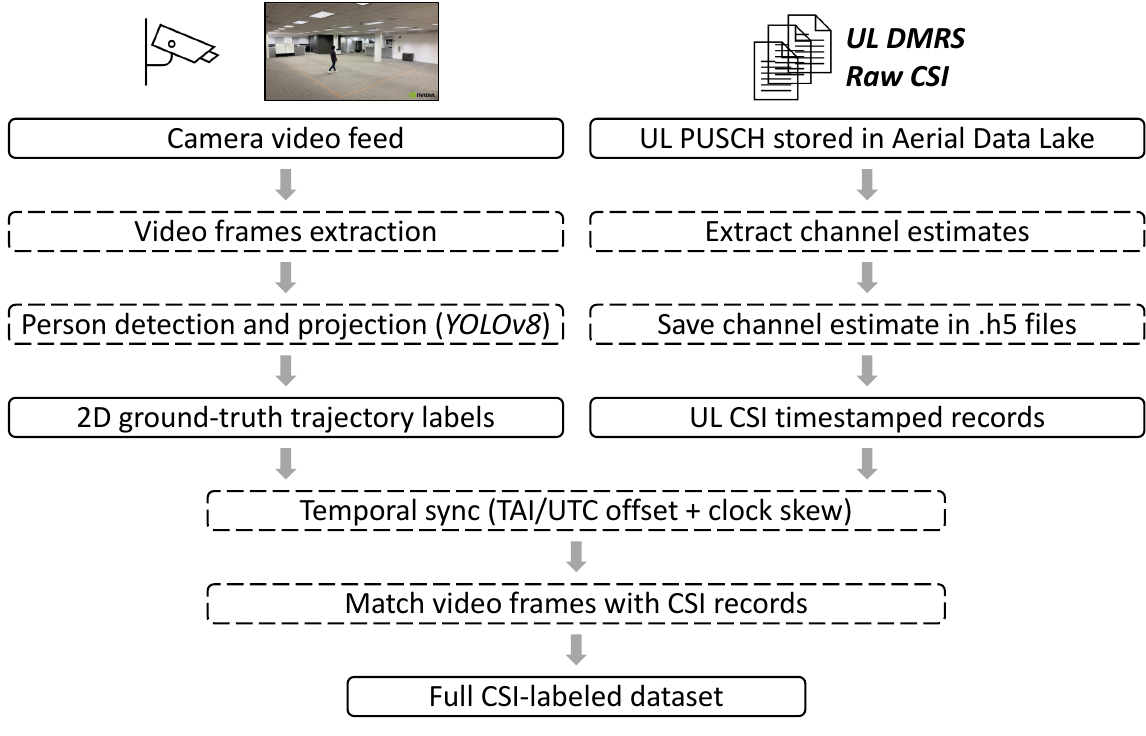}
  \caption{cuSense data collection and labeling pipeline: camera-based ground-truth extraction (left) and UL CSI record collection (right) are temporally synchronized to produce the labeled dataset for training, validation, and testing.}
  \label{fig:cusense-labeling}
\end{figure}

\textbf{Temporal Video and Sensing Synchronization.}
As shown in Figure~\ref{fig:cusense-labeling}, to build the complete dataset for training and testing, we collect \gls{ul} \gls{csi} data (i.e., \gls{dmrs} \gls{hest}) together with a camera video recording the scene at a fixed frame rate, used to generate ground-truth 2D trajectories of the person walking. Synchronizing \gls{csi} data and video frames requires handling both time-reference differences and clock skew between devices.
\gls{csi} records from \gls{adl} use \gls{tai} timestamps, while video frames from the camera (an iPhone in our experiments, though any commercial camera could be used) employ standard \gls{utc} UNIX timestamps. \gls{tai} is a continuous atomic time scale that differs from \gls{utc} by a constant offset $\Delta_{\text{leap}}$ ($37$~s at the time of our experiments), such that $t_{\text{TAI}} = t_{\text{UTC}} + \Delta_{\text{leap}}$ due to the leap seconds introduced since 1972 to account for Earth's rotation variations.
In addition to this systematic offset, we compensate for residual clock skew between the camera and the \gls{gh} server with respect to a common \gls{ntp} server. The \gls{gnb} maintains microsecond-level timing accuracy through \gls{ptp} synchronization with the grandmaster clock, making its contribution to the skew negligible, while the camera achieves millisecond-level precision, which we estimate during the measurement campaigns.
In a second step, we run an offline pipeline that parses the video file, extracts per-frame timestamps and images, and aligns them with the \gls{csi} after compensating for the $37$~s \gls{tai} offset and the measured server–camera clock skew. We assign each \gls{csi} record to the closest video frame within a threshold of half the frame period, generating tightly synchronized \gls{csi}–video pairs, where a single frame may correspond to multiple \gls{csi} records.

\textbf{Ground-Truth Extraction.}
To extract the ground-truth locations, we derive the 2D person coordinates from the camera stream using an offline computer-vision pipeline based on YOLOv8~\cite{yolov8}. For each video frame, we run a person detector and retain the bounding box with the highest confidence (our experiments are limited to a single walking subject). The box centroid $(u,v)$ in image coordinates is then projected onto the floor plane using a planar homography $H$ estimated from the manually annotated image corners of the rectangular area of interest in the environment. This generates a continuous trajectory $\{(x_t,y_t)\}$ in meters, sampled at the camera frame rate.
Each \gls{csi} record then inherits a physical position $(x_t,y_t)$, since it is associated with the closest video frame in time. For training, we discretize the floor plan into an $H \times W$ grid and quantize each position to the corresponding cell $(i,j)$, producing a one-hot label $P_{\text{one-hot}}(i,j)$ for that \gls{csi} sample. These one-hot labels are then converted into smoothed target distributions $P_{Y}$ using the Gaussian label-smoothing procedure described in Stage~3.

\subsection{Performance Results}
\label{subsec:cusense-results}

We now evaluate cuSense on the dataset described in Section~\ref{subsec:cusense-data}, focusing on three aspects: (i) localization accuracy against the ground-truth; (ii) comparison with \gls{3gpp} sensing service categories; and (iii) real-time inference capability within our dApp framework.

\textbf{Dataset and Evaluation Setup.}
The cuSense model is trained on \gls{csi} data from 5 runs with a moving target\blue{ (runs 2, 4, 6, 8, and 10 of the dataset)}, for a total of $362{,}318$ slot samples each with 4 antennas and 3 \gls{dmrs} symbols. We use a temporal-random split strategy of 80\% training, 10\% validation, 10\% test as shown in the examples of Figure~\ref{fig:cusense_dataset}. Validation and test sets are two contiguous temporal blocks randomly positioned within each run's timeline, rather than randomly sampling individual points. This design choice is critical because consecutive \gls{csi} values may be highly correlated, leading to overfitting and poor generalization to unseen trajectories. 
\begin{figure}[tp]
  \centering
  \setlength\fwidth{\linewidth}
  \setlength\fheight{\linewidth}
  \input{figures_tex/cusense_dataset}
  \caption{Temporal-random split with contiguous validation/test blocks within training runs and fully unseen runs.}
  \label{fig:cusense_dataset}
\end{figure}
The background template $H_{\text{B}}$ is computed from a single dedicated run of $54{,}400$ samples in a static environment with no moving target\blue{ (run 7)}. We also evaluate cuSense generalization across two independent unseen runs (Unseen-Run~1 and Unseen-Run~2\blue{, runs 12 and 14}) collected separately from all training, testing, or validation data.
We measure localization performance using standard \gls{rmse}, median, and success rate within different thresholds over all unseen \gls{csi} records. Additionally, the inference pipeline applies a temporal arithmetic mean over 10 consecutive predictions from stage~3 processing output for noise reduction, followed by a Kalman filter tracking with process noise $Q = 10^{-5}$ and measurement noise base $R_{\text{base}} = 750$ to produce temporally coherent trajectory estimates. \blue{Both steps are causal, i.e., each output uses only the current and past predictions, so they can also run in real time.}

\textbf{Localization Accuracy.}
Table~\ref{tab:cusense_accuracy} summarizes the localization performance on the unseen runs. cuSense achieves a mean localization error of \blue{$73.3$}~cm with a median of \blue{$54.9$}~cm, demonstrating sub-meter accuracy for the majority of predictions. The \gls{rmse} of \blue{$95$}~cm reflects occasional outliers in challenging multipath conditions. Notably, performance is consistent across both test campaigns (\blue{$72.4$ and $74.3$}~cm), indicating robust generalization to unseen measurement campaigns in the same environment\blue{, while the accuracy also reflects the trained model, the quality and synchronization of the camera-derived labels, and the specific environment}.

\begin{table}[htbp]
    \centering
    \caption{cuSense localization accuracy on unseen runs\blue{, after averaging and Kalman filtering}.}
    \label{tab:cusense_accuracy}
    \small
    \setlength{\tabcolsep}{2pt}
    \begin{tabular}{lccc}
        \toprule
        \textbf{Metric} & \textbf{Unseen-Run 1} & \textbf{Unseen-Run 2} & \blue{\textbf{Combined}} \\
        \midrule
        Samples & \blue{23,168} & \blue{23,249} & \blue{46,417} \\
        Mean Error [cm] & \blue{72.4} & \blue{74.3} & \blue{73.3} \\
        Median Error [cm] & \blue{54.9} & \blue{54.9} & \blue{54.9} \\
        Std. Dev. [cm] & \blue{62.2} & \blue{58.4} & \blue{60.3} \\
        RMSE [cm] & \blue{95.5} & \blue{94.4} & \blue{95.0} \\
        \blue{p90 Error [cm]} & \blue{156.2} & \blue{147.3} & \blue{148.3} \\
        \blue{p95 Error [cm]} & \blue{214.7} & \blue{203.2} & \blue{209.7} \\
        \blue{p99 Error [cm]} & \blue{286.3} & \blue{299.6} & \blue{297.1} \\
        \bottomrule
    \end{tabular}
\end{table}

\textbf{3GPP Sensing Service Categories.}
We benchmark these results against the 3GPP Release 19 \blue{sensing requirements} defined for 5G wireless~\cite{3gpp22137}. \blue{For object detection and tracking, TS~22.137 defines four service categories, each combining a positioning accuracy at a confidence level of 95\% with a maximum latency from the triggering event to the sensing result, a refreshing rate of the results, and missed detection and false alarm probabilities.}
\blue{Figure~\ref{fig:cusense_accuracy} shows the share of predictions within each threshold and the \gls{cdf} of the localization error on the two unseen runs for the raw, averaged, and Kalman-filtered predictions, with the accuracy thresholds of Categories~4, 3, and 2 ($0.5$, $1$, and $2$~m). On the combined runs, $42.6\%$, $78.6\%$, and $94.4\%$ of the Kalman-filtered predictions fall within these thresholds, and the error at 95\% confidence is $2.10$~m. This meets the $10$~m of Category~1 and narrowly misses the $2$~m requirement of Category~2. The results are consistent across the two runs, with $2.15$ and $2.03$~m at 95\% confidence,} despite these runs being entirely excluded from training.
\blue{The latency and the refreshing rate meet all four categories. In the offline evaluation of the recorded runs, cuSense produces one prediction per \gls{tdd} period, i.e., every $5$~ms, each available after an inference time of at most a few milliseconds (see below), well below the strictest requirements, i.e., a latency of $100$~ms for Categories~3 and 4 and $50$~ms between results for Category~3. Since the averaging and the Kalman filter are causal, their delay is already included in the error. Therefore, only the accuracy limits cuSense to Category~1. Without smoothing, the error at 95\% confidence grows to $3.00$~m, and to $2.69$~m with the averaging alone. Missed detection and false alarm are not evaluated, since cuSense assumes that a target is present and has no detection stage.}
\begin{figure}[tbp]
    \centering
    \input{figures_tex/cusense_cdf}
    \caption{\blue{cuSense localization error on the two unseen runs for the raw, averaged, and Kalman-filtered predictions, with (a) the share of predictions within the TS~22.137 accuracy thresholds, and (b) the \acrshort{cdf} with the error at 95\% confidence.}}
    \label{fig:cusense_accuracy}
\end{figure}

\textbf{Trajectory Tracking.}
Figure~\ref{fig:cusense_trajectory} visualizes the X and Y position estimates over time for a representative test segment (\blue{one prediction every $25$~ms shown} for clarity), comparing raw model predictions, temporally averaged predictions, Kalman-filtered output, and ground truth. The raw predictions exhibit significant frame-to-frame jitter from measurement noise.
Temporal averaging reduces this variability, while the Kalman filter produces smooth, physically plausible trajectories that closely track the ground truth. The multi-stage refinement provides cumulative error reduction of \blue{about $30\%$} compared to raw predictions, with the Kalman filter successfully handling both slow movements and rapid direction changes.
\begin{figure}[htbp]
  \centering
  \input{figures_tex/run12_three_align_trajectory_includable}
  \caption{Trajectory tracking comparison of an unseen run segment for the X and Y axes.}
  \label{fig:cusense_trajectory}
\end{figure}

\textbf{Real-Time dApp Inference Performance.}
\label{subsec:cuSense-inference}
\blue{We evaluate the cuSense dApp live on the testbed of Section~\ref{subsec:setup}. The dApp subscribes to the \gls{ul} channel estimates and, as in Stage~2, averages the slots received within $\Delta t = 2$~ms, i.e., the three \gls{ul} slots of a \gls{tdd} period, before the preprocessing and the inference. After each inference, a cooldown of $15$~ms yields one prediction every $20$~ms, which we added for the slowest serving option. Without the cooldown, the native option predicts in every \gls{tdd} period, i.e., every $5$~ms as in the offline evaluation, with the same latency. The dApp returns each prediction as is, and the averaging and Kalman filtering of the offline evaluation can be applied by the consumer of the results, where they take about $85~\mu$s per prediction on one Grace core. We measure the latency from the notification of the third slot (step~2 in Table~\ref{tab:datapath}) to the end of its inference.}

\blue{Figure~\ref{fig:cusense-serving} compares four serving options for the same pipeline, from the original Triton deployment to a native one. In the first two, a Triton Python model buffers the slots, runs the averaging and the preprocessing, and calls the \gls{cnn} as a TorchScript or a TensorRT model. In the third, the dApp buffers the slots and a single TensorRT engine in Triton runs both the preprocessing and the \gls{cnn}. The fourth runs the preprocessing as CUDA kernels and the \gls{cnn} with TensorRT directly in the dApp, without Triton. It also reads the slots in place from the \gls{shm}, which saves about $40~\mu$s with \gls{mig} compared to copying them into its own window.}
\blue{On the GH200 with \gls{mig}, the p50 latency drops from $2138~\mu$s with the original deployment to $241~\mu$s with the native option ($249~\mu$s at p99), of which about $35~\mu$s are spent in the E3 path. \gls{mps} gives similar results, while time-slicing slows down every option, up to $5156~\mu$s. On the DGX Spark, the two Python options are faster than on the GH200, as their \gls{cpu} work runs on the faster cores, while the fused and native options are $1.3$ to $1.5$ times slower, consistent with Section~\ref{subsec:spark}.}
\blue{To compare with an optimized \gls{cpu} implementation, we run the same preprocessing and \gls{cnn} offline on recorded windows with ONNX Runtime on dedicated Grace cores. With 16 cores, it takes $2694~\mu$s at p50, $4.8$ times the $565~\mu$s of the fused \gls{gpu} pipeline run in the same way on the \gls{mig} instance and $7.6$ times on the DGX Spark with 4 cores. On the GH200, adding cores beyond 8 brings little gain. The \gls{gpu} options also leave these cores free for the \gls{ran} and other processes.}
\blue{Finally, we run cuSense with the native option together with PRB Power with the Triton C~API, both subscribed to the same E3 Agent and sharing the GH200 with \gls{mps}. The cuSense latency grows from $275$ to $316~\mu$s at p50, while PRB Power is unaffected. The E3 path of both grows by about $8~\mu$s, as the copy to host \gls{shm} (step~3) includes the streams of both dApps and the E3 Agent serves two subscribers, consistent with Section~\ref{subsec:scaling}.}
\begin{figure}[tbp]
    \centering
    \input{figures_tex/cusense_serving}
    \caption{\blue{cuSense latency from the notification of the third slot to the end of its inference per serving option and \gls{gpu} sharing mode. Bars and labels give the p50, lines the p99.}}
    \label{fig:cusense-serving}
\end{figure}

\section{Related Work}
\label{sec:related}

\begin{table*}[t]
\centering
\caption{\blue{Comparison with frameworks for real-time RAN programmability. Latencies are as reported by each work and are not measured in the same way.}}
\label{tab:related}
\scriptsize
\setlength{\tabcolsep}{2.5pt}
\begin{tabular*}{\textwidth}{@{\extracolsep{\fill}}lllllll@{}}
\toprule
\textbf{Framework} & \textbf{RAN stack} & \textbf{Exposed data} & \textbf{Placement} & \textbf{Compute} & \textbf{Interface} & \textbf{Reported latency} \\
\midrule
OAI dApps~\cite{lacava2025dapps} & OAI & I/Q samples & Co-located container & CPU & E3 & $<450~\mu$s mean loop \\
OCUDU E3Controller~\cite{feraudo2026ocudu} & OCUDU & Fronthaul I/Q, spectrum & Sidecar (jbpf hooks) & CPU & E3 & $115$--$120~\mu$s p50 to dApp \\
EdgeRIC~\cite{ko2024edgeric,edgeric-ocudu} & srsRAN, OCUDU & L2, FAPI, fronthaul packets & Separate processes & CPU & ZMQ, jbpf & $51$--$114~\mu$s p50 report to action \\
Janus~\cite{xenofonjanus} & FlexRAN, OAI & RAN function state & In-process codelets & CPU & eBPF & Bounded codelet runtime \\
RANBooster~\cite{foukas2025ranbooster} & O-RAN 7.2 DU/RU & Fronthaul packets & Fronthaul middlebox & CPU & Fronthaul & $0.3$--$6~\mu$s per packet \\
CUDA OCUDU~\cite{oshea2026realtime,oshea2026platform} & OCUDU & PHY tensors, MAC state & In-process or separate & GPU & C ABI, E3 & $5.2~\mu$s p99.9 in-process \\
\midrule
\textbf{This work} & Aerial L1 & I/Q, \gls{hest}, PUSCH, SRS, PHY metrics & Separate container & GPU or CPU & E3 & $123~\mu$s p50 loop with inference \\
\bottomrule
\end{tabular*}
\end{table*}

\textbf{Open and AI-Native RAN Testbeds.}
Several open and AI-native \gls{5g}/O-RAN testbeds have been proposed to accelerate experimentation with data-driven RAN control. \cite{oaic} provides an open-source O-RAN platform with non-RT, near-RT, and a proposed real-time \gls{ric} (zApps) for AI-based RAN management. \cite{ford2025sim} introduces a digital-twin-based AI-RAN development workflow and an NVIDIA \gls{atb}-based testbed for \gls{ai} \gls{phy} evaluation with commercial \glspl{ue}. Other platforms such as POWDER, COSMOS, Colosseum, and X5G enable at-scale experimentation with open RAN and GPU-accelerated \gls{phy} functions~\cite{breen2020powder,chen2023open-access,bonati2021colosseum,villa2025tmc}.
In parallel, the AI-RAN Alliance and 3GPP Release~19 studies on \gls{isac} and \gls{ai} for RAN functions position AI-native RAN and \gls{isac} as key components of \gls{5g}-Advanced and \gls{6g} evolution~\cite{airan,3gpp22137}.
Our work builds on this ecosystem and focuses on integrating \glspl{dapp} with \gls{atb} and exposing on-\gls{gpu} \gls{phy} telemetry for real-time applications, such as spectrum sharing and \gls{isac}.

\textbf{\blue{dApps and E3 Implementations.}}
Recent work extends O-RAN toward real-time and user-plane control following initial visions and implementations (see Section~\ref{sec:background})~\cite{D_Oro_2022,ngrg-dapp-1,ngrg-dapp-2,lacava2025dapps}. \blue{Besides the OAI framework of~\cite{lacava2025dapps}, the E3 interface has been brought to OCUDU through a sidecar controller~\cite{feraudo2026ocudu} that supports the same dApps as NVIDIA Aerial and OAI, a step toward dApps that run across \gls{ran} stacks in the spirit of the open interfaces of O-RAN. Containerized dApps have also been characterized on commodity servers, where the \gls{cpu} becomes the bottleneck as their number grows~\cite{boeira2026performance}.} Follow-up work demonstrates dApps for CPU power management, spectrum classification, \blue{and uplink scheduling~\cite{crespo2025energy,olimpieri2025libiqrealtimespectrumclassification,elkael2026auguste}, for} interference detection on GPU-accelerated \gls{5g} \glspl{phy}~\cite{neasamoni2025interforan}\blue{, and for switching between AI-based and conventional \gls{phy} processing at slot granularity~\cite{santhi2026arches}. The last two build on earlier versions of the framework presented in this paper.}

\textbf{\blue{Real-Time RAN Programmability.}}
\blue{Other frameworks target different points in the RAN programmability space, and Table~\ref{tab:related} compares them with ours.} Janus~\cite{xenofonjanus} embeds sandboxed eBPF codelets directly into vRAN functions, enabling flexible telemetry collection and real-time control with strict safety guarantees at microsecond timescales, but is tightly coupled to the RAN process and CPU-centric.
EdgeRIC~\cite{ko2024edgeric} co-locates a real-time \gls{ric} with the \gls{du}\blue{, and its release on OCUDU with jbpf~\cite{edgeric-ocudu} runs CPU-based $\mu$Apps as separate processes with observations and actions tied to specific \glspl{tti}, an architecture close to dApps, while its hooks also reach FAPI messages and fronthaul packets. Our framework instead exposes the per-slot \gls{phy} data produced on the \gls{gpu} to dApps that run inference on the same \gls{gpu}, next to the \gls{ran}}. OAIC/zApps~\cite{oaic} propose decomposing RAN functions into real-time microservices at the \gls{ric} level, targeting sub-$10$~ms control, but with a different architecture from third-party dApps consuming telemetry. RANBooster~\cite{foukas2025ranbooster} introduces programmable middleboxes in the fronthaul between \gls{du} and \gls{ru}, enabling use cases such as distributed MIMO and PRB monitoring, but operates outside the baseband processing pipeline. \blue{On a CUDA-accelerated OCUDU \gls{du}, DeepSig groups dApps into three classes, i.e., inline \gls{phy} processing, bounded control that answers the \gls{mac} within a deadline, and asynchronous observation and advisory~\cite{oshea2026realtime}, which they implement in an open dApp platform~\cite{oshea2026platform}. Their measurements show that, with the \gls{du} on the air, only in-process paths meet the tightest deadlines, while asynchronous use cases remain feasible out of process. Inline accelerators such as \glspl{fpga} speed up fixed \gls{phy} functions, but do not offer a programmable host for third-party \gls{ai} models. \gls{ai} models can also run inside the \gls{phy} pipeline itself~\cite{cohenarazi2025aerial,santhi2026arches}, where they replace a \gls{phy} function rather than consume its outputs, and ARCHES combines the two approaches, with a dApp that selects between AI-based and conventional channel estimators.}
Our work can be seen as complementary to these systems and targets a different point in the design space by exposing high-dimensional, per-slot \gls{phy} telemetry from a \gls{gpu}-native stack via \gls{shm} and enabling co-located \gls{gpu}-accelerated \gls{ai} inference through out-of-process dApps decoupled from the \gls{gnb} and vendor-specific \gls{phy} kernels\blue{, which share the \gls{gpu} with the \gls{ran} through \gls{mig} or \gls{mps}}.

\textbf{ISAC and CSI-based Sensing.}
\gls{isac} has been widely studied as a key enabler for beyond-\gls{5g} and \gls{6g} networks, with recent surveys on its architectures, limits, and open challenges~\cite{isac2024lu,isac2022liu}. In \gls{5g}, recent work has explored communication-centric \gls{isac} schemes for cooperative target localization and \gls{ul}-collaborative sensing using \gls{ofdm}/\gls{nr} signals~\cite{zhang2025isac,huang2025ul}, as well as passive radar sensing and \gls{csi}-based localization using \gls{nr} reference signals~\cite{dwivedi2024radar,ruan2022hiloc}. Earlier Wi-Fi sensing systems~\cite{adib2014tracking,adib2013wifi,wang2019iccv} showed that commodity Wi-Fi \gls{csi} can enable fine-grained localization and person perception. \blue{A related line of work localizes the \gls{ue} itself, with \gls{srs}-based estimates exported to xApps through OAI and FlexRIC~\cite{bouknana2025oran} or processed in real time on OAI~\cite{hsieh2026srs}, and with public \gls{csi} datasets for user positioning and channel charting collected on an NVIDIA testbed~\cite{wiesmayr2025csi}. cuSense instead localizes a person who carries no device, so its accuracy is not directly comparable to these results.}
Our work is complementary to this literature: rather than using Wi-Fi or standalone receivers, we implement \gls{csi}-based indoor localization as an uplink-collaborative \gls{isac} service on a \gls{3gpp}-compliant, GPU-accelerated \gls{5g} \gls{gnb}, exposed as a programmable dApp within an O-RAN/AI-RAN framework without dedicated sensing hardware\blue{, and we release the collected measurements as a public dataset (Section~\ref{subsec:cusense-data})}.

\section{Conclusions and Future Work}
\label{sec:conclusion}

In this work, we designed, implemented, and evaluated a GPU-accelerated framework for real-time O-RAN dApps on NVIDIA \gls{atb} that exposes \blue{\gls{phy} data} via \gls{shm} and an E3-based interface, supporting \blue{four} inference engines and multiple \gls{ai} backends. \blue{The framework adds about $84~\mu$s per slot, leaving most of the time until the next \gls{ul} slot to the dApp logic.} We profiled the framework across two \gls{gpu} platforms\blue{, with different \gls{gpu} sharing modes, and with an increasing number of cells and dApps. These results also guide developers in deploying their dApps, from the choice of the inference engine, which trades latency for flexibility, to the \gls{gpu} sharing mode, where \gls{mig} and \gls{mps} keep the processing within hundreds of microseconds while time-slicing adds about $0.5$~ms.} On top of this framework, we realized cuSense, an uplink \gls{dmrs} \gls{csi}-based indoor localization dApp operating in real time on a production-grade \gls{5g} network without dedicated sensing hardware\blue{, with a live inference latency of $241~\mu$s}. Experiments across independent unseen runs show sub-meter accuracy\blue{, with a mean error of $73$~cm,} and consistent performance, indicating robust generalization within the same environment.

Future framework improvements include broader telemetry coverage, wider bandwidth and antenna configurations\blue{, up to massive \gls{mimo},} and data-path optimizations\blue{, such as a binary encoding of the E3 messages, fewer copies along the data path, and a return path through \gls{shm} for large payloads from dApps to the \gls{ran}, as well as other \gls{gpu} sharing mechanisms, such as CUDA green contexts}. The current \gls{ul}-collaborative bistatic architecture with a transmitting \gls{ue} offers a practical deployment model but also limits spatial diversity and relies on environment-specific calibration and the availability of collaborative devices. These constraints motivate future work on dynamic environments, multi-cell configurations, multi-target detection, and sensing modes that reduce or eliminate the need for explicit collaboration.

The dApp framework is released as open source\blue{, together with the cuSense dataset}, providing the community with a reference design and inference pipelines for developing \gls{gpu}-accelerated real-time applications for \gls{ai}-native \gls{ran}.

\balance
\bibliographystyle{IEEEtran}
\bibliography{biblio/sample-base}

\begin{IEEEbiography}[{\includegraphics[width=1in,height=1.25in,clip,keepaspectratio]{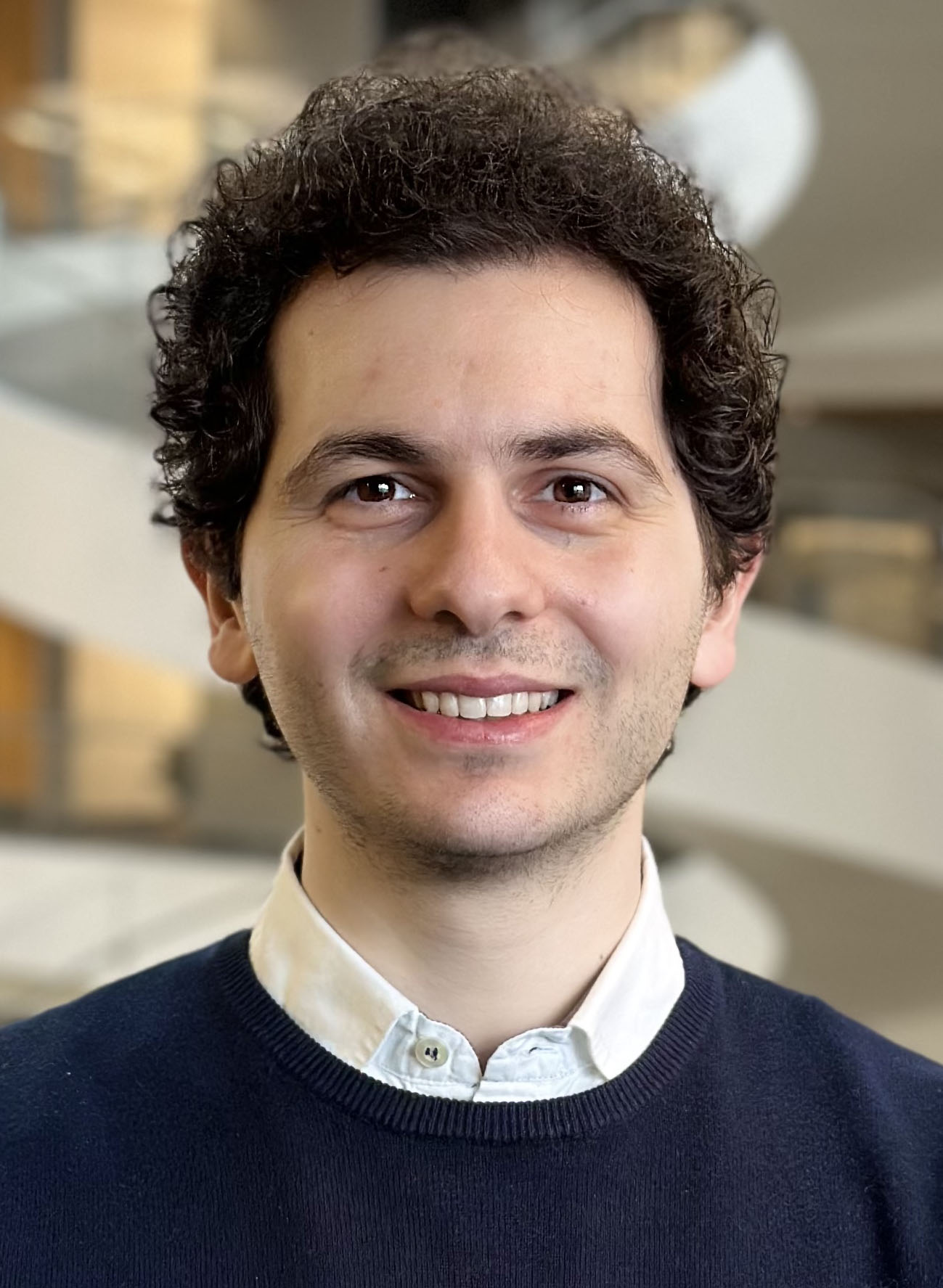}}]{Davide Villa} received his B.S. in Computer Engineering from the University of Pisa, Italy, in 2015, and his M.S. in Embedded Computing Systems from Sant’Anna School of Advanced Studies and the University of Pisa, Italy, cum Laude, in 2018. From 2018 to 2020, he worked as a Research Scientist in the Embedded Systems and Networks Group at Raytheon Technologies (former United Technologies Research Center) in Cork, Ireland. He received his Ph.D. in Computer Engineering from the Institute for Intelligent Networked Systems at Northeastern University, Boston, USA, in 2025, under the guidance of Prof. Tommaso Melodia. During his Ph.D., he interned at Samsung Research America in 2024 and NVIDIA in 2025. He joined NVIDIA as a Senior Software Engineer in 2026. His research interests include 5G and beyond cellular networks, O-RAN, channel characterization, and software-defined networking for wireless testbeds.
\end{IEEEbiography}

\vskip -2\baselineskip plus -1fil

\begin{IEEEbiography}[{\includegraphics[width=1in,height=1.25in,clip,keepaspectratio]{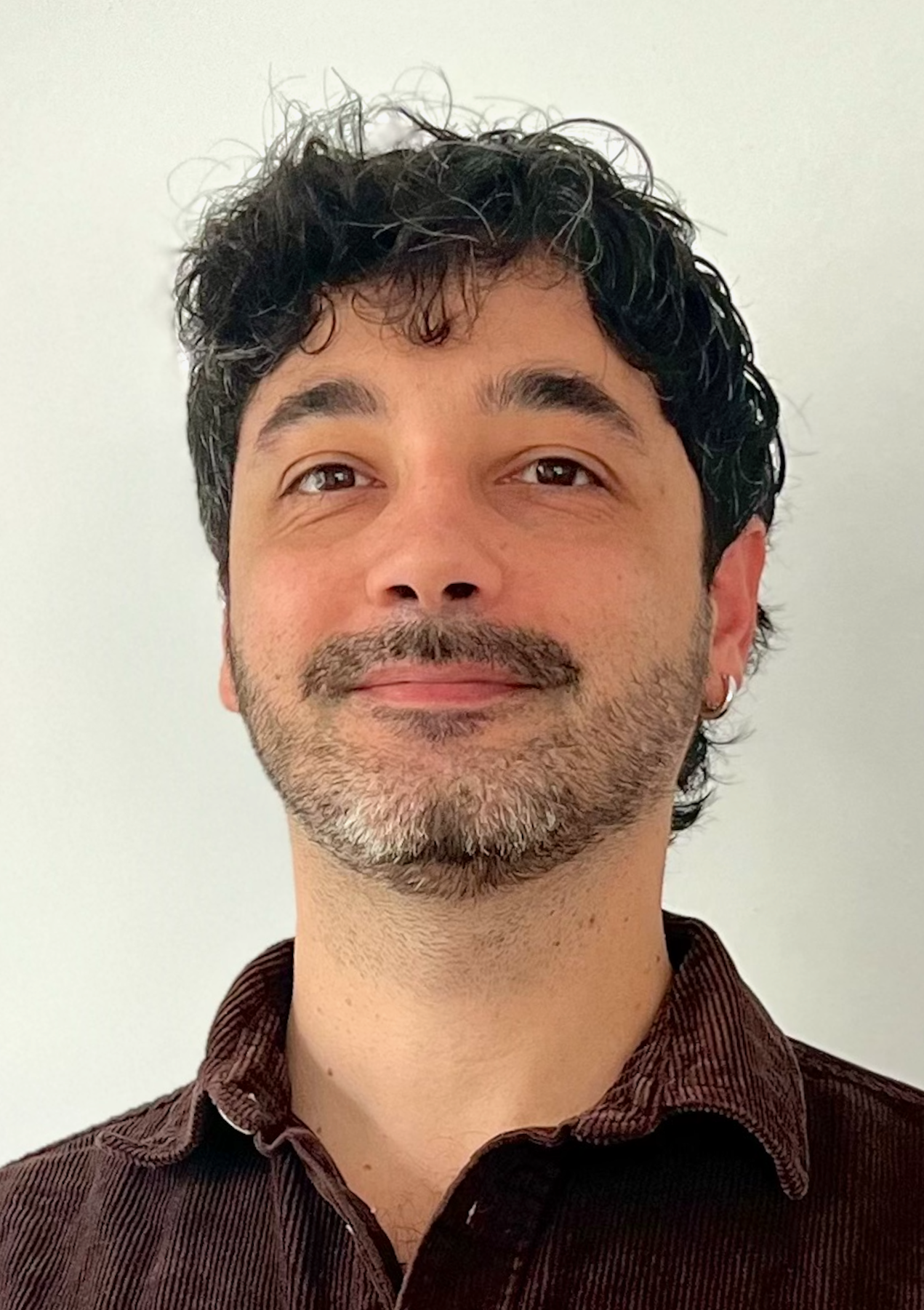}}]{Mauro Belgiovine} received his B.S. and M.S. in Computer Science from the University of Bologna, Italy, in 2013 and 2017, respectively. He received his Ph.D. degree in Computer Engineering from Northeastern University, Boston (MA), in 2025 under the guidance of Prof. Kaushik Chowdhury. He joined NVIDIA full-time in 2025 to continue his research on deep learning applications to wireless communication, digital twins, semantic communications, and generative AI.
\end{IEEEbiography}

\vskip -2\baselineskip plus -1fil

\begin{IEEEbiographynophoto}{Nicholas Hedberg} is a Senior Engineer in the Public Sector at NVIDIA in Zurich, Switzerland, having joined in October 2021. He brings extensive experience from his previous tenure at Viasat Inc., where he was a System Engineering Team Lead in Lausanne, focusing on cutting-edge phased array antennas for satellite communications. Prior roles at Viasat in Carlsbad involved significant contributions to mobile terminal Verilog modules and ASIC development. Nicholas holds a BS in Physics and a BA in Economics from UC San Diego (2003-2007).
\end{IEEEbiographynophoto}

\vskip -2\baselineskip plus -1fil

\begin{IEEEbiography}[{\includegraphics[width=1in,height=1.25in,clip,keepaspectratio]{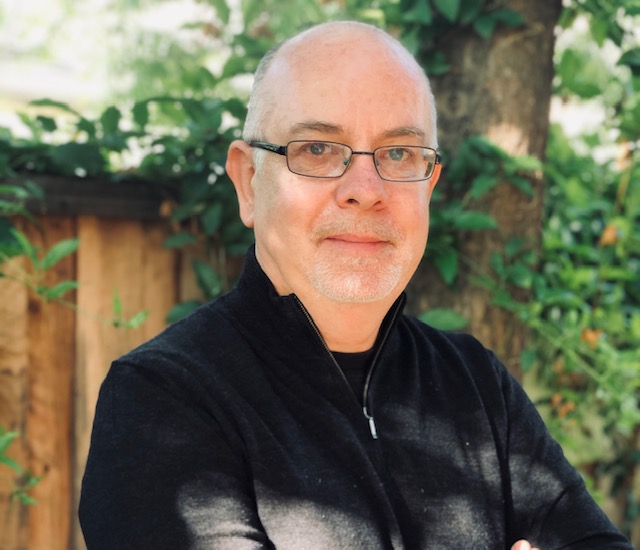}}]{Chris Dick} is a Senior Distinguished Engineer at NVIDIA, working on the application of artificial intelligence and machine learning to 5G and 6G wireless. He is part of the team developing algorithms, system architecture, design methodologies, and real-time over-the-air testbeds that enable the path from AI/ML simulation flows to real-time operation. His research is in the area of 6G architecture, ML model architecture, channel coding, design flows for GPU signal processing systems, and digital front-end technology for cellular systems, with a particular emphasis on ML algorithms for RF signal processing. From 1998 to 2020, he was a Fellow and the DSP Chief Architect at Xilinx, and before that he was a tenured academic in Melbourne, Australia, for 13 years. He has over 250 publications and over 120 patents. He chairs the AI-for-RAN Working Group of the AI-RAN Alliance and serves on the Board of Directors of the OpenAirInterface Software Alliance. In 2018, he received the IEEE Communications Society Award for Advances in Communication for research in the area of full-duplex wireless communication.
\end{IEEEbiography}

\vskip -2\baselineskip plus -1fil

\begin{IEEEbiography}[{\includegraphics[width=1in,height=1.1in,clip,keepaspectratio]{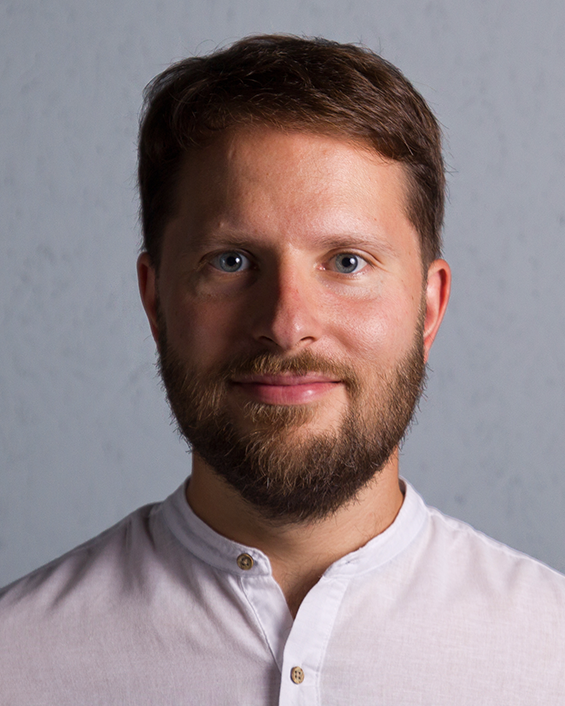}}]{Michele Polese} is a Research Assistant Professor at Northeastern University and the Technical Director of the Institute for Intelligent Networked Systems. He is also the chair of the AI-and-RAN working group of the AI-RAN Alliance and a co-founder of zTouch Networks, Inc. He received his Ph.D. from the Department of Information Engineering of the University of Padova in 2020, and then joined Northeastern University as a research scientist and part-time lecturer. During his Ph.D., he visited New York University (NYU), AT\&T Labs in Bedminster, NJ, and Northeastern University. His research interests are in the end-to-end design and testing of open, programmable, intelligent, and secure next-generation cellular networks, including Open RAN, AI-RAN, and 6G. He has contributed to O-RAN technical specifications and submitted responses to multiple FCC and NTIA notices of inquiry and requests for comments. He is an editor for the IEEE Transactions on Mobile Computing, IEEE Communications Magazine, and Computer Networks, was TPC co-chair of European Wireless 2025, WONS 2024, WiNTECH 2024, and WNS3 2021-2022, among others, and received the 2022 Mario Gerla Award for Research in Computer Science. He holds multiple patents on Open RAN and led the establishment of the Open6G testing center at Northeastern.
\end{IEEEbiography}

\vskip -2\baselineskip plus -1fil

\begin{IEEEbiography}[{\includegraphics[width=1in,height=1.25in,clip,keepaspectratio]{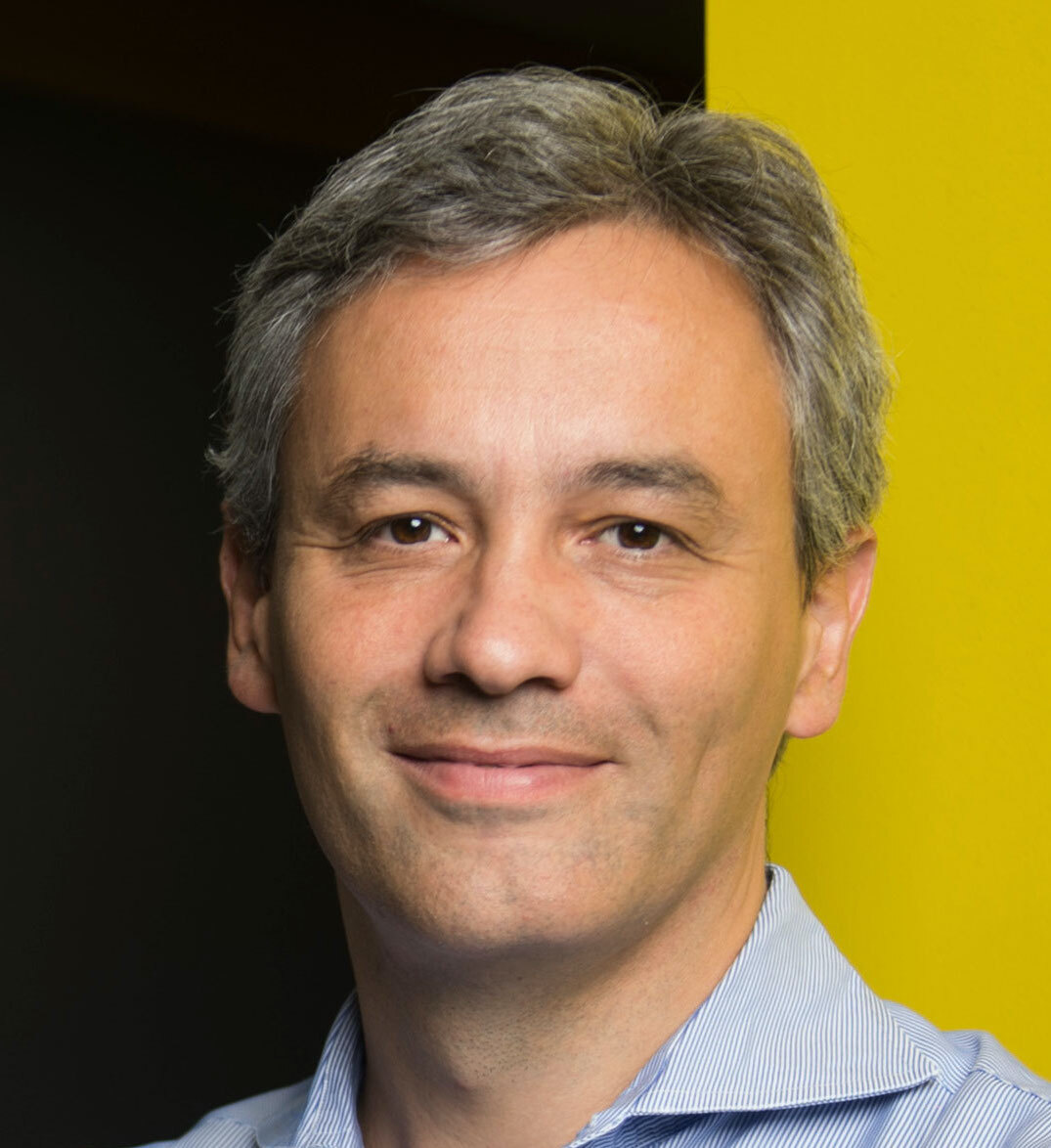}}]{Tommaso Melodia}
is the William Lincoln Smith Professor in the Department of Electrical and Computer Engineering at Northeastern University in Boston, and the Founding Director of the Institute for Intelligent Networked Systems. He received his Laurea (integrated B.S. and M.S.) from the University of Rome ``La Sapienza'' and his Ph.D. in Electrical and Computer Engineering from the Georgia Institute of Technology in 2007. He co-founded the AI-RAN Alliance, where he serves on the Executive Board, founded Open6G, and directs Colosseum. From 2017 to 2025, he was the Director of Research for the PAWR Project Office, a \$100M public-private partnership to establish four city-scale platforms for wireless research. He is the Editor-in-Chief of Computer Networks (Elsevier) and has served as Technical Program Committee Chair for IEEE INFOCOM 2018 and General Chair for IEEE SECON 2019, ACM Nanocom 2019, and ACM WUWnet 2014. He is a recipient of the National Science Foundation CAREER award, an ACM Distinguished Speaker, and a Fellow of the IEEE, the ACM, and the National Academy of Inventors. He has authored more than 370 peer-reviewed publications.
\end{IEEEbiography}

\end{document}

%% file: acronyms.tex
\newacronym{aoa}{AoA}{Angle of Arrival}
\newacronym{cuphy}{cuPHY}{CUDA Physical layer}
\newacronym{cuda}{CUDA}{Compute Unified Device Architecture}
\newacronym{cubb}{cuBB}{CUDA Baseband}
\newacronym{onnx}{ONNX}{Open Neural Network Exchange}
\newacronym{ldpc}{LDPC}{Low Density Parity-Check Code}
\newacronym{crc}{CRC}{Cyclic Redundancy Check}
\newacronym{dnn}{DNN}{Deep Neural Network}
\newacronym{ort}{ORT}{ONNX Runtime}
\newacronym{tensorrt}{TensorRT}{Tensor RealTime}
\newacronym{cnn}{CNN}{Convolutional Neural Network}
\newacronym{cudnn}{cuDNN}{CUDA Deep Neural Network library}
\newacronym{cublas}{cuBLAS}{CUDA Basic Linear Algebra Subroutines}
\newacronym{hdf5}{HDF5}{Hierarchical Data Format version 5}
\newacronym{rmr}{RMR}{RIC Message Router}
\newacronym{3gpp}{3GPP}{3rd Generation Partnership Project}
\newacronym{4g}{4G}{4th generation}
\newacronym{5g}{5G}{fifth generation}
\newacronym{6g}{6G}{sixth generation}
\newacronym{5gc}{5GC}{5G Core}
\newacronym{adc}{ADC}{Analog to Digital Converter}
\newacronym{adl}{ADL}{Aerial Data Lake}
\newacronym{aerpaw}{AERPAW}{Aerial Experimentation and Research Platform for Advanced Wireless}
\newacronym{ai}{AI}{Artificial Intelligence}
\newacronym{aimd}{AIMD}{Additive Increase Multiplicative Decrease}
\newacronym{am}{AM}{Acknowledged Mode}
\newacronym{amc}{AMC}{Adaptive Modulation and Coding}
\newacronym{amf}{AMF}{Access and Mobility Management Function}
\newacronym{aops}{AOPS}{Adaptive Order Prediction Scheduling}
\newacronym{api}{API}{Application Programming Interface}
\newacronym{apn}{APN}{Access Point Name}
\newacronym{aqm}{AQM}{Active Queue Management}
\newacronym{arc-ota}{ARC-OTA}{Aerial RAN CoLab Over-the-Air}
\newacronym{atb}{ATB}{Aerial Testbed}
\newacronym{ausf}{AUSF}{Authentication Server Function}
\newacronym{avc}{AVC}{Advanced Video Coding}
\newacronym{awgn}{AWGN}{Additive White Gaussian Noise}
\newacronym{balia}{BALIA}{Balanced Link Adaptation Algorithm}
\newacronym{bbu}{BBU}{Base Band Unit}
\newacronym{bdp}{BDP}{Bandwidth-Delay Product}
\newacronym{ber}{BER}{Bit Error Rate}
\newacronym{bf}{BF}{Beamforming}
\newacronym{bler}{BLER}{Block Error Rate}
\newacronym{brr}{BRR}{Bayesian Ridge Regressor}
\newacronym{bsr}{BSR}{Buffer Status Report}
\newacronym{bs}{BS}{Base Station}
\newacronym{bpsk}{BPSK}{Binary Phase-shift keying}
\newacronym{bss}{BSS}{Business Support System}
\newacronym{ca}{CA}{Carrier Aggregation}
\newacronym{caas}{CaaS}{Connectivity-as-a-Service}
\newacronym{cb}{CB}{Code Block}
\newacronym{cc}{CC}{Congestion Control}
\newacronym{ccid}{CCID}{Congestion Control ID}
\newacronym{cco}{CC}{Carrier Component}
\newacronym{cdd}{CDD}{Cyclic Delay Diversity}
\newacronym{cdf}{CDF}{Cumulative Distribution Function}
\newacronym{cdn}{CDN}{Content Distribution Network}
\newacronym{cir}{CIR}{Channel Impulse Response}
\newacronym{cn}{CN}{Core Network}
\newacronym{codel}{CoDel}{Controlled Delay Management}
\newacronym{comac}{COMAC}{Converged Multi-Access and Core}
\newacronym{cord}{CORD}{Central Office Re-architected as a Datacenter}
\newacronym{cornet}{CORNET}{COgnitive Radio NETwork}
\newacronym{cosmos}{COSMOS}{Cloud Enhanced Open Software Defined Mobile Wireless Testbed for City-Scale Deployment}
\newacronym{cots}{COTS}{Commercial Off-the-Shelf}
\newacronym{cp}{CP}{Control Plane}
\newacronym{cpu}{CPU}{Central Processing Unit}
\newacronym{cqi}{CQI}{Channel Quality Information}
\newacronym{cr}{CR}{Cognitive Radio}
\newacronym{cran}{CRAN}{Cloud \gls{ran}}
\newacronym{crs}{CRS}{Cell Reference Signal}
\newacronym{csi}{CSI}{Channel State Information}
\newacronym{csirs}{CSI-RS}{Channel State Information - Reference Signal}
\newacronym{cu}{CU}{Central Unit}
\newacronym{d2tcp}{D$^2$TCP}{Deadline-aware Data center TCP}
\newacronym{d3}{D$^3$}{Deadline-Driven Delivery}
\newacronym{dac}{DAC}{Digital to Analog Converter}
\newacronym{dag}{DAG}{Directed Acyclic Graph}
\newacronym{dapp}{dApp}{distributed Application}
\newacronym{darpa}{DARPA}{Defense Advanced Research Projects Agency}
\newacronym{das}{DAS}{Distributed Antenna System}
\newacronym{dash}{DASH}{Dynamic Adaptive Streaming over HTTP}
\newacronym{dc}{DC}{Direct Current}
\newacronym{dccp}{DCCP}{Datagram Congestion Control Protocol}
\newacronym{dce}{DCE}{Direct Code Execution}
\newacronym{dci}{DCI}{Downlink Control Information}
\newacronym{dcl}{DCL}{Dear Colleague Letter}
\newacronym{dctcp}{DCTCP}{Data Center TCP}
\newacronym{dl}{DL}{Downlink}
\newacronym{dma}{DMA}{Direct Memory Access}
\newacronym{dmr}{DMR}{Deadline Miss Ratio}
\newacronym{dmrs}{DMRS}{Demodulation Reference Signal}
\newacronym{dpu}{DPU}{Data Processing Unit}
\newacronym{drlcc}{DRL-CC}{Deep Reinforcement Learning Congestion Control}
\newacronym{drs}{DRS}{Discovery Reference Signal}
\newacronym{du}{DU}{Distributed Unit}
\newacronym{e2e}{E2E}{end-to-end}
\newacronym{e2sm}{E2SM}{E2 Service Model}
\newacronym{e3ap}{E3AP}{E3 Application Protocol}
\newacronym{e3sm}{E3SM}{E3 Service Model}
\newacronym{ecaas}{ECaaS}{Edge-Cloud-as-a-Service}
\newacronym{ecn}{ECN}{Explicit Congestion Notification}
\newacronym{edf}{EDF}{Earliest Deadline First}
\newacronym{eirp}{EIRP}{Effective Isotropic Radiated Power}
\newacronym{em}{EM}{Electro-Magnetic}
\newacronym{embb}{eMBB}{Enhanced Mobile Broadband}
\newacronym{empower}{EMPOWER}{EMpowering transatlantic PlatfOrms for advanced WirEless Research}
\newacronym{enb}{eNB}{evolved Node Base}
\newacronym{endc}{EN-DC}{E-UTRAN-\gls{nr} \gls{dc}}
\newacronym{epc}{EPC}{Evolved Packet Core}
\newacronym{eps}{EPS}{Evolved Packet System}
\newacronym{es}{ES}{Edge Server}
\newacronym{etsi}{ETSI}{European Telecommunications Standards Institute}
\newacronym[firstplural=Estimated Times of Arrival (ETAs)]{eta}{ETA}{Estimated Time of Arrival}
\newacronym{eutran}{E-UTRAN}{Evolved Universal Terrestrial Access Network}
\newacronym{faas}{FaaS}{Function-as-a-Service}
\newacronym{fapi}{FAPI}{Functional Application Platform Interface}
\newacronym{fcc}{FCC}{Federal Communications Commission}
\newacronym{fdd}{FDD}{Frequency Division Duplexing}
\newacronym{fdm}{FDM}{Frequency Division Multiplexing}
\newacronym{fdma}{FDMA}{Frequency Division Multiple Access}
\newacronym{fed4fire}{FED4FIRE+}{Federation 4 Future Internet Research and Experimentation Plus}
\newacronym{fir}{FIR}{Finite Impulse Response}
\newacronym{fit}{FIT}{Future \acrlong{iot}}
\newacronym{fpga}{FPGA}{Field Programmable Gate Array}
\newacronym{fps}{FPS}{Frames per second}
\newacronym{fr1}{FR1}{Frequency Range 1}
\newacronym{fr2}{FR2}{Frequency Range 2}
\newacronym{fs}{FS}{Fast Switching}
\newacronym{fscc}{FSCC}{Flow Sharing Congestion Control}
\newacronym{ftp}{FTP}{File Transfer Protocol}
\newacronym{fw}{FW}{Flow Window}
\newacronym{ga128}{Ga}{Golay Sequence type A}
\newacronym{ge}{GE}{Gaussian Elimination}
\newacronym{gh}{GH}{Grace Hopper}
\newacronym{glfsr}{GLFSR}{Galois Linear Feedback Shift Register}
\newacronym{gnb}{gNB}{Next Generation Node B}
\newacronym{gold}{Gold}{Gold}
\newacronym{gop}{GOP}{Group of Pictures}
\newacronym{gpr}{GPR}{Gaussian Process Regressor}
\newacronym{gpu}{GPU}{Graphics Processing Unit}
\newacronym{grpc}{gRPC}{gRPC Remote Procedure Calls}
\newacronym{gtp}{GTP}{GPRS Tunneling Protocol}
\newacronym{gtpc}{GTP-C}{GPRS Tunnelling Protocol Control Plane}
\newacronym{gtpu}{GTP-U}{GPRS Tunnelling Protocol User Plane}
\newacronym{gtpv2c}{GTPv2-C}{\gls{gtp} v2 - Control}
\newacronym{gw}{GW}{Gateway}
\newacronym{harq}{HARQ}{Hybrid Automatic Repeat Request}
\newacronym{hest}{$\hat{\mathbf{H}}$}{Channel Estimates}
\newacronym{hetnet}{HetNet}{Heterogeneous Network}
\newacronym{hh}{HH}{Hard Handover}
\newacronym{hol}{HOL}{Head-of-Line}
\newacronym{hqf}{HQF}{Highest-quality-first}
\newacronym{hss}{HSS}{Home Subscription Server}
\newacronym{http}{HTTP}{HyperText Transfer Protocol}
\newacronym{ia}{IA}{Initial Access}
\newacronym{iab}{IAB}{Integrated Access and Backhaul}
\newacronym{ic}{IC}{Incident Command}
\newacronym{ietf}{IETF}{Internet Engineering Task Force}
\newacronym{ifw}{IFW}{Interference Free Window}
\newacronym{imsi}{IMSI}{International Mobile Subscriber Identity}
\newacronym{imt}{IMT}{International Mobile Telecommunication}
\newacronym{iot}{IoT}{Internet of Things}
\newacronym{ip}{IP}{Internet Protocol}
\newacronym{iq}{I/Q}{In-phase and Quadrature}
\newacronym{isac}{ISAC}{Integrated Sensing and Communication}
\newacronym{itu}{ITU}{International Telecommunication Union}
\newacronym{iw}{IW}{Interference Whitening}
\newacronym{jit}{JIT}{Just-In-Time}
\newacronym{kl}{KL}{Kullback–Leibler}
\newacronym{kpi}{KPI}{Key Performance Indicator}
\newacronym{kpm}{KPM}{Key Performance Measurement}
\newacronym{kvm}{KVM}{Kernel-based Virtual Machine}
\newacronym{leo}{LEO}{Low Earth Orbit}
\newacronym{los}{LOS}{Line-of-Sight}
\newacronym{ls}{LS}{Loosely Synchronised}
\newacronym{lsm}{LSM}{Link-to-System Mapping}
\newacronym{lstm}{LSTM}{Long Short Term Memory}
\newacronym{lte}{LTE}{Long Term Evolution}
\newacronym{lxc}{LXC}{Linux Container}
\newacronym{m2m}{M2M}{Machine to Machine}
\newacronym{mac}{MAC}{Medium Access Control}
\newacronym{manet}{MANET}{Mobile Ad Hoc Network}
\newacronym{mano}{MANO}{Management and Orchestration}
\newacronym{mc}{MC}{Multi-Connectivity}
\newacronym{mcc}{MCC}{Mobile Cloud Computing}
\newacronym{mchem}{MCHEM}{Massive Channel Emulator}
\newacronym{mcs}{MCS}{Modulation and Coding Scheme}
\newacronym{mec}{MEC}{Multi-access Edge Computing}
\newacronym{mec2}{MEC}{Mobile Edge Cloud}
\newacronym{mfc}{MFC}{Mobile Fog Computing}
\newacronym{mi}{MI}{Mutual Information}
\newacronym{mig}{MIG}{Multi-Instance GPU}
\newacronym{mib}{MIB}{Master Information Block}
\newacronym{miesm}{MIESM}{Mutual Information Based Effective SINR}
\newacronym{mimo}{MIMO}{Multiple Input, Multiple Output}
\newacronym{mgen}{MGEN}{Multi-Generator}
\newacronym{ml}{ML}{Machine Learning}
\newacronym{mlp}{MLP}{Multilayer Perceptron}
\newacronym{mlr}{MLR}{Maximum-local-rate}
\newacronym[plural=\gls{mme}s,firstplural=Mobility Management Entities (MMEs)]{mme}{MME}{Mobility Management Entity}
\newacronym{mmtc}{mMTC}{Massive Machine-Type Communications}
\newacronym{mmwave}{mmWave}{millimeter wave}
\newacronym{mpdccp}{MP-DCCP}{Multipath Datagram Congestion Control Protocol}
\newacronym{mps}{MPS}{Multi-Process Service}
\newacronym{mptcp}{MPTCP}{Multipath TCP}
\newacronym{mr}{MR}{Maximum Rate}
\newacronym{mrdc}{MR-DC}{Multi \gls{rat} \gls{dc}}
\newacronym{mse}{MSE}{Mean Square Error}
\newacronym{mss}{MSS}{Maximum Segment Size}
\newacronym{mt}{MT}{Mobile Termination}
\newacronym{mtd}{MTD}{Machine-Type Device}
\newacronym{mtu}{MTU}{Maximum Transmission Unit}
\newacronym{mumimo}{MU-MIMO}{Multi-user \gls{mimo}}
\newacronym{mvno}{MVNO}{Mobile Virtual Network Operator}
\newacronym{nalu}{NALU}{Network Abstraction Layer Unit}
\newacronym{nas}{NAS}{Network Attached Storage}
\newacronym{nbiot}{NB-IoT}{Narrow Band IoT}
\newacronym{nfv}{NFV}{Network Function Virtualization}
\newacronym{nfvi}{NFVI}{Network Function Virtualization Infrastructure}
\newacronym{nic}{NIC}{Network Interface Card}
\newacronym{nlos}{NLOS}{Non-Line-of-Sight}
\newacronym{now}{NOW}{Non Overlapping Window}
\newacronym{nn}{NN}{Neural Network}
\newacronym{nrdz}{NRDZ}{National Radio Dynamic Zone}
\newacronym{nsf}{NSF}{National Science Foundation}
\newacronym{nsm}{NSM}{Network Service Mesh}
\newacronym[type=hidden]{nr}{NR}{New Radio}
\newacronym{nrf}{NRF}{Network Repository Function}
\newacronym{nsa}{NSA}{Non Stand Alone}
\newacronym{nse}{NSE}{Network Slicing Engine}
\newacronym{nssf}{NSSF}{Network Slice Selection Function}
\newacronym{ntp}{NTP}{Network Time Protocol}
\newacronym{o2i}{O2I}{Outdoor to Indoor}
\newacronym{oai}{OAI}{OpenAirInterface}
\newacronym{oaic}{OAIC}{Open AI Cellular}
\newacronym{oaicn}{OAI-CN}{\gls{oai} \acrlong{cn}}
\newacronym{oairan}{OAI-RAN}{\acrlong{oai} \acrlong{ran}}
\newacronym{oam}{OAM}{Operations, Administration and Maintenance}
\newacronym[plural=\gls{obu}s,firstplural=Onboard Units (OBUs)]{obu}{OBU}{Onboard Unit}
\newacronym{odc}{ODC}{ORAN Development Company}
\newacronym{ofdm}{OFDM}{Orthogonal Frequency Division Multiplexing}
\newacronym{olia}{OLIA}{Opportunistic Linked Increase Algorithm}
\newacronym{omec}{OMEC}{Open Mobile Evolved Core}
\newacronym{onap}{ONAP}{Open Network Automation Platform}
\newacronym{onf}{ONF}{Open Networking Foundation}
\newacronym{onos}{ONOS}{Open Networking Operating System}
\newacronym{oom}{OOM}{\gls{onap} Operations Manager}
\newacronym{opnfv}{OPNFV}{Open Platform for \gls{nfv}}
\newacronym{orbit}{ORBIT}{Open-Access Research Testbed for Next-Generation Wireless Networks}
\newacronym{os}{OS}{Operating System}
\newacronym{osc}{OSC}{O-RAN Software Community}
\newacronym{osm}{OSM}{Open Street Map}
\newacronym{oss}{OSS}{Operations Support System}
\newacronym{pa}{PA}{Position-aware}
\newacronym{pase}{PASE}{Prioritization, Arbitration, and Self-adjusting Endpoints}
\newacronym{pawr}{PAWR}{Platforms for Advanced Wireless Research}
\newacronym{pbch}{PBCH}{Physical Broadcast Channel}
\newacronym{pci}{PCI}{Peripheral Component Interconnect}
\newacronym{pcef}{PCEF}{Policy and Charging Enforcement Function}
\newacronym{pcfich}{PCFICH}{Physical Control Format Indicator Channel}
\newacronym{pcrf}{PCRF}{Policy and Charging Rules Function}
\newacronym{pdcch}{PDCCH}{Physical Downlink Control Channel}
\newacronym{pdcp}{PDCP}{Packet Data Convergence Protocol}
\newacronym{pdsch}{PDSCH}{Physical Downlink Shared Channel}
\newacronym{pdu}{PDU}{Packet Data Unit}
\newacronym{pdp}{PDP}{Power Delay Profile}
\newacronym{pf}{PF}{Proportional Fair}
\newacronym{pgw}{PGW}{Packet Gateway}
\newacronym{ph}{PH}{Power Headroom}
\newacronym{phich}{PHICH}{Physical Hybrid ARQ Indicator Channel}
\newacronym{phy}{PHY}{Physical}
\newacronym{pl}{PL}{Path Loss}
\newacronym{pmch}{PMCH}{Physical Multicast Channel}
\newacronym{pmi}{PMI}{Precoding Matrix Indicators}
\newacronym{powder}{POWDER}{Platform for Open Wireless Data-driven Experimental Research}
\newacronym{ppo}{PPO}{Proximal Policy Optimization}
\newacronym{ppp}{PPP}{Poisson Point Process}
\newacronym{prach}{PRACH}{Physical Random Access Channel}
\newacronym{prb}{PRB}{Physical Resource Block}
\newacronym{psnr}{PSNR}{Peak Signal to Noise Ratio}
\newacronym{pss}{PSS}{Primary Synchronization Signal}
\newacronym{pucch}{PUCCH}{Physical Uplink Control Channel}
\newacronym{pusch}{PUSCH}{Physical Uplink Shared Channel}
\newacronym{qam}{QAM}{Quadrature Amplitude Modulation}
\newacronym{qci}{QCI}{\gls{qos} Class Identifier}
\newacronym{qoe}{QoE}{Quality of Experience}
\newacronym{qos}{QoS}{Quality of Service}
\newacronym{qtgui}{QT-GUI}{QT Graphical User Interface}
\newacronym{qsfp28}{QSFP28}{Quad Small Form-factor Pluggable 28}
\newacronym{quic}{QUIC}{Quick UDP Internet Connections}
\newacronym{rach}{RACH}{Random Access Channel}
\newacronym{ran}{RAN}{Radio Access Network}
\newacronym[firstplural=Radio Access Technologies (RATs)]{rat}{RAT}{Radio Access Technology}
\newacronym{rb}{RB}{Resource Block}
\newacronym{rcn}{RCN}{Research Coordination Network}
\newacronym{rdma}{RDMA}{Remote Direct Memory Access}
\newacronym{rec}{REC}{Radio Edge Cloud}
\newacronym{red}{RED}{Random Early Detection}
\newacronym{renew}{RENEW}{Reconfigurable Eco-system for Next-generation End-to-end Wireless}
\newacronym{rf}{RF}{Radio Frequency}
\newacronym{rfc}{RFC}{Request for Comments}
\newacronym{rfr}{RFR}{Random Forest Regressor}
\newacronym{ric}{RIC}{RAN Intelligent Controller}
\newacronym{near-rt-ric}{near-RT-RIC}{near-RT-\gls{ric}}
\newacronym{non-rt}{non-RT}{non-Real-Time}
\newacronym{near-rt}{near-RT}{near-Real-Time}
\newacronym{rlc}{RLC}{Radio Link Control}
\newacronym{rlf}{RLF}{Radio Link Failure}
\newacronym{rlnc}{RLNC}{Random Linear Network Coding}
\newacronym{rmse}{RMSE}{Root Mean Squared Error}
\newacronym{rnis}{RNIS}{Radio Network Information Service}
\newacronym{rr}{RR}{Round Robin}
\newacronym{rrc}{RRC}{Radio Resource Control}
\newacronym{rrm}{RRM}{Radio Resource Management}
\newacronym{rru}{RRU}{Remote Radio Unit}
\newacronym{rs}{RS}{Remote Server}
\newacronym{rsrp}{RSRP}{Reference Signal Received Power}
\newacronym{rsrq}{RSRQ}{Reference Signal Received Quality}
\newacronym{rss}{RSS}{Received Signal Strength}
\newacronym{rssi}{RSSI}{Received Signal Strength Indicator}
\newacronym{rsu}{RSU}{Road-Side Unit}
\newacronym{rt}{RT}{Real-Time}
\newacronym{rtt}{RTT}{Round Trip Time}
\newacronym{ru}{RU}{Radio Unit}
\newacronym{rw}{RW}{Receive Window}
\newacronym{rx}{RX}{Receiver}
\newacronym{s1ap}{S1AP}{S1 Application Protocol}
\newacronym{sa}{SA}{standalone}
\newacronym{sack}{SACK}{Selective Acknowledgment}
\newacronym{sap}{SAP}{Service Access Point}
\newacronym{sas}{SAS}{Spectrum Access System}
\newacronym{sc2}{SC2}{Spectrum Collaboration Challenge}
\newacronym{scef}{SCEF}{Service Capability Exposure Function}
\newacronym{sch}{SCH}{Secondary Cell Handover}
\newacronym{scoot}{SCOOT}{Split Cycle Offset Optimization Technique}
\newacronym{sfp+}{SFP+}{Small Form-factor Pluggable Plus}
\newacronym{sctp}{SCTP}{Stream Control Transmission Protocol}
\newacronym{sdap}{SDAP}{Service Data Adaptation Protocol}
\newacronym{sd}{SD}{Standard Deviation}
\newacronym{sdk}{SDK}{Software Development Kit}
\newacronym{sdm}{SDM}{Space Division Multiplexing}
\newacronym{sdma}{SDMA}{Spatial Division Multiple Access}
\newacronym{sdn}{SDN}{Software-defined Networking}
\newacronym{sdr}{SDR}{Software-defined Radio}
\newacronym{seba}{SEBA}{SDN-Enabled Broadband Access}
\newacronym{sgsn}{SGSN}{Serving GPRS Support Node}
\newacronym{sgw}{SGW}{Service Gateway}
\newacronym{shm}{SHM}{Shared Memory}
\newacronym{si}{SI}{Study Item}
\newacronym{sib}{SIB}{Secondary Information Block}
\newacronym{sinr}{SINR}{Signal to Interference plus Noise Ratio}
\newacronym{sip}{SIP}{Session Initiation Protocol}
\newacronym{siso}{SISO}{Single Input, Single Output}
\newacronym{sla}{SLA}{Service Level Agreement}
\newacronym{sm}{SM}{Service Model}
\newacronym{smf}{SMF}{Session Management Function}
\newacronym{smo}{SMO}{Service Management and Orchestration}
\newacronym{sms}{SMS}{Short Message Service}
\newacronym{smsgmsc}{SMS-GMSC}{\gls{sms}-Gateway}
\newacronym{snr}{SNR}{Signal-to-Noise Ratio}
\newacronym{son}{SON}{Self-Organizing Network}
\newacronym{sptcp}{SPTCP}{Single Path TCP}
\newacronym{srb}{SRB}{Service Radio Bearer}
\newacronym{srn}{SRN}{Standard Radio Node}
\newacronym{srs}{SRS}{Sounding Reference Signal}
\newacronym{ss}{SS}{Synchronization Signal}
\newacronym{sss}{SSS}{Secondary Synchronization Signal}
\newacronym{st}{ST}{Spanning Tree}
\newacronym{svc}{SVC}{Scalable Video Coding}
\newacronym{synce}{SyncE}{Synchronous Ethernet}
\newacronym{tai}{TAI}{International Atomic Time}
\newacronym{tb}{TB}{Transport Block}
\newacronym{tcp}{TCP}{Transmission Control Protocol}
\newacronym{tdd}{TDD}{Time Division Duplexing}
\newacronym{tdm}{TDM}{Time Division Multiplexing}
\newacronym{tdma}{TDMA}{Time Division Multiple Access}
\newacronym{tf}{TF}{Tensorflow}
\newacronym{tfl}{TfL}{Transport for London}
\newacronym{tfrc}{TFRC}{TCP-Friendly Rate Control}
\newacronym{tft}{TFT}{Traffic Flow Template}
\newacronym{tgen}{TGEN}{Traffic Generator}
\newacronym{tip}{TIP}{Telecom Infra Project}
\newacronym{tm}{TM}{Transparent Mode}
\newacronym{to}{TO}{Telco Operator}
\newacronym{toa}{ToA}{Time of Arrival}
\newacronym{tl}{TL}{Transfer Learning}
\newacronym{tr}{TR}{Technical Report}
\newacronym{trt}{TRT}{TensorRT}
\newacronym{trp}{TRP}{Transmitter Receiver Pair}
\newacronym{ts}{TS}{Technical Specification}
\newacronym{tti}{TTI}{Transmission Time Interval}
\newacronym{ttt}{TTT}{Time-to-Trigger}
\newacronym{tx}{TX}{Transmitter}
\newacronym{uas}{UAS}{Unmanned Aerial System}
\newacronym{uav}{UAV}{Unmanned Aerial Vehicle}
\newacronym{udm}{UDM}{Unified Data Management}
\newacronym{udp}{UDP}{User Datagram Protocol}
\newacronym{udr}{UDR}{Unified Data Repository}
\newacronym{ue}{UE}{User Equipment}
\newacronym{uhd}{UHD}{\gls{usrp} Hardware Driver}
\newacronym{ul}{UL}{uplink}
\newacronym{um}{UM}{Unacknowledged Mode}
\newacronym{uml}{UML}{Unified Modeling Language}
\newacronym{upa}{UPA}{Uniform Planar Array}
\newacronym{upf}{UPF}{User Plane Function}
\newacronym{urllc}{URLLC}{Ultra Reliable and Low Latency Communication}
\newacronym{usa}{U.S.}{United States}
\newacronym{usim}{USIM}{Universal Subscriber Identity Module}
\newacronym{usrp}{USRP}{Universal Software Radio Peripheral}
\newacronym{utc}{UTC}{Coordinated Universal Time}
\newacronym{vim}{VIM}{Virtualization Infrastructure Manager}
\newacronym{vlan}{VLAN}{Virtual Local Area Network}
\newacronym{vm}{VM}{Virtual Machine}
\newacronym{vnf}{VNF}{Virtual Network Function}
\newacronym{volte}{VoLTE}{Voice over \gls{lte}}
\newacronym{voltha}{VOLTHA}{Virtual OLT HArdware Abstraction}
\newacronym{vr}{VR}{Virtual Reality}
\newacronym{vran}{vRAN}{Virtualized \gls{ran}}
\newacronym{vss}{VSS}{Video Streaming Server}
\newacronym{wbf}{WBF}{Wired Bias Function}
\newacronym{wf}{WF}{Wired-first}
\newacronym{wi}{WI}{Wireless InSite}
\newacronym{wlan}{WLAN}{Wireless Local Area Network}
\newacronym{zmq}{ZMQ}{ZeroMQ}
\newacronym{pnf}{PNF}{Physical Network Function}
\newacronym{drl}{DRL}{Deep Reinforcement Learning}
\newacronym{mtc}{MTC}{Machine-type Communications}
\newacronym{v2x}{V2X}{Vehicle-to-everything}
\newacronym{cast}{\textit{CaST}}{Channel emulation generator and Sounder Toolchain}
\newacronym{abr}{ABR}{Adaptive Bitrate Streaming}
\newacronym{arc}{ARC}{Aerial RAN CoLab}
\newacronym{dsp}{DSP}{Digital Signal Processing}
\newacronym{ota}{OTA}{Over-the-Air}
\newacronym{bom}{BoM}{Bill of Materials}
\newacronym{frand}{FRAND}{Fair, Reasonable, And Non-Discriminatory}
\newacronym{nvipc}{NVIPC}{NVIDIA Inter-Process Communication}
\newacronym{hdr}{HDR}{High Dynamic Range}
\newacronym{ipc}{IPC}{Inter-Process Communication}
\newacronym{uci}{UCI}{Uplink Control Indication}
\newacronym{cbrs}{CBRS}{Citizen Broadband Radio Service}
\newacronym{ptp}{PTP}{Precision Time Protocol}
\newacronym{scf}{SCF}{Small Cell Forum}
\newacronym{re}{RE}{Resource Element}
\newacronym{fp16}{FP16}{Float16}
\newacronym{fp32}{FP32}{Float32}
\newacronym{int32}{INT32}{32-bit Integer}
\newacronym{tdl}{TDL}{Tapped Delay Line}
\newacronym{fh}{FH}{Front-haul}
\newacronym{ta}{TA}{Timing Advance}
\newacronym{cfo}{CFO}{Carrier Frequency Offset}
\newacronym{sir}{SIR}{Signal to Interference Ratio}
\newacronym{mmse}{MMSE}{Minimum Mean Square Error}
\newacronym{irc}{IRC}{Interference Rejection Combining}

%% file: figures_tex/datapath_bars.tex
\begin{tikzpicture}
\definecolor{s2}{HTML}{F0E442}
\definecolor{s3}{HTML}{E69F00}
\definecolor{s4}{HTML}{56B4E9}
\definecolor{s5}{HTML}{009E73}
\definecolor{s6}{HTML}{0072B2}
\definecolor{s7}{HTML}{CC79A7}
\begin{axis}[
width=\fwidth, height=\fheight, scale only axis,
xmin=0, xmax=150, ymin=-0.5, ymax=6.70,
axis y line*=left, axis x line*=bottom, xtick={0,25,50,75,100,125,150},
xticklabel style={font=\scriptsize}, yticklabel style={font=\footnotesize},
xlabel={Latency [$\mu$s]}, xlabel style={font=\footnotesize},
ytick={5.80,4.80,3.80,2.80,1.00,0.00},
yticklabels={{$\hat{\mathbf{H}}$},{I/Q},{All streams},{SRS},{MPS},{MIG}},
clip=false, legend columns=4, legend cell align=left,
legend style={font=\scriptsize, draw=none, at={(0.5,1.02)}, anchor=south, column sep=3pt},
legend image code/.code={\draw[#1] (0cm,-0.08cm) rectangle (0.3cm,0.08cm);},
]
\addlegendimage{pattern=north east lines, pattern color=black!55, draw=black!55}
\addlegendentry{1 GPU$\rightarrow$host}
\addlegendimage{fill=s2, draw=none}
\addlegendentry{2 Notify}
\addlegendimage{fill=s3, draw=none}
\addlegendentry{3 Copy to SHM}
\addlegendimage{empty legend}
\addlegendentry{}
\addlegendimage{fill=s4, draw=none}
\addlegendentry{4 E3 indication}
\addlegendimage{fill=s5, draw=none}
\addlegendentry{5 Processing}
\addlegendimage{fill=s6, draw=none}
\addlegendentry{6 E3 control}
\addlegendimage{fill=s7, draw=none}
\addlegendentry{7 Apply}
\draw[pattern=north east lines, pattern color=black!55, draw=black!55] (axis cs:0,5.490) rectangle (axis cs:26.0,6.110);
\node[font=\tiny, fill=white, inner sep=0.5pt] at (axis cs:13.0,5.800) {26};
\fill[s2] (axis cs:26.00,5.490) rectangle (axis cs:28.20,6.110);
\fill[s3] (axis cs:28.20,5.490) rectangle (axis cs:36.20,6.110);
\fill[s4] (axis cs:36.20,5.490) rectangle (axis cs:55.80,6.110);
\node[font=\tiny, text=black] at (axis cs:46.00,5.800) {20};
\fill[s5] (axis cs:55.80,5.490) rectangle (axis cs:56.20,6.110);
\fill[s6] (axis cs:56.20,5.490) rectangle (axis cs:69.10,6.110);
\node[font=\tiny, text=white] at (axis cs:62.65,5.800) {13};
\fill[s7] (axis cs:69.10,5.490) rectangle (axis cs:72.10,6.110);
\node[font=\scriptsize, anchor=west] at (axis cs:73.60,5.800) {\textbf{72} / $\leq$88};
\draw[pattern=north east lines, pattern color=black!55, draw=black!55] (axis cs:0,4.490) rectangle (axis cs:26.0,5.110);
\node[font=\tiny, fill=white, inner sep=0.5pt] at (axis cs:13.0,4.800) {26};
\fill[s2] (axis cs:26.00,4.490) rectangle (axis cs:28.00,5.110);
\fill[s3] (axis cs:28.00,4.490) rectangle (axis cs:42.70,5.110);
\node[font=\tiny, text=black] at (axis cs:35.35,4.800) {15};
\fill[s4] (axis cs:42.70,4.490) rectangle (axis cs:64.60,5.110);
\node[font=\tiny, text=black] at (axis cs:53.65,4.800) {22};
\fill[s5] (axis cs:64.60,4.490) rectangle (axis cs:65.00,5.110);
\fill[s6] (axis cs:65.00,4.490) rectangle (axis cs:78.80,5.110);
\node[font=\tiny, text=white] at (axis cs:71.90,4.800) {14};
\fill[s7] (axis cs:78.80,4.490) rectangle (axis cs:81.90,5.110);
\node[font=\scriptsize, anchor=west] at (axis cs:83.40,4.800) {\textbf{82} / $\leq$97};
\draw[pattern=north east lines, pattern color=black!55, draw=black!55] (axis cs:0,3.490) rectangle (axis cs:26.0,4.110);
\node[font=\tiny, fill=white, inner sep=0.5pt] at (axis cs:13.0,3.800) {26};
\fill[s2] (axis cs:26.00,3.490) rectangle (axis cs:28.20,4.110);
\fill[s3] (axis cs:28.20,3.490) rectangle (axis cs:50.40,4.110);
\node[font=\tiny, text=black] at (axis cs:39.30,3.800) {22};
\fill[s4] (axis cs:50.40,3.490) rectangle (axis cs:75.00,4.110);
\node[font=\tiny, text=black] at (axis cs:62.70,3.800) {25};
\fill[s5] (axis cs:75.00,3.490) rectangle (axis cs:75.40,4.110);
\fill[s6] (axis cs:75.40,3.490) rectangle (axis cs:87.00,4.110);
\node[font=\tiny, text=white] at (axis cs:81.20,3.800) {12};
\fill[s7] (axis cs:87.00,3.490) rectangle (axis cs:90.10,4.110);
\node[font=\scriptsize, anchor=west] at (axis cs:91.60,3.800) {\textbf{90} / $\leq$107};
\draw[pattern=north east lines, pattern color=black!55, draw=black!55] (axis cs:0,2.490) rectangle (axis cs:15.5,3.110);
\node[font=\tiny, fill=white, inner sep=0.5pt] at (axis cs:7.75,2.800) {16};
\fill[s2] (axis cs:15.50,2.490) rectangle (axis cs:17.30,3.110);
\fill[s3] (axis cs:17.30,2.490) rectangle (axis cs:39.00,3.110);
\node[font=\tiny, text=black] at (axis cs:28.15,2.800) {22};
\fill[s4] (axis cs:39.00,2.490) rectangle (axis cs:64.50,3.110);
\node[font=\tiny, text=black] at (axis cs:51.75,2.800) {26};
\fill[s5] (axis cs:64.50,2.490) rectangle (axis cs:64.90,3.110);
\fill[s6] (axis cs:64.90,2.490) rectangle (axis cs:76.80,3.110);
\node[font=\tiny, text=white] at (axis cs:70.85,2.800) {12};
\fill[s7] (axis cs:76.80,2.490) rectangle (axis cs:79.80,3.110);
\node[font=\scriptsize, anchor=west] at (axis cs:81.30,2.800) {\textbf{80} / $\leq$92};
\draw[pattern=north east lines, pattern color=black!55, draw=black!55] (axis cs:0,0.690) rectangle (axis cs:26.0,1.310);
\node[font=\tiny, fill=white, inner sep=0.5pt] at (axis cs:13.0,1.000) {26};
\fill[s2] (axis cs:26.00,0.690) rectangle (axis cs:28.00,1.310);
\fill[s3] (axis cs:28.00,0.690) rectangle (axis cs:41.40,1.310);
\node[font=\tiny, text=black] at (axis cs:34.70,1.000) {13};
\fill[s4] (axis cs:41.40,0.690) rectangle (axis cs:65.20,1.310);
\node[font=\tiny, text=black] at (axis cs:53.30,1.000) {24};
\fill[s5] (axis cs:65.20,0.690) rectangle (axis cs:103.90,1.310);
\node[font=\tiny, text=white] at (axis cs:84.55,1.000) {39};
\fill[s6] (axis cs:103.90,0.690) rectangle (axis cs:119.20,1.310);
\node[font=\tiny, text=white] at (axis cs:111.55,1.000) {15};
\fill[s7] (axis cs:119.20,0.690) rectangle (axis cs:122.30,1.310);
\node[font=\scriptsize, anchor=west] at (axis cs:123.80,1.000) {\textbf{122} / $\leq$141};
\draw[pattern=north east lines, pattern color=black!55, draw=black!55] (axis cs:0,-0.310) rectangle (axis cs:40.4,0.310);
\node[font=\tiny, fill=white, inner sep=0.5pt] at (axis cs:20.2,0.000) {40};
\fill[s2] (axis cs:40.40,-0.310) rectangle (axis cs:42.30,0.310);
\fill[s3] (axis cs:42.30,-0.310) rectangle (axis cs:55.80,0.310);
\node[font=\tiny, text=black] at (axis cs:49.05,0.000) {14};
\fill[s4] (axis cs:55.80,-0.310) rectangle (axis cs:78.40,0.310);
\node[font=\tiny, text=black] at (axis cs:67.10,0.000) {23};
\fill[s5] (axis cs:78.40,-0.310) rectangle (axis cs:121.50,0.310);
\node[font=\tiny, text=white] at (axis cs:99.95,0.000) {43};
\fill[s6] (axis cs:121.50,-0.310) rectangle (axis cs:137.20,0.310);
\node[font=\tiny, text=white] at (axis cs:129.35,0.000) {16};
\fill[s7] (axis cs:137.20,-0.310) rectangle (axis cs:140.20,0.310);
\node[font=\scriptsize, anchor=west] at (axis cs:141.70,0.000) {\textbf{140} / $\leq$162};
\node[font=\scriptsize\itshape, anchor=south west, inner sep=1pt] at (axis cs:0,6.130) {No processing};
\node[font=\scriptsize\itshape, anchor=south west, inner sep=1pt] at (axis cs:0,1.330) {Native TensorRT, I/Q};
\end{axis}
\end{tikzpicture}

%% file: figures_tex/scaling_bars.tex
\begin{tikzpicture}
\definecolor{s2}{HTML}{F0E442}
\definecolor{s3}{HTML}{E69F00}
\definecolor{s4}{HTML}{56B4E9}
\definecolor{s5}{HTML}{009E73}
\definecolor{s6}{HTML}{0072B2}
\definecolor{s7}{HTML}{CC79A7}
\begin{axis}[
width=\fwidth, height=\fheight, scale only axis,
xmin=0, xmax=500, ymin=-15.15, ymax=0.55,
axis y line*=left, axis x line*=bottom, xtick={0,100,200,300,400,500},
xticklabel style={font=\scriptsize}, yticklabel style={font=\scriptsize},
xlabel={Latency [$\mu$s]}, xlabel style={font=\footnotesize},
ytick={-0.95,-1.95,-2.95,-3.95,-6.15,-7.15,-8.15,-9.15,-11.35,-12.35,-13.35,-14.35},
yticklabels={{40 MHz, 2 Rx},{40 MHz, 4 Rx},{100 MHz, 2 Rx},{100 MHz, 4 Rx},{1 cell (3)},{2 cells (6)},{4 cells (6)},{10 cells (9)},{1 dApp},{2 dApps},{4 dApps},{10 dApps}},
clip=false, legend columns=2, legend cell align={left},
legend style={font=\scriptsize, draw=none, fill=none, at={(axis cs:545,-0.42)}, anchor=north east, row sep=-1pt, column sep=3pt},
legend image code/.code={\draw[#1] (0cm,-0.08cm) rectangle (0.3cm,0.08cm);},
]
\addlegendimage{pattern=north east lines, pattern color=black!55, draw=black!55}
\addlegendentry{1 GPU$\rightarrow$host}
\addlegendimage{fill=s2, draw=none}
\addlegendentry{2 Notify}
\addlegendimage{fill=s3, draw=none}
\addlegendentry{3 Copy}
\addlegendimage{fill=s4, draw=none}
\addlegendentry{4 Indication}
\addlegendimage{fill=s6, draw=none}
\addlegendentry{6 Control}
\addlegendimage{fill=s7, draw=none}
\addlegendentry{7 Apply}
\node[font=\scriptsize\itshape, anchor=west, inner sep=1pt] at (axis cs:0,0.000) {Bandwidth and antennas (all PUSCH streams, 1 cell)};
\draw[pattern=north east lines, pattern color=black!55, draw=black!55] (axis cs:0,-1.260) rectangle (axis cs:19.4,-0.640);
\node[font=\tiny, fill=white, inner sep=0.5pt] at (axis cs:9.7,-0.950) {19};
\fill[s2] (axis cs:19.40,-1.260) rectangle (axis cs:21.60,-0.640);
\fill[s3] (axis cs:21.60,-1.260) rectangle (axis cs:30.40,-0.640);
\fill[s4] (axis cs:30.40,-1.260) rectangle (axis cs:53.50,-0.640);
\node[font=\tiny, text=black] at (axis cs:41.95,-0.950) {23};
\fill[s5] (axis cs:53.50,-1.260) rectangle (axis cs:53.90,-0.640);
\fill[s6] (axis cs:53.90,-1.260) rectangle (axis cs:65.50,-0.640);
\fill[s7] (axis cs:65.50,-1.260) rectangle (axis cs:68.50,-0.640);
\node[font=\scriptsize, anchor=west] at (axis cs:71.50,-0.950) {\textbf{69} / $\leq$85};
\draw[pattern=north east lines, pattern color=black!55, draw=black!55] (axis cs:0,-2.260) rectangle (axis cs:19.5,-1.640);
\node[font=\tiny, fill=white, inner sep=0.5pt] at (axis cs:9.75,-1.950) {20};
\fill[s2] (axis cs:19.50,-2.260) rectangle (axis cs:21.70,-1.640);
\fill[s3] (axis cs:21.70,-2.260) rectangle (axis cs:36.00,-1.640);
\fill[s4] (axis cs:36.00,-2.260) rectangle (axis cs:58.40,-1.640);
\node[font=\tiny, text=black] at (axis cs:47.20,-1.950) {22};
\fill[s5] (axis cs:58.40,-2.260) rectangle (axis cs:58.80,-1.640);
\fill[s6] (axis cs:58.80,-2.260) rectangle (axis cs:70.70,-1.640);
\fill[s7] (axis cs:70.70,-2.260) rectangle (axis cs:73.70,-1.640);
\node[font=\scriptsize, anchor=west] at (axis cs:76.70,-1.950) {\textbf{74} / $\leq$91};
\draw[pattern=north east lines, pattern color=black!55, draw=black!55] (axis cs:0,-3.260) rectangle (axis cs:21.1,-2.640);
\node[font=\tiny, fill=white, inner sep=0.5pt] at (axis cs:10.55,-2.950) {21};
\fill[s2] (axis cs:21.10,-3.260) rectangle (axis cs:23.30,-2.640);
\fill[s3] (axis cs:23.30,-3.260) rectangle (axis cs:36.50,-2.640);
\fill[s4] (axis cs:36.50,-3.260) rectangle (axis cs:61.70,-2.640);
\node[font=\tiny, text=black] at (axis cs:49.10,-2.950) {25};
\fill[s5] (axis cs:61.70,-3.260) rectangle (axis cs:62.10,-2.640);
\fill[s6] (axis cs:62.10,-3.260) rectangle (axis cs:74.60,-2.640);
\fill[s7] (axis cs:74.60,-3.260) rectangle (axis cs:77.90,-2.640);
\node[font=\scriptsize, anchor=west] at (axis cs:80.90,-2.950) {\textbf{78} / $\leq$96};
\draw[pattern=north east lines, pattern color=black!55, draw=black!55] (axis cs:0,-4.260) rectangle (axis cs:26.7,-3.640);
\node[font=\tiny, fill=white, inner sep=0.5pt] at (axis cs:13.35,-3.950) {27};
\fill[s2] (axis cs:26.70,-4.260) rectangle (axis cs:28.90,-3.640);
\fill[s3] (axis cs:28.90,-4.260) rectangle (axis cs:51.10,-3.640);
\node[font=\tiny, text=black] at (axis cs:40.00,-3.950) {22};
\fill[s4] (axis cs:51.10,-4.260) rectangle (axis cs:75.70,-3.640);
\node[font=\tiny, text=black] at (axis cs:63.40,-3.950) {25};
\fill[s5] (axis cs:75.70,-4.260) rectangle (axis cs:76.10,-3.640);
\fill[s6] (axis cs:76.10,-4.260) rectangle (axis cs:87.70,-3.640);
\fill[s7] (axis cs:87.70,-4.260) rectangle (axis cs:90.80,-3.640);
\node[font=\scriptsize, anchor=west] at (axis cs:93.80,-3.950) {\textbf{91} / $\leq$107};
\node[font=\scriptsize\itshape, anchor=west, inner sep=1pt] at (axis cs:0,-5.200) {Cells per E3 Agent (all PUSCH streams, RU emulator)};
\draw[pattern=north east lines, pattern color=black!55, draw=black!55] (axis cs:0,-6.460) rectangle (axis cs:24.0,-5.840);
\node[font=\tiny, fill=white, inner sep=0.5pt] at (axis cs:12.0,-6.150) {24};
\fill[s2] (axis cs:24.00,-6.460) rectangle (axis cs:26.30,-5.840);
\fill[s3] (axis cs:26.30,-6.460) rectangle (axis cs:53.80,-5.840);
\node[font=\tiny, text=black] at (axis cs:40.05,-6.150) {28};
\fill[s4] (axis cs:53.80,-6.460) rectangle (axis cs:78.10,-5.840);
\node[font=\tiny, text=black] at (axis cs:65.95,-6.150) {24};
\fill[s5] (axis cs:78.10,-6.460) rectangle (axis cs:78.50,-5.840);
\fill[s6] (axis cs:78.50,-6.460) rectangle (axis cs:90.00,-5.840);
\fill[s7] (axis cs:90.00,-6.460) rectangle (axis cs:93.00,-5.840);
\node[font=\scriptsize, anchor=west] at (axis cs:96.00,-6.150) {\textbf{93} / $\leq$117};
\draw[pattern=north east lines, pattern color=black!55, draw=black!55] (axis cs:0,-7.460) rectangle (axis cs:42.3,-6.840);
\node[font=\tiny, fill=white, inner sep=0.5pt] at (axis cs:21.15,-7.150) {42};
\fill[s2] (axis cs:42.30,-7.460) rectangle (axis cs:44.60,-6.840);
\fill[s3] (axis cs:44.60,-7.460) rectangle (axis cs:75.00,-6.840);
\node[font=\tiny, text=black] at (axis cs:59.80,-7.150) {30};
\fill[s4] (axis cs:75.00,-7.460) rectangle (axis cs:105.40,-6.840);
\node[font=\tiny, text=black] at (axis cs:90.20,-7.150) {30};
\fill[s5] (axis cs:105.40,-7.460) rectangle (axis cs:105.80,-6.840);
\fill[s6] (axis cs:105.80,-7.460) rectangle (axis cs:117.90,-6.840);
\fill[s7] (axis cs:117.90,-7.460) rectangle (axis cs:121.10,-6.840);
\node[font=\scriptsize, anchor=west] at (axis cs:124.10,-7.150) {\textbf{122} / $\leq$155};
\draw[pattern=north east lines, pattern color=black!55, draw=black!55] (axis cs:0,-8.460) rectangle (axis cs:88.0,-7.840);
\node[font=\tiny, fill=white, inner sep=0.5pt] at (axis cs:44.0,-8.150) {88};
\fill[s2] (axis cs:88.00,-8.460) rectangle (axis cs:90.30,-7.840);
\fill[s3] (axis cs:90.30,-8.460) rectangle (axis cs:150.70,-7.840);
\node[font=\tiny, text=black] at (axis cs:120.50,-8.150) {60};
\fill[s4] (axis cs:150.70,-8.460) rectangle (axis cs:192.90,-7.840);
\node[font=\tiny, text=black] at (axis cs:171.80,-8.150) {42};
\fill[s5] (axis cs:192.90,-8.460) rectangle (axis cs:193.30,-7.840);
\fill[s6] (axis cs:193.30,-8.460) rectangle (axis cs:206.80,-7.840);
\fill[s7] (axis cs:206.80,-8.460) rectangle (axis cs:210.80,-7.840);
\node[font=\scriptsize, anchor=west] at (axis cs:213.80,-8.150) {\textbf{211} / $\leq$258};
\draw[pattern=north east lines, pattern color=black!55, draw=black!55] (axis cs:0,-9.460) rectangle (axis cs:187.3,-8.840);
\node[font=\tiny, fill=white, inner sep=0.5pt] at (axis cs:93.65,-9.150) {187};
\fill[s2] (axis cs:187.30,-9.460) rectangle (axis cs:189.60,-8.840);
\fill[s3] (axis cs:189.60,-9.460) rectangle (axis cs:314.70,-8.840);
\node[font=\tiny, text=black] at (axis cs:252.15,-9.150) {125};
\fill[s4] (axis cs:314.70,-9.460) rectangle (axis cs:383.10,-8.840);
\node[font=\tiny, text=black] at (axis cs:348.90,-9.150) {68};
\fill[s5] (axis cs:383.10,-9.460) rectangle (axis cs:383.50,-8.840);
\fill[s6] (axis cs:383.50,-9.460) rectangle (axis cs:395.40,-8.840);
\fill[s7] (axis cs:395.40,-9.460) rectangle (axis cs:398.40,-8.840);
\node[font=\scriptsize, anchor=west] at (axis cs:401.40,-9.150) {\textbf{399} / $\leq$438};
\node[font=\scriptsize\itshape, anchor=west, inner sep=1pt] at (axis cs:0,-10.400) {dApps per E3 Agent (I/Q, last dApp served)};
\draw[pattern=north east lines, pattern color=black!55, draw=black!55] (axis cs:0,-11.660) rectangle (axis cs:25.2,-11.040);
\node[font=\tiny, fill=white, inner sep=0.5pt] at (axis cs:12.6,-11.350) {25};
\fill[s2] (axis cs:25.20,-11.660) rectangle (axis cs:27.30,-11.040);
\fill[s3] (axis cs:27.30,-11.660) rectangle (axis cs:42.60,-11.040);
\node[font=\tiny, text=black] at (axis cs:34.95,-11.350) {15};
\fill[s4] (axis cs:42.60,-11.660) rectangle (axis cs:62.50,-11.040);
\node[font=\tiny, text=black] at (axis cs:52.55,-11.350) {20};
\fill[s5] (axis cs:62.50,-11.660) rectangle (axis cs:62.90,-11.040);
\fill[s6] (axis cs:62.90,-11.660) rectangle (axis cs:76.00,-11.040);
\fill[s7] (axis cs:76.00,-11.660) rectangle (axis cs:79.00,-11.040);
\node[font=\scriptsize, anchor=west] at (axis cs:82.00,-11.350) {\textbf{79} / $\leq$95};
\draw[pattern=north east lines, pattern color=black!55, draw=black!55] (axis cs:0,-12.660) rectangle (axis cs:25.4,-12.040);
\node[font=\tiny, fill=white, inner sep=0.5pt] at (axis cs:12.7,-12.350) {25};
\fill[s2] (axis cs:25.40,-12.660) rectangle (axis cs:27.50,-12.040);
\fill[s3] (axis cs:27.50,-12.660) rectangle (axis cs:42.80,-12.040);
\node[font=\tiny, text=black] at (axis cs:35.15,-12.350) {15};
\fill[s4] (axis cs:42.80,-12.660) rectangle (axis cs:66.00,-12.040);
\node[font=\tiny, text=black] at (axis cs:54.40,-12.350) {23};
\fill[s5] (axis cs:66.00,-12.660) rectangle (axis cs:66.40,-12.040);
\fill[s6] (axis cs:66.40,-12.660) rectangle (axis cs:86.90,-12.040);
\node[font=\tiny, text=white] at (axis cs:76.65,-12.350) {20};
\fill[s7] (axis cs:86.90,-12.660) rectangle (axis cs:89.70,-12.040);
\node[font=\scriptsize, anchor=west] at (axis cs:92.70,-12.350) {\textbf{90} / $\leq$107};
\draw[pattern=north east lines, pattern color=black!55, draw=black!55] (axis cs:0,-13.660) rectangle (axis cs:25.8,-13.040);
\node[font=\tiny, fill=white, inner sep=0.5pt] at (axis cs:12.9,-13.350) {26};
\fill[s2] (axis cs:25.80,-13.660) rectangle (axis cs:28.00,-13.040);
\fill[s3] (axis cs:28.00,-13.660) rectangle (axis cs:43.30,-13.040);
\node[font=\tiny, text=black] at (axis cs:35.65,-13.350) {15};
\fill[s4] (axis cs:43.30,-13.660) rectangle (axis cs:71.10,-13.040);
\node[font=\tiny, text=black] at (axis cs:57.20,-13.350) {28};
\fill[s5] (axis cs:71.10,-13.660) rectangle (axis cs:71.50,-13.040);
\fill[s6] (axis cs:71.50,-13.660) rectangle (axis cs:104.20,-13.040);
\node[font=\tiny, text=white] at (axis cs:87.85,-13.350) {33};
\fill[s7] (axis cs:104.20,-13.660) rectangle (axis cs:106.80,-13.040);
\node[font=\scriptsize, anchor=west] at (axis cs:109.80,-13.350) {\textbf{107} / $\leq$128};
\draw[pattern=north east lines, pattern color=black!55, draw=black!55] (axis cs:0,-14.660) rectangle (axis cs:26.9,-14.040);
\node[font=\tiny, fill=white, inner sep=0.5pt] at (axis cs:13.45,-14.350) {27};
\fill[s2] (axis cs:26.90,-14.660) rectangle (axis cs:29.10,-14.040);
\fill[s3] (axis cs:29.10,-14.660) rectangle (axis cs:44.40,-14.040);
\node[font=\tiny, text=black] at (axis cs:36.75,-14.350) {15};
\fill[s4] (axis cs:44.40,-14.660) rectangle (axis cs:93.90,-14.040);
\node[font=\tiny, text=black] at (axis cs:69.15,-14.350) {50};
\fill[s5] (axis cs:93.90,-14.660) rectangle (axis cs:94.30,-14.040);
\fill[s6] (axis cs:94.30,-14.660) rectangle (axis cs:186.00,-14.040);
\node[font=\tiny, text=white] at (axis cs:140.15,-14.350) {92};
\fill[s7] (axis cs:186.00,-14.660) rectangle (axis cs:188.70,-14.040);
\node[font=\scriptsize, anchor=west] at (axis cs:191.70,-14.350) {\textbf{190} / $\leq$209};
\end{axis}
\end{tikzpicture}

%% file: figures_tex/engines_bars.tex
\begin{tikzpicture}
\definecolor{e0}{HTML}{9467BD}
\definecolor{e1}{HTML}{1F77B4}
\definecolor{e2}{HTML}{FF7F0E}
\definecolor{e3}{HTML}{2CA02C}
\begin{axis}[
name=p0, at={(0.000\fwidth,0)}, anchor=south west,
width=\fwidth, height=\fheight, scale only axis,
xmin=-0.55, xmax=4.55, ymin=0, ymax=1150,
axis y line*=left, axis x line*=bottom, clip=false,
xtick={0,1,2,3,4}, xticklabels={TRT,ONNX,LibTorch,Torch,NumPy},
xticklabel style={font=\scriptsize}, yticklabel style={font=\scriptsize},
ytick={0,250,500,750,1000}, xtick style={draw=none}, yticklabel style={/pgf/number format/1000 sep={}},
ymajorgrids, grid style={black!12, line width=0.3pt},
xlabel={(a) MIG}, xlabel style={font=\footnotesize},
ylabel={Step 5 latency [$\mu$s]}, ylabel style={font=\footnotesize},
]
\draw[gray, densely dashed, thin] (axis cs:-0.55,500) -- (axis cs:4.55,500);
\draw[draw=black, line width=0.2pt, fill=e0, postaction={pattern=crosshatch, pattern color=black!70}] (axis cs:-0.400,0) rectangle (axis cs:-0.200,40.6);
\draw[black, line width=0.4pt] (axis cs:-0.300,40.6) -- (axis cs:-0.300,49.5) (axis cs:-0.350,49.5) -- (axis cs:-0.250,49.5);
\node[font=\fontsize{6pt}{7pt}\selectfont, rotate=90, anchor=west, inner sep=1pt] at (axis cs:-0.300,57.5) {41};
\draw[draw=black, line width=0.2pt, fill=e1, postaction={pattern=north east lines, pattern color=black!70}] (axis cs:-0.200,0) rectangle (axis cs:0.000,133.3);
\draw[black, line width=0.4pt] (axis cs:-0.100,133.3) -- (axis cs:-0.100,224.0) (axis cs:-0.150,224.0) -- (axis cs:-0.050,224.0);
\node[font=\fontsize{6pt}{7pt}\selectfont, rotate=90, anchor=west, inner sep=1pt] at (axis cs:-0.100,232.0) {133};
\draw[draw=black, line width=0.2pt, fill=e2, postaction={pattern=north west lines, pattern color=black!70}] (axis cs:0.000,0) rectangle (axis cs:0.200,261.2);
\draw[black, line width=0.4pt] (axis cs:0.100,261.2) -- (axis cs:0.100,384.9) (axis cs:0.050,384.9) -- (axis cs:0.150,384.9);
\node[font=\fontsize{6pt}{7pt}\selectfont, rotate=90, anchor=west, inner sep=1pt] at (axis cs:0.100,392.9) {261};
\draw[draw=black, line width=0.2pt, fill=e3, postaction={pattern=horizontal lines, pattern color=black!70}] (axis cs:0.200,0) rectangle (axis cs:0.400,240.1);
\draw[black, line width=0.4pt] (axis cs:0.300,240.1) -- (axis cs:0.300,264.6) (axis cs:0.250,264.6) -- (axis cs:0.350,264.6);
\node[font=\fontsize{6pt}{7pt}\selectfont, rotate=90, anchor=west, inner sep=1pt] at (axis cs:0.300,272.6) {240};
\draw[draw=black, line width=0.2pt, fill=e1, postaction={pattern=north east lines, pattern color=black!70}] (axis cs:0.800,0) rectangle (axis cs:1.000,164.3);
\draw[black, line width=0.4pt] (axis cs:0.900,164.3) -- (axis cs:0.900,200.1) (axis cs:0.850,200.1) -- (axis cs:0.950,200.1);
\node[font=\fontsize{6pt}{7pt}\selectfont, rotate=90, anchor=west, inner sep=1pt] at (axis cs:0.900,208.1) {164};
\draw[draw=black, line width=0.2pt, fill=e2, postaction={pattern=north west lines, pattern color=black!70}] (axis cs:1.000,0) rectangle (axis cs:1.200,299.8);
\draw[black, line width=0.4pt] (axis cs:1.100,299.8) -- (axis cs:1.100,364.4) (axis cs:1.050,364.4) -- (axis cs:1.150,364.4);
\node[font=\fontsize{6pt}{7pt}\selectfont, rotate=90, anchor=west, inner sep=1pt] at (axis cs:1.100,372.4) {300};
\draw[draw=black, line width=0.2pt, fill=e3, postaction={pattern=horizontal lines, pattern color=black!70}] (axis cs:1.200,0) rectangle (axis cs:1.400,305.1);
\draw[black, line width=0.4pt] (axis cs:1.300,305.1) -- (axis cs:1.300,335.6) (axis cs:1.250,335.6) -- (axis cs:1.350,335.6);
\node[font=\fontsize{6pt}{7pt}\selectfont, rotate=90, anchor=west, inner sep=1pt] at (axis cs:1.300,343.6) {305};
\draw[draw=black, line width=0.2pt, fill=e1, postaction={pattern=north east lines, pattern color=black!70}] (axis cs:1.800,0) rectangle (axis cs:2.000,189.5);
\draw[black, line width=0.4pt] (axis cs:1.900,189.5) -- (axis cs:1.900,244.5) (axis cs:1.850,244.5) -- (axis cs:1.950,244.5);
\node[font=\fontsize{6pt}{7pt}\selectfont, rotate=90, anchor=west, inner sep=1pt] at (axis cs:1.900,252.5) {190};
\draw[draw=black, line width=0.2pt, fill=e2, postaction={pattern=north west lines, pattern color=black!70}] (axis cs:2.000,0) rectangle (axis cs:2.200,320.6);
\draw[black, line width=0.4pt] (axis cs:2.100,320.6) -- (axis cs:2.100,393.8) (axis cs:2.050,393.8) -- (axis cs:2.150,393.8);
\node[font=\fontsize{6pt}{7pt}\selectfont, rotate=90, anchor=west, inner sep=1pt] at (axis cs:2.100,401.8) {321};
\draw[draw=black, line width=0.2pt, fill=e1, postaction={pattern=north east lines, pattern color=black!70}] (axis cs:2.800,0) rectangle (axis cs:3.000,384.7);
\draw[black, line width=0.4pt] (axis cs:2.900,384.7) -- (axis cs:2.900,466.6) (axis cs:2.850,466.6) -- (axis cs:2.950,466.6);
\node[font=\fontsize{6pt}{7pt}\selectfont, rotate=90, anchor=west, inner sep=1pt] at (axis cs:2.900,474.6) {385};
\draw[draw=black, line width=0.2pt, fill=e2, postaction={pattern=north west lines, pattern color=black!70}] (axis cs:3.000,0) rectangle (axis cs:3.200,531.3);
\draw[black, line width=0.4pt] (axis cs:3.100,531.3) -- (axis cs:3.100,635.6) (axis cs:3.050,635.6) -- (axis cs:3.150,635.6);
\node[font=\fontsize{6pt}{7pt}\selectfont, rotate=90, anchor=west, inner sep=1pt] at (axis cs:3.100,643.6) {531};
\draw[draw=black, line width=0.2pt, fill=e3, postaction={pattern=horizontal lines, pattern color=black!70}] (axis cs:3.200,0) rectangle (axis cs:3.400,311.8);
\draw[black, line width=0.4pt] (axis cs:3.300,311.8) -- (axis cs:3.300,345.3) (axis cs:3.250,345.3) -- (axis cs:3.350,345.3);
\node[font=\fontsize{6pt}{7pt}\selectfont, rotate=90, anchor=west, inner sep=1pt] at (axis cs:3.300,353.3) {312};
\draw[draw=black, line width=0.2pt, fill=e1, postaction={pattern=north east lines, pattern color=black!70}] (axis cs:3.800,0) rectangle (axis cs:4.000,577.1);
\draw[black, line width=0.4pt] (axis cs:3.900,577.1) -- (axis cs:3.900,742.7) (axis cs:3.850,742.7) -- (axis cs:3.950,742.7);
\node[font=\fontsize{6pt}{7pt}\selectfont, rotate=90, anchor=west, inner sep=1pt] at (axis cs:3.900,750.7) {577};
\draw[draw=black, line width=0.2pt, fill=e2, postaction={pattern=north west lines, pattern color=black!70}] (axis cs:4.000,0) rectangle (axis cs:4.200,732.7);
\draw[black, line width=0.4pt] (axis cs:4.100,732.7) -- (axis cs:4.100,905.6) (axis cs:4.050,905.6) -- (axis cs:4.150,905.6);
\node[font=\fontsize{6pt}{7pt}\selectfont, rotate=90, anchor=west, inner sep=1pt] at (axis cs:4.100,913.6) {733};
\draw[draw=black, line width=0.2pt, fill=e3, postaction={pattern=horizontal lines, pattern color=black!70}] (axis cs:4.200,0) rectangle (axis cs:4.400,780.7);
\draw[black, line width=0.4pt] (axis cs:4.300,780.7) -- (axis cs:4.300,821.7) (axis cs:4.250,821.7) -- (axis cs:4.350,821.7);
\node[font=\fontsize{6pt}{7pt}\selectfont, rotate=90, anchor=west, inner sep=1pt] at (axis cs:4.300,829.7) {781};
\end{axis}
\begin{axis}[
name=p1, at={(1.060\fwidth,0)}, anchor=south west,
width=\fwidth, height=\fheight, scale only axis,
xmin=-0.55, xmax=4.55, ymin=0, ymax=1150,
axis y line*=left, axis x line*=bottom, clip=false,
xtick={0,1,2,3,4}, xticklabels={TRT,ONNX,LibTorch,Torch,NumPy},
xticklabel style={font=\scriptsize}, yticklabel style={font=\scriptsize},
ytick={0,250,500,750,1000}, xtick style={draw=none}, yticklabel style={/pgf/number format/1000 sep={}},
ymajorgrids, grid style={black!12, line width=0.3pt},
xlabel={(b) MPS}, xlabel style={font=\footnotesize},
yticklabels={},
]
\draw[gray, densely dashed, thin] (axis cs:-0.55,500) -- (axis cs:4.55,500);
\draw[draw=black, line width=0.2pt, fill=e0, postaction={pattern=crosshatch, pattern color=black!70}] (axis cs:-0.400,0) rectangle (axis cs:-0.200,36.4);
\draw[black, line width=0.4pt] (axis cs:-0.300,36.4) -- (axis cs:-0.300,41.5) (axis cs:-0.350,41.5) -- (axis cs:-0.250,41.5);
\node[font=\fontsize{6pt}{7pt}\selectfont, rotate=90, anchor=west, inner sep=1pt] at (axis cs:-0.300,49.5) {36};
\draw[draw=black, line width=0.2pt, fill=e1, postaction={pattern=north east lines, pattern color=black!70}] (axis cs:-0.200,0) rectangle (axis cs:0.000,133.0);
\draw[black, line width=0.4pt] (axis cs:-0.100,133.0) -- (axis cs:-0.100,270.0) (axis cs:-0.150,270.0) -- (axis cs:-0.050,270.0);
\node[font=\fontsize{6pt}{7pt}\selectfont, rotate=90, anchor=west, inner sep=1pt] at (axis cs:-0.100,278.0) {133};
\draw[draw=black, line width=0.2pt, fill=e2, postaction={pattern=north west lines, pattern color=black!70}] (axis cs:0.000,0) rectangle (axis cs:0.200,257.6);
\draw[black, line width=0.4pt] (axis cs:0.100,257.6) -- (axis cs:0.100,413.1) (axis cs:0.050,413.1) -- (axis cs:0.150,413.1);
\node[font=\fontsize{6pt}{7pt}\selectfont, rotate=90, anchor=west, inner sep=1pt] at (axis cs:0.100,421.1) {258};
\draw[draw=black, line width=0.2pt, fill=e3, postaction={pattern=horizontal lines, pattern color=black!70}] (axis cs:0.200,0) rectangle (axis cs:0.400,226.0);
\draw[black, line width=0.4pt] (axis cs:0.300,226.0) -- (axis cs:0.300,250.8) (axis cs:0.250,250.8) -- (axis cs:0.350,250.8);
\node[font=\fontsize{6pt}{7pt}\selectfont, rotate=90, anchor=west, inner sep=1pt] at (axis cs:0.300,258.8) {226};
\draw[draw=black, line width=0.2pt, fill=e1, postaction={pattern=north east lines, pattern color=black!70}] (axis cs:0.800,0) rectangle (axis cs:1.000,153.4);
\draw[black, line width=0.4pt] (axis cs:0.900,153.4) -- (axis cs:0.900,205.0) (axis cs:0.850,205.0) -- (axis cs:0.950,205.0);
\node[font=\fontsize{6pt}{7pt}\selectfont, rotate=90, anchor=west, inner sep=1pt] at (axis cs:0.900,213.0) {153};
\draw[draw=black, line width=0.2pt, fill=e2, postaction={pattern=north west lines, pattern color=black!70}] (axis cs:1.000,0) rectangle (axis cs:1.200,283.6);
\draw[black, line width=0.4pt] (axis cs:1.100,283.6) -- (axis cs:1.100,344.9) (axis cs:1.050,344.9) -- (axis cs:1.150,344.9);
\node[font=\fontsize{6pt}{7pt}\selectfont, rotate=90, anchor=west, inner sep=1pt] at (axis cs:1.100,352.9) {284};
\draw[draw=black, line width=0.2pt, fill=e3, postaction={pattern=horizontal lines, pattern color=black!70}] (axis cs:1.200,0) rectangle (axis cs:1.400,290.5);
\draw[black, line width=0.4pt] (axis cs:1.300,290.5) -- (axis cs:1.300,322.5) (axis cs:1.250,322.5) -- (axis cs:1.350,322.5);
\node[font=\fontsize{6pt}{7pt}\selectfont, rotate=90, anchor=west, inner sep=1pt] at (axis cs:1.300,330.5) {290};
\draw[draw=black, line width=0.2pt, fill=e1, postaction={pattern=north east lines, pattern color=black!70}] (axis cs:1.800,0) rectangle (axis cs:2.000,183.2);
\draw[black, line width=0.4pt] (axis cs:1.900,183.2) -- (axis cs:1.900,224.3) (axis cs:1.850,224.3) -- (axis cs:1.950,224.3);
\node[font=\fontsize{6pt}{7pt}\selectfont, rotate=90, anchor=west, inner sep=1pt] at (axis cs:1.900,232.3) {183};
\draw[draw=black, line width=0.2pt, fill=e2, postaction={pattern=north west lines, pattern color=black!70}] (axis cs:2.000,0) rectangle (axis cs:2.200,315.9);
\draw[black, line width=0.4pt] (axis cs:2.100,315.9) -- (axis cs:2.100,396.3) (axis cs:2.050,396.3) -- (axis cs:2.150,396.3);
\node[font=\fontsize{6pt}{7pt}\selectfont, rotate=90, anchor=west, inner sep=1pt] at (axis cs:2.100,404.3) {316};
\draw[draw=black, line width=0.2pt, fill=e1, postaction={pattern=north east lines, pattern color=black!70}] (axis cs:2.800,0) rectangle (axis cs:3.000,370.8);
\draw[black, line width=0.4pt] (axis cs:2.900,370.8) -- (axis cs:2.900,467.9) (axis cs:2.850,467.9) -- (axis cs:2.950,467.9);
\node[font=\fontsize{6pt}{7pt}\selectfont, rotate=90, anchor=west, inner sep=1pt] at (axis cs:2.900,475.9) {371};
\draw[draw=black, line width=0.2pt, fill=e2, postaction={pattern=north west lines, pattern color=black!70}] (axis cs:3.000,0) rectangle (axis cs:3.200,515.3);
\draw[black, line width=0.4pt] (axis cs:3.100,515.3) -- (axis cs:3.100,624.2) (axis cs:3.050,624.2) -- (axis cs:3.150,624.2);
\node[font=\fontsize{6pt}{7pt}\selectfont, rotate=90, anchor=west, inner sep=1pt] at (axis cs:3.100,632.2) {515};
\draw[draw=black, line width=0.2pt, fill=e3, postaction={pattern=horizontal lines, pattern color=black!70}] (axis cs:3.200,0) rectangle (axis cs:3.400,297.8);
\draw[black, line width=0.4pt] (axis cs:3.300,297.8) -- (axis cs:3.300,329.2) (axis cs:3.250,329.2) -- (axis cs:3.350,329.2);
\node[font=\fontsize{6pt}{7pt}\selectfont, rotate=90, anchor=west, inner sep=1pt] at (axis cs:3.300,337.2) {298};
\draw[draw=black, line width=0.2pt, fill=e1, postaction={pattern=north east lines, pattern color=black!70}] (axis cs:3.800,0) rectangle (axis cs:4.000,583.1);
\draw[black, line width=0.4pt] (axis cs:3.900,583.1) -- (axis cs:3.900,766.4) (axis cs:3.850,766.4) -- (axis cs:3.950,766.4);
\node[font=\fontsize{6pt}{7pt}\selectfont, rotate=90, anchor=west, inner sep=1pt] at (axis cs:3.900,774.4) {583};
\draw[draw=black, line width=0.2pt, fill=e2, postaction={pattern=north west lines, pattern color=black!70}] (axis cs:4.000,0) rectangle (axis cs:4.200,740.0);
\draw[black, line width=0.4pt] (axis cs:4.100,740.0) -- (axis cs:4.100,919.3) (axis cs:4.050,919.3) -- (axis cs:4.150,919.3);
\node[font=\fontsize{6pt}{7pt}\selectfont, rotate=90, anchor=west, inner sep=1pt] at (axis cs:4.100,927.3) {740};
\draw[draw=black, line width=0.2pt, fill=e3, postaction={pattern=horizontal lines, pattern color=black!70}] (axis cs:4.200,0) rectangle (axis cs:4.400,798.0);
\draw[black, line width=0.4pt] (axis cs:4.300,798.0) -- (axis cs:4.300,849.5) (axis cs:4.250,849.5) -- (axis cs:4.350,849.5);
\node[font=\fontsize{6pt}{7pt}\selectfont, rotate=90, anchor=west, inner sep=1pt] at (axis cs:4.300,857.5) {798};
\end{axis}
\begin{axis}[
name=p2, at={(2.120\fwidth,0)}, anchor=south west,
width=\fwidth, height=\fheight, scale only axis,
xmin=-0.55, xmax=4.55, ymin=0, ymax=1150,
axis y line*=left, axis x line*=bottom, clip=false,
xtick={0,1,2,3,4}, xticklabels={TRT,ONNX,LibTorch,Torch,NumPy},
xticklabel style={font=\scriptsize}, yticklabel style={font=\scriptsize},
ytick={0,250,500,750,1000}, xtick style={draw=none}, yticklabel style={/pgf/number format/1000 sep={}},
ymajorgrids, grid style={black!12, line width=0.3pt},
xlabel={(c) Time-slicing}, xlabel style={font=\footnotesize},
yticklabels={},
]
\draw[gray, densely dashed, thin] (axis cs:-0.55,500) -- (axis cs:4.55,500);
\node[font=\tiny, text=gray, anchor=south east] at (axis cs:4.55,500) {0.5~ms};
\draw[draw=black, line width=0.2pt, fill=e0, postaction={pattern=crosshatch, pattern color=black!70}] (axis cs:-0.400,0) rectangle (axis cs:-0.200,502.0);
\draw[black, line width=0.4pt] (axis cs:-0.300,502.0) -- (axis cs:-0.300,716.9) (axis cs:-0.350,716.9) -- (axis cs:-0.250,716.9);
\node[font=\fontsize{6pt}{7pt}\selectfont, rotate=90, anchor=west, inner sep=1pt] at (axis cs:-0.300,724.9) {502};
\draw[draw=black, line width=0.2pt, fill=e1, postaction={pattern=north east lines, pattern color=black!70}] (axis cs:-0.200,0) rectangle (axis cs:0.000,511.1);
\draw[black, line width=0.4pt] (axis cs:-0.100,511.1) -- (axis cs:-0.100,812.2) (axis cs:-0.150,812.2) -- (axis cs:-0.050,812.2);
\node[font=\fontsize{6pt}{7pt}\selectfont, rotate=90, anchor=west, inner sep=1pt] at (axis cs:-0.100,820.2) {511};
\draw[draw=black, line width=0.2pt, fill=e2, postaction={pattern=north west lines, pattern color=black!70}] (axis cs:0.000,0) rectangle (axis cs:0.200,493.1);
\draw[black, line width=0.4pt] (axis cs:0.100,493.1) -- (axis cs:0.100,860.6) (axis cs:0.050,860.6) -- (axis cs:0.150,860.6);
\node[font=\fontsize{6pt}{7pt}\selectfont, rotate=90, anchor=west, inner sep=1pt] at (axis cs:0.100,868.6) {493};
\draw[draw=black, line width=0.2pt, fill=e3, postaction={pattern=horizontal lines, pattern color=black!70}] (axis cs:0.200,0) rectangle (axis cs:0.400,494.0);
\draw[black, line width=0.4pt] (axis cs:0.300,494.0) -- (axis cs:0.300,744.5) (axis cs:0.250,744.5) -- (axis cs:0.350,744.5);
\node[font=\fontsize{6pt}{7pt}\selectfont, rotate=90, anchor=west, inner sep=1pt] at (axis cs:0.300,752.5) {494};
\draw[draw=black, line width=0.2pt, fill=e1, postaction={pattern=north east lines, pattern color=black!70}] (axis cs:0.800,0) rectangle (axis cs:1.000,562.2);
\draw[black, line width=0.4pt] (axis cs:0.900,562.2) -- (axis cs:0.900,778.9) (axis cs:0.850,778.9) -- (axis cs:0.950,778.9);
\node[font=\fontsize{6pt}{7pt}\selectfont, rotate=90, anchor=west, inner sep=1pt] at (axis cs:0.900,786.9) {562};
\draw[draw=black, line width=0.2pt, fill=e2, postaction={pattern=north west lines, pattern color=black!70}] (axis cs:1.000,0) rectangle (axis cs:1.200,547.5);
\draw[black, line width=0.4pt] (axis cs:1.100,547.5) -- (axis cs:1.100,830.2) (axis cs:1.050,830.2) -- (axis cs:1.150,830.2);
\node[font=\fontsize{6pt}{7pt}\selectfont, rotate=90, anchor=west, inner sep=1pt] at (axis cs:1.100,838.2) {548};
\draw[draw=black, line width=0.2pt, fill=e3, postaction={pattern=horizontal lines, pattern color=black!70}] (axis cs:1.200,0) rectangle (axis cs:1.400,587.7);
\draw[black, line width=0.4pt] (axis cs:1.300,587.7) -- (axis cs:1.300,803.0) (axis cs:1.250,803.0) -- (axis cs:1.350,803.0);
\node[font=\fontsize{6pt}{7pt}\selectfont, rotate=90, anchor=west, inner sep=1pt] at (axis cs:1.300,811.0) {588};
\draw[draw=black, line width=0.2pt, fill=e1, postaction={pattern=north east lines, pattern color=black!70}] (axis cs:1.800,0) rectangle (axis cs:2.000,512.5);
\draw[black, line width=0.4pt] (axis cs:1.900,512.5) -- (axis cs:1.900,746.3) (axis cs:1.850,746.3) -- (axis cs:1.950,746.3);
\node[font=\fontsize{6pt}{7pt}\selectfont, rotate=90, anchor=west, inner sep=1pt] at (axis cs:1.900,754.3) {512};
\draw[draw=black, line width=0.2pt, fill=e2, postaction={pattern=north west lines, pattern color=black!70}] (axis cs:2.000,0) rectangle (axis cs:2.200,499.8);
\draw[black, line width=0.4pt] (axis cs:2.100,499.8) -- (axis cs:2.100,801.4) (axis cs:2.050,801.4) -- (axis cs:2.150,801.4);
\node[font=\fontsize{6pt}{7pt}\selectfont, rotate=90, anchor=west, inner sep=1pt] at (axis cs:2.100,809.4) {500};
\draw[draw=black, line width=0.2pt, fill=e1, postaction={pattern=north east lines, pattern color=black!70}] (axis cs:2.800,0) rectangle (axis cs:3.000,511.9);
\draw[black, line width=0.4pt] (axis cs:2.900,511.9) -- (axis cs:2.900,745.8) (axis cs:2.850,745.8) -- (axis cs:2.950,745.8);
\node[font=\fontsize{6pt}{7pt}\selectfont, rotate=90, anchor=west, inner sep=1pt] at (axis cs:2.900,753.8) {512};
\draw[draw=black, line width=0.2pt, fill=e2, postaction={pattern=north west lines, pattern color=black!70}] (axis cs:3.000,0) rectangle (axis cs:3.200,533.0);
\draw[black, line width=0.4pt] (axis cs:3.100,533.0) -- (axis cs:3.100,800.3) (axis cs:3.050,800.3) -- (axis cs:3.150,800.3);
\node[font=\fontsize{6pt}{7pt}\selectfont, rotate=90, anchor=west, inner sep=1pt] at (axis cs:3.100,808.3) {533};
\draw[draw=black, line width=0.2pt, fill=e3, postaction={pattern=horizontal lines, pattern color=black!70}] (axis cs:3.200,0) rectangle (axis cs:3.400,505.3);
\draw[black, line width=0.4pt] (axis cs:3.300,505.3) -- (axis cs:3.300,789.4) (axis cs:3.250,789.4) -- (axis cs:3.350,789.4);
\node[font=\fontsize{6pt}{7pt}\selectfont, rotate=90, anchor=west, inner sep=1pt] at (axis cs:3.300,797.4) {505};
\draw[draw=black, line width=0.2pt, fill=e1, postaction={pattern=north east lines, pattern color=black!70}] (axis cs:3.800,0) rectangle (axis cs:4.000,593.6);
\draw[black, line width=0.4pt] (axis cs:3.900,593.6) -- (axis cs:3.900,755.3) (axis cs:3.850,755.3) -- (axis cs:3.950,755.3);
\node[font=\fontsize{6pt}{7pt}\selectfont, rotate=90, anchor=west, inner sep=1pt] at (axis cs:3.900,763.3) {594};
\draw[draw=black, line width=0.2pt, fill=e2, postaction={pattern=north west lines, pattern color=black!70}] (axis cs:4.000,0) rectangle (axis cs:4.200,739.4);
\draw[black, line width=0.4pt] (axis cs:4.100,739.4) -- (axis cs:4.100,905.2) (axis cs:4.050,905.2) -- (axis cs:4.150,905.2);
\node[font=\fontsize{6pt}{7pt}\selectfont, rotate=90, anchor=west, inner sep=1pt] at (axis cs:4.100,913.2) {739};
\draw[draw=black, line width=0.2pt, fill=e3, postaction={pattern=horizontal lines, pattern color=black!70}] (axis cs:4.200,0) rectangle (axis cs:4.400,796.5);
\draw[black, line width=0.4pt] (axis cs:4.300,796.5) -- (axis cs:4.300,843.2) (axis cs:4.250,843.2) -- (axis cs:4.350,843.2);
\node[font=\fontsize{6pt}{7pt}\selectfont, rotate=90, anchor=west, inner sep=1pt] at (axis cs:4.300,851.2) {796};
\end{axis}
\node[font=\scriptsize, anchor=south] at ([yshift=4pt]p1.north) {\tikz[baseline=-0.6ex]\draw[draw=black, line width=0.2pt, fill=e0, postaction={pattern=crosshatch, pattern color=black!70}] (0,-0.08) rectangle (0.3,0.08);~Native TRT \quad \tikz[baseline=-0.6ex]\draw[draw=black, line width=0.2pt, fill=e1, postaction={pattern=north east lines, pattern color=black!70}] (0,-0.08) rectangle (0.3,0.08);~Triton C API \quad \tikz[baseline=-0.6ex]\draw[draw=black, line width=0.2pt, fill=e2, postaction={pattern=north west lines, pattern color=black!70}] (0,-0.08) rectangle (0.3,0.08);~Triton gRPC \quad \tikz[baseline=-0.6ex]\draw[draw=black, line width=0.2pt, fill=e3, postaction={pattern=horizontal lines, pattern color=black!70}] (0,-0.08) rectangle (0.3,0.08);~Python};
\end{tikzpicture}

%% file: figures_tex/spark_path.tex
\begin{tikzpicture}
\definecolor{s2}{HTML}{F0E442}
\definecolor{s3}{HTML}{E69F00}
\definecolor{s4}{HTML}{56B4E9}
\definecolor{s5}{HTML}{009E73}
\definecolor{s6}{HTML}{0072B2}
\definecolor{s7}{HTML}{CC79A7}
\definecolor{e0}{HTML}{9467BD}
\definecolor{e1}{HTML}{1F77B4}
\definecolor{e2}{HTML}{1F77B4}
\definecolor{e3}{HTML}{FF7F0E}
\definecolor{e4}{HTML}{2CA02C}
\definecolor{e5}{HTML}{1F77B4}
\begin{axis}[
name=pa, at={(0,0)}, anchor=south west, width=0.62\linewidth, height=0.36\linewidth, scale only axis,
xmin=0, xmax=200, ymin=-5.60, ymax=0.5, clip=false,
axis y line*=left, axis x line*=bottom, xtick={0,50,100,150,200},
xticklabel style={font=\scriptsize}, yticklabel style={font=\scriptsize},
xlabel={Latency [$\mu$s]}, xlabel style={font=\footnotesize},
ytick={-0.85,-1.85,-3.90,-4.90}, yticklabels={GH200,DGX Spark,GH200,DGX Spark},
]
\node[font=\scriptsize\itshape, anchor=west, inner sep=1pt] at (axis cs:0,0.00) {No processing};
\draw[pattern=north east lines, pattern color=black!55, draw=black!55] (axis cs:0,-1.16) rectangle (axis cs:25.4,-0.54);
\node[font=\tiny, fill=white, inner sep=0.5pt] at (axis cs:12.7,-0.85) {25};
\fill[s2] (axis cs:25.40,-1.16) rectangle (axis cs:27.40,-0.54);
\fill[s3] (axis cs:27.40,-1.16) rectangle (axis cs:52.80,-0.54);
\node[font=\tiny, text=black] at (axis cs:40.10,-0.85) {25};
\fill[s4] (axis cs:52.80,-1.16) rectangle (axis cs:73.10,-0.54);
\node[font=\tiny, text=black] at (axis cs:62.95,-0.85) {20};
\fill[s5] (axis cs:73.10,-1.16) rectangle (axis cs:73.50,-0.54);
\fill[s6] (axis cs:73.50,-1.16) rectangle (axis cs:87.00,-0.54);
\fill[s7] (axis cs:87.00,-1.16) rectangle (axis cs:90.00,-0.54);
\node[font=\scriptsize, anchor=west] at (axis cs:92.0,-0.85) {\textbf{90} / $\leq$106};
\draw[pattern=north east lines, pattern color=black!55, draw=black!55] (axis cs:0,-2.16) rectangle (axis cs:69.5,-1.54);
\node[font=\tiny, fill=white, inner sep=0.5pt] at (axis cs:34.8,-1.85) {70};
\draw[draw=s2, line width=0.3pt, fill=white, postaction={pattern=crosshatch dots, pattern color=s2}] (axis cs:69.50,-2.16) rectangle (axis cs:71.80,-1.54);
\draw[draw=s3, line width=0.3pt, fill=white, postaction={pattern=crosshatch dots, pattern color=s3}] (axis cs:71.80,-2.16) rectangle (axis cs:91.80,-1.54);
\node[font=\tiny, text=black, fill=white, inner sep=0.5pt] at (axis cs:81.80,-1.85) {20};
\draw[draw=s4, line width=0.3pt, fill=white, postaction={pattern=crosshatch dots, pattern color=s4}] (axis cs:91.80,-2.16) rectangle (axis cs:113.80,-1.54);
\node[font=\tiny, text=black, fill=white, inner sep=0.5pt] at (axis cs:102.80,-1.85) {22};
\draw[draw=s5, line width=0.3pt, fill=white, postaction={pattern=crosshatch dots, pattern color=s5}] (axis cs:113.80,-2.16) rectangle (axis cs:114.10,-1.54);
\draw[draw=s6, line width=0.3pt, fill=white, postaction={pattern=crosshatch dots, pattern color=s6}] (axis cs:114.10,-2.16) rectangle (axis cs:130.00,-1.54);
\node[font=\tiny, text=black, fill=white, inner sep=0.5pt] at (axis cs:122.05,-1.85) {16};
\draw[draw=s7, line width=0.3pt, fill=white, postaction={pattern=crosshatch dots, pattern color=s7}] (axis cs:130.00,-2.16) rectangle (axis cs:134.40,-1.54);
\node[font=\scriptsize, anchor=west] at (axis cs:136.4,-1.85) {\textbf{135} / $\leq$152};
\node[font=\scriptsize\itshape, anchor=west, inner sep=1pt] at (axis cs:0,-3.05) {Native TRT, MPS};
\draw[pattern=north east lines, pattern color=black!55, draw=black!55] (axis cs:0,-4.21) rectangle (axis cs:25.8,-3.59);
\node[font=\tiny, fill=white, inner sep=0.5pt] at (axis cs:12.9,-3.90) {26};
\fill[s2] (axis cs:25.80,-4.21) rectangle (axis cs:27.80,-3.59);
\fill[s3] (axis cs:27.80,-4.21) rectangle (axis cs:51.30,-3.59);
\node[font=\tiny, text=black] at (axis cs:39.55,-3.90) {24};
\fill[s4] (axis cs:51.30,-4.21) rectangle (axis cs:72.70,-3.59);
\node[font=\tiny, text=black] at (axis cs:62.00,-3.90) {21};
\fill[s5] (axis cs:72.70,-4.21) rectangle (axis cs:110.70,-3.59);
\node[font=\tiny, text=white] at (axis cs:91.70,-3.90) {38};
\fill[s6] (axis cs:110.70,-4.21) rectangle (axis cs:124.90,-3.59);
\fill[s7] (axis cs:124.90,-4.21) rectangle (axis cs:128.00,-3.59);
\node[font=\scriptsize, anchor=west] at (axis cs:130.0,-3.90) {\textbf{128} / $\leq$147};
\draw[pattern=north east lines, pattern color=black!55, draw=black!55] (axis cs:0,-5.21) rectangle (axis cs:66.8,-4.59);
\node[font=\tiny, fill=white, inner sep=0.5pt] at (axis cs:33.4,-4.90) {67};
\draw[draw=s2, line width=0.3pt, fill=white, postaction={pattern=crosshatch dots, pattern color=s2}] (axis cs:66.80,-5.21) rectangle (axis cs:69.10,-4.59);
\draw[draw=s3, line width=0.3pt, fill=white, postaction={pattern=crosshatch dots, pattern color=s3}] (axis cs:69.10,-5.21) rectangle (axis cs:87.90,-4.59);
\node[font=\tiny, text=black, fill=white, inner sep=0.5pt] at (axis cs:78.50,-4.90) {19};
\draw[draw=s4, line width=0.3pt, fill=white, postaction={pattern=crosshatch dots, pattern color=s4}] (axis cs:87.90,-5.21) rectangle (axis cs:109.90,-4.59);
\node[font=\tiny, text=black, fill=white, inner sep=0.5pt] at (axis cs:98.90,-4.90) {22};
\draw[draw=s5, line width=0.3pt, fill=white, postaction={pattern=crosshatch dots, pattern color=s5}] (axis cs:109.90,-5.21) rectangle (axis cs:159.20,-4.59);
\node[font=\tiny, text=black, fill=white, inner sep=0.5pt] at (axis cs:134.55,-4.90) {49};
\draw[draw=s6, line width=0.3pt, fill=white, postaction={pattern=crosshatch dots, pattern color=s6}] (axis cs:159.20,-5.21) rectangle (axis cs:179.00,-4.59);
\node[font=\tiny, text=black, fill=white, inner sep=0.5pt] at (axis cs:169.10,-4.90) {20};
\draw[draw=s7, line width=0.3pt, fill=white, postaction={pattern=crosshatch dots, pattern color=s7}] (axis cs:179.00,-5.21) rectangle (axis cs:183.40,-4.59);
\node[font=\scriptsize, anchor=west] at (axis cs:185.4,-4.90) {\textbf{187} / $\leq$212};
\end{axis}
\node[font=\scriptsize, anchor=south, align=center] at ([yshift=4pt]pa.north) {\tikz[baseline=-0.6ex]\draw[draw=black, fill=black!45] (0,-0.08) rectangle (0.3,0.08);~GH200 \quad \tikz[baseline=-0.6ex]\draw[draw=black!60, fill=white, postaction={pattern=crosshatch dots, pattern color=black!60}] (0,-0.08) rectangle (0.3,0.08);~DGX Spark \\ Steps: \tikz[baseline=-0.6ex]\draw[pattern=north east lines, pattern color=black!55, draw=black!55] (0,-0.08) rectangle (0.3,0.08);~1 \quad \tikz[baseline=-0.6ex]\fill[s2] (0,-0.08) rectangle (0.3,0.08);~2 \quad \tikz[baseline=-0.6ex]\fill[s3] (0,-0.08) rectangle (0.3,0.08);~3 \quad \tikz[baseline=-0.6ex]\fill[s4] (0,-0.08) rectangle (0.3,0.08);~4 \quad \tikz[baseline=-0.6ex]\fill[s5] (0,-0.08) rectangle (0.3,0.08);~5 \quad \tikz[baseline=-0.6ex]\fill[s6] (0,-0.08) rectangle (0.3,0.08);~6 \quad \tikz[baseline=-0.6ex]\fill[s7] (0,-0.08) rectangle (0.3,0.08);~7};
\end{tikzpicture}

%% file: figures_tex/spark_step5.tex
\begin{tikzpicture}
\definecolor{s2}{HTML}{F0E442}
\definecolor{s3}{HTML}{E69F00}
\definecolor{s4}{HTML}{56B4E9}
\definecolor{s5}{HTML}{009E73}
\definecolor{s6}{HTML}{0072B2}
\definecolor{s7}{HTML}{CC79A7}
\definecolor{e0}{HTML}{9467BD}
\definecolor{e1}{HTML}{1F77B4}
\definecolor{e2}{HTML}{1F77B4}
\definecolor{e3}{HTML}{FF7F0E}
\definecolor{e4}{HTML}{2CA02C}
\definecolor{e5}{HTML}{1F77B4}
\begin{axis}[
name=pb, at={(0,0)}, anchor=south west, width=0.40\linewidth, height=0.45\linewidth, scale only axis,
xmin=-0.55, xmax=5.55, ymin=0, ymax=1500, clip=false,
axis y line*=left, axis x line*=bottom, xtick style={draw=none},
xtick={0,1,2,3,4,5},
xticklabels={{Native TRT},{C API TRT},{C API ONNX},{gRPC TRT},{Python TRT},{C API NumPy}},
xticklabel style={font=\scriptsize, rotate=40, anchor=north east, inner sep=1pt}, yticklabel style={font=\scriptsize},
ytick={0,250,500,750,1000,1250,1500}, yticklabel style={/pgf/number format/1000 sep={}},
ymajorgrids, grid style={black!12, line width=0.3pt},
xlabel={(a) MPS}, xlabel style={font=\footnotesize},
ylabel={Step 5 latency [$\mu$s]}, ylabel style={font=\footnotesize},
]
\draw[gray, densely dashed, thin] (axis cs:-0.55,500) -- (axis cs:5.55,500);
\draw[draw=black, line width=0.2pt, fill=e0, postaction={pattern=crosshatch, pattern color=black!70}] (axis cs:-0.360,0) rectangle (axis cs:0.000,36.4);
\draw[black, line width=0.4pt] (axis cs:-0.180,36.4) -- (axis cs:-0.180,41.5) (axis cs:-0.270,41.5) -- (axis cs:-0.090,41.5);
\node[font=\fontsize{6pt}{7pt}\selectfont, rotate=90, anchor=west, inner sep=1pt] at (axis cs:-0.180,56.5) {36};
\draw[draw=black, line width=0.2pt, fill=e1, postaction={pattern=north east lines, pattern color=black!70}] (axis cs:0.640,0) rectangle (axis cs:1.000,133.0);
\draw[black, line width=0.4pt] (axis cs:0.820,133.0) -- (axis cs:0.820,270.0) (axis cs:0.730,270.0) -- (axis cs:0.910,270.0);
\node[font=\fontsize{6pt}{7pt}\selectfont, rotate=90, anchor=west, inner sep=1pt] at (axis cs:0.820,285.0) {133};
\draw[draw=black, line width=0.2pt, fill=e2, postaction={pattern=north east lines, pattern color=black!70}] (axis cs:1.640,0) rectangle (axis cs:2.000,153.4);
\draw[black, line width=0.4pt] (axis cs:1.820,153.4) -- (axis cs:1.820,205.0) (axis cs:1.730,205.0) -- (axis cs:1.910,205.0);
\node[font=\fontsize{6pt}{7pt}\selectfont, rotate=90, anchor=west, inner sep=1pt] at (axis cs:1.820,220.0) {153};
\draw[draw=black, line width=0.2pt, fill=e3, postaction={pattern=north west lines, pattern color=black!70}] (axis cs:2.640,0) rectangle (axis cs:3.000,257.6);
\draw[black, line width=0.4pt] (axis cs:2.820,257.6) -- (axis cs:2.820,413.1) (axis cs:2.730,413.1) -- (axis cs:2.910,413.1);
\node[font=\fontsize{6pt}{7pt}\selectfont, rotate=90, anchor=west, inner sep=1pt] at (axis cs:2.820,428.1) {258};
\draw[draw=black, line width=0.2pt, fill=e4, postaction={pattern=horizontal lines, pattern color=black!70}] (axis cs:3.640,0) rectangle (axis cs:4.000,226.0);
\draw[black, line width=0.4pt] (axis cs:3.820,226.0) -- (axis cs:3.820,250.8) (axis cs:3.730,250.8) -- (axis cs:3.910,250.8);
\node[font=\fontsize{6pt}{7pt}\selectfont, rotate=90, anchor=west, inner sep=1pt] at (axis cs:3.820,265.8) {226};
\draw[draw=black, line width=0.2pt, fill=e5, postaction={pattern=north east lines, pattern color=black!70}] (axis cs:4.640,0) rectangle (axis cs:5.000,583.1);
\draw[black, line width=0.4pt] (axis cs:4.820,583.1) -- (axis cs:4.820,766.4) (axis cs:4.730,766.4) -- (axis cs:4.910,766.4);
\node[font=\fontsize{6pt}{7pt}\selectfont, rotate=90, anchor=west, inner sep=1pt] at (axis cs:4.820,781.4) {583};
\draw[draw=e0, line width=0.5pt, fill=white, postaction={pattern=crosshatch, pattern color=e0}] (axis cs:0.000,0) rectangle (axis cs:0.360,58.0);
\draw[black, line width=0.4pt] (axis cs:0.180,58.0) -- (axis cs:0.180,75.9) (axis cs:0.090,75.9) -- (axis cs:0.270,75.9);
\node[font=\fontsize{6pt}{7pt}\selectfont, rotate=90, anchor=west, inner sep=1pt] at (axis cs:0.180,90.9) {58};
\draw[draw=e1, line width=0.5pt, fill=white, postaction={pattern=north east lines, pattern color=e1}] (axis cs:1.000,0) rectangle (axis cs:1.360,134.5);
\draw[black, line width=0.4pt] (axis cs:1.180,134.5) -- (axis cs:1.180,212.4) (axis cs:1.090,212.4) -- (axis cs:1.270,212.4);
\node[font=\fontsize{6pt}{7pt}\selectfont, rotate=90, anchor=west, inner sep=1pt] at (axis cs:1.180,227.4) {134};
\draw[draw=e2, line width=0.5pt, fill=white, postaction={pattern=north east lines, pattern color=e2}] (axis cs:2.000,0) rectangle (axis cs:2.360,256.0);
\draw[black, line width=0.4pt] (axis cs:2.180,256.0) -- (axis cs:2.180,321.5) (axis cs:2.090,321.5) -- (axis cs:2.270,321.5);
\node[font=\fontsize{6pt}{7pt}\selectfont, rotate=90, anchor=west, inner sep=1pt] at (axis cs:2.180,336.5) {256};
\draw[draw=e3, line width=0.5pt, fill=white, postaction={pattern=north west lines, pattern color=e3}] (axis cs:3.000,0) rectangle (axis cs:3.360,248.7);
\draw[black, line width=0.4pt] (axis cs:3.180,248.7) -- (axis cs:3.180,323.2) (axis cs:3.090,323.2) -- (axis cs:3.270,323.2);
\node[font=\fontsize{6pt}{7pt}\selectfont, rotate=90, anchor=west, inner sep=1pt] at (axis cs:3.180,338.2) {249};
\draw[draw=e4, line width=0.5pt, fill=white, postaction={pattern=horizontal lines, pattern color=e4}] (axis cs:4.000,0) rectangle (axis cs:4.360,153.0);
\draw[black, line width=0.4pt] (axis cs:4.180,153.0) -- (axis cs:4.180,184.0) (axis cs:4.090,184.0) -- (axis cs:4.270,184.0);
\node[font=\fontsize{6pt}{7pt}\selectfont, rotate=90, anchor=west, inner sep=1pt] at (axis cs:4.180,199.0) {153};
\draw[draw=e5, line width=0.5pt, fill=white, postaction={pattern=north east lines, pattern color=e5}] (axis cs:5.000,0) rectangle (axis cs:5.360,860.4);
\draw[black, line width=0.4pt] (axis cs:5.180,860.4) -- (axis cs:5.180,1195.5) (axis cs:5.090,1195.5) -- (axis cs:5.270,1195.5);
\node[font=\fontsize{6pt}{7pt}\selectfont, rotate=90, anchor=west, inner sep=1pt] at (axis cs:5.180,1210.5) {860};
\end{axis}
\begin{axis}[
name=pc, at={($(pb.south east)+(0.03\linewidth,0)$)}, anchor=south west, width=0.40\linewidth, height=0.45\linewidth, scale only axis,
xmin=-0.55, xmax=5.55, ymin=0, ymax=1500, clip=false,
axis y line*=left, axis x line*=bottom, xtick style={draw=none},
xtick={0,1,2,3,4,5},
xticklabels={{Native TRT},{C API TRT},{C API ONNX},{gRPC TRT},{Python TRT},{C API NumPy}},
xticklabel style={font=\scriptsize, rotate=40, anchor=north east, inner sep=1pt}, yticklabel style={font=\scriptsize},
ytick={0,250,500,750,1000,1250,1500}, yticklabel style={/pgf/number format/1000 sep={}},
ymajorgrids, grid style={black!12, line width=0.3pt},
xlabel={(b) Time-slicing}, xlabel style={font=\footnotesize},
yticklabels={},
]
\draw[gray, densely dashed, thin] (axis cs:-0.55,500) -- (axis cs:5.55,500);
\draw[draw=black, line width=0.2pt, fill=e0, postaction={pattern=crosshatch, pattern color=black!70}] (axis cs:-0.360,0) rectangle (axis cs:0.000,502.0);
\draw[black, line width=0.4pt] (axis cs:-0.180,502.0) -- (axis cs:-0.180,716.9) (axis cs:-0.270,716.9) -- (axis cs:-0.090,716.9);
\node[font=\fontsize{6pt}{7pt}\selectfont, rotate=90, anchor=west, inner sep=1pt] at (axis cs:-0.180,731.9) {502};
\draw[draw=black, line width=0.2pt, fill=e1, postaction={pattern=north east lines, pattern color=black!70}] (axis cs:0.640,0) rectangle (axis cs:1.000,511.1);
\draw[black, line width=0.4pt] (axis cs:0.820,511.1) -- (axis cs:0.820,812.2) (axis cs:0.730,812.2) -- (axis cs:0.910,812.2);
\node[font=\fontsize{6pt}{7pt}\selectfont, rotate=90, anchor=west, inner sep=1pt] at (axis cs:0.820,827.2) {511};
\draw[draw=black, line width=0.2pt, fill=e2, postaction={pattern=north east lines, pattern color=black!70}] (axis cs:1.640,0) rectangle (axis cs:2.000,562.2);
\draw[black, line width=0.4pt] (axis cs:1.820,562.2) -- (axis cs:1.820,778.9) (axis cs:1.730,778.9) -- (axis cs:1.910,778.9);
\node[font=\fontsize{6pt}{7pt}\selectfont, rotate=90, anchor=west, inner sep=1pt] at (axis cs:1.820,793.9) {562};
\draw[draw=black, line width=0.2pt, fill=e3, postaction={pattern=north west lines, pattern color=black!70}] (axis cs:2.640,0) rectangle (axis cs:3.000,493.1);
\draw[black, line width=0.4pt] (axis cs:2.820,493.1) -- (axis cs:2.820,860.6) (axis cs:2.730,860.6) -- (axis cs:2.910,860.6);
\node[font=\fontsize{6pt}{7pt}\selectfont, rotate=90, anchor=west, inner sep=1pt] at (axis cs:2.820,875.6) {493};
\draw[draw=black, line width=0.2pt, fill=e4, postaction={pattern=horizontal lines, pattern color=black!70}] (axis cs:3.640,0) rectangle (axis cs:4.000,494.0);
\draw[black, line width=0.4pt] (axis cs:3.820,494.0) -- (axis cs:3.820,744.5) (axis cs:3.730,744.5) -- (axis cs:3.910,744.5);
\node[font=\fontsize{6pt}{7pt}\selectfont, rotate=90, anchor=west, inner sep=1pt] at (axis cs:3.820,759.5) {494};
\draw[draw=black, line width=0.2pt, fill=e5, postaction={pattern=north east lines, pattern color=black!70}] (axis cs:4.640,0) rectangle (axis cs:5.000,593.6);
\draw[black, line width=0.4pt] (axis cs:4.820,593.6) -- (axis cs:4.820,755.3) (axis cs:4.730,755.3) -- (axis cs:4.910,755.3);
\node[font=\fontsize{6pt}{7pt}\selectfont, rotate=90, anchor=west, inner sep=1pt] at (axis cs:4.820,770.3) {594};
\draw[draw=e0, line width=0.5pt, fill=white, postaction={pattern=crosshatch, pattern color=e0}] (axis cs:0.000,0) rectangle (axis cs:0.360,434.0);
\draw[black, line width=0.4pt] (axis cs:0.180,434.0) -- (axis cs:0.180,662.0) (axis cs:0.090,662.0) -- (axis cs:0.270,662.0);
\node[font=\fontsize{6pt}{7pt}\selectfont, rotate=90, anchor=west, inner sep=1pt] at (axis cs:0.180,677.0) {434};
\draw[draw=e1, line width=0.5pt, fill=white, postaction={pattern=north east lines, pattern color=e1}] (axis cs:1.000,0) rectangle (axis cs:1.360,447.0);
\draw[black, line width=0.4pt] (axis cs:1.180,447.0) -- (axis cs:1.180,1291.0) (axis cs:1.090,1291.0) -- (axis cs:1.270,1291.0);
\node[font=\fontsize{6pt}{7pt}\selectfont, rotate=90, anchor=west, inner sep=1pt] at (axis cs:1.180,1306.0) {447};
\draw[draw=e2, line width=0.5pt, fill=white, postaction={pattern=north east lines, pattern color=e2}] (axis cs:2.000,0) rectangle (axis cs:2.360,625.0);
\draw[black, line width=0.4pt] (axis cs:2.180,625.0) -- (axis cs:2.180,829.0) (axis cs:2.090,829.0) -- (axis cs:2.270,829.0);
\node[font=\fontsize{6pt}{7pt}\selectfont, rotate=90, anchor=west, inner sep=1pt] at (axis cs:2.180,844.0) {625};
\draw[draw=e3, line width=0.5pt, fill=white, postaction={pattern=north west lines, pattern color=e3}] (axis cs:3.000,0) rectangle (axis cs:3.360,528.0);
\draw[black, line width=0.4pt] (axis cs:3.180,528.0) -- (axis cs:3.180,1328.0) (axis cs:3.090,1328.0) -- (axis cs:3.270,1328.0);
\node[font=\fontsize{6pt}{7pt}\selectfont, rotate=90, anchor=west, inner sep=1pt] at (axis cs:3.180,1343.0) {528};
\draw[draw=e4, line width=0.5pt, fill=white, postaction={pattern=horizontal lines, pattern color=e4}] (axis cs:4.000,0) rectangle (axis cs:4.360,460.0);
\draw[black, line width=0.4pt] (axis cs:4.180,460.0) -- (axis cs:4.180,675.0) (axis cs:4.090,675.0) -- (axis cs:4.270,675.0);
\node[font=\fontsize{6pt}{7pt}\selectfont, rotate=90, anchor=west, inner sep=1pt] at (axis cs:4.180,690.0) {460};
\draw[draw=e5, line width=0.5pt, fill=white, postaction={pattern=north east lines, pattern color=e5}] (axis cs:5.000,0) rectangle (axis cs:5.360,876.0);
\draw[black, line width=0.4pt] (axis cs:5.180,876.0) -- (axis cs:5.180,1302.0) (axis cs:5.090,1302.0) -- (axis cs:5.270,1302.0);
\node[font=\fontsize{6pt}{7pt}\selectfont, rotate=90, anchor=west, inner sep=1pt] at (axis cs:5.180,1317.0) {876};
\end{axis}
\node[font=\scriptsize, anchor=south] at ([yshift=4pt]$(pb.north)!0.5!(pc.north)$) {\tikz[baseline=-0.6ex]\draw[draw=black, fill=black!45] (0,-0.08) rectangle (0.3,0.08);~GH200 \quad \tikz[baseline=-0.6ex]\draw[draw=black!60, fill=white, postaction={pattern=north east lines, pattern color=black!60}] (0,-0.08) rectangle (0.3,0.08);~DGX Spark};
\end{tikzpicture}

%% file: figures_tex/cusense_resnet.tex
\begin{tikzpicture}[
    node distance=0.15cm,
    >=Stealth,
    scale=0.825, transform shape,
    input/.style={rectangle, rounded corners=3pt, minimum width=1.2cm, minimum height=0.6cm,
        draw=inputcolor!80!black, fill=inputcolor!30, font=\scriptsize\bfseries, text=inputcolor!80!black},
    conv/.style={rectangle, rounded corners=2pt, minimum width=1.8cm, minimum height=0.45cm,
        draw=convcolor!80!black, fill=convcolor!25, font=\tiny, text=convcolor!80!black},
    bn/.style={rectangle, rounded corners=2pt, minimum width=1.8cm, minimum height=0.35cm,
        draw=bncolor!80!black, fill=bncolor!25, font=\tiny, text=bncolor!80!black},
    activation/.style={rectangle, rounded corners=2pt, minimum width=1.8cm, minimum height=0.35cm,
        draw=relucolor!80!black, fill=relucolor!20, font=\tiny, text=relucolor!80!black},
    pool/.style={rectangle, rounded corners=2pt, minimum width=1.8cm, minimum height=0.45cm,
        draw=poolcolor!80!black, fill=poolcolor!25, font=\tiny, text=poolcolor!80!black},
    fc/.style={rectangle, rounded corners=2pt, minimum width=1.8cm, minimum height=0.45cm,
        draw=fccolor!80!black, fill=fccolor!25, font=\tiny, text=fccolor!80!black},
    dropout/.style={rectangle, rounded corners=2pt, minimum width=1.8cm, minimum height=0.35cm,
        draw=dropcolor!80!black, fill=dropcolor!25, font=\tiny, text=dropcolor!80!black},
    softmax/.style={rectangle, rounded corners=2pt, minimum width=1.8cm, minimum height=0.45cm,
        draw=softmaxcolor!80!black, fill=softmaxcolor!25, font=\tiny, text=softmaxcolor!80!black},
    output/.style={rectangle, rounded corners=3pt, minimum width=1.2cm, minimum height=0.6cm,
        draw=outputcolor!80!black, fill=outputcolor!30, font=\scriptsize\bfseries, text=outputcolor!80!black},
    resblock/.style={rectangle, rounded corners=5pt, draw=arrowcolor!50, dashed, fill=resblockcolor!50, inner sep=5pt},
    arrow/.style={->, very thin, arrowcolor},
    bigarrow/.style={->, thin, arrowcolor},
    dimtext/.style={font=\tiny\itshape, gray!70},
]

\def\centerY{0}

\node[input, label={[font=\tiny\bfseries, yshift=1pt]above:CSI-based Input}] (input) at (0, \centerY - 0.26) {\textbf{$X_{\mathrm{t}}$}};
\node[dimtext, below=0.15cm of input] {$[B, A, K_v]$};

\node[conv] (conv0) at (2.2, \centerY + 0.66) {Conv1d(4$\to$64, k=7)};
\node[bn, below=0.2cm of conv0] (bn0) {BatchNorm1d};
\node[activation, below=0.2cm of bn0] (relu0) {ReLU};
\node[pool, below=0.2cm of relu0] (pool0) {MaxPool1d(2)};
\begin{scope}[on background layer]
    \node[resblock, fit=(conv0)(bn0)(relu0)(pool0), label={[font=\tiny\bfseries, yshift=1pt]above:Initial Conv}] (initblock) {};
\end{scope}
\node[dimtext, below=0.15cm of initblock] {$[B, 64, K_v/2]$};

\node[conv] (conv1a) at (5.1, \centerY + 1.26) {Conv1d(64$\to$128, s=2)};
\node[bn, below=0.2cm of conv1a] (bn1a) {BatchNorm1d};
\node[activation, below=0.2cm of bn1a] (relu1a) {ReLU};
\node[conv, below=0.2cm of relu1a] (conv1b) {Conv1d(128$\to$128)};
\node[bn, below=0.2cm of conv1b] (bn1b) {BatchNorm1d};
\node[activation, below=0.2cm of bn1b] (relu1b) {ReLU};
\begin{scope}[on background layer]
    \node[resblock, fit=(conv1a)(bn1a)(relu1a)(conv1b)(bn1b)(relu1b), label={[font=\tiny\bfseries, yshift=1pt]above:Conv Block 1}] (block1) {};
\end{scope}
\node[dimtext, below=0.15cm of block1] {$[B, 128, K_v/4]$};

\node[conv] (conv2a) at (8.1, \centerY + 1.26) {Conv1d(128$\to$256, s=2)};
\node[bn, below=0.2cm of conv2a] (bn2a) {BatchNorm1d};
\node[activation, below=0.2cm of bn2a] (relu2a) {ReLU};
\node[conv, below=0.2cm of relu2a] (conv2b) {Conv1d(256$\to$256)};
\node[bn, below=0.2cm of conv2b] (bn2b) {BatchNorm1d};
\node[activation, below=0.2cm of bn2b] (relu2b) {ReLU};
\begin{scope}[on background layer]
    \node[resblock, fit=(conv2a)(bn2a)(relu2a)(conv2b)(bn2b)(relu2b), label={[font=\tiny\bfseries, yshift=1pt]above:Conv Block 2}] (block2) {};
\end{scope}
\node[dimtext, below=0.15cm of block2] {$[B, 256, K_v/8]$};

\node[conv] (conv3a) at (11.1, \centerY + 1.26) {Conv1d(256$\to$512, s=2)};
\node[bn, below=0.2cm of conv3a] (bn3a) {BatchNorm1d};
\node[activation, below=0.2cm of bn3a] (relu3a) {ReLU};
\node[conv, below=0.2cm of relu3a] (conv3b) {Conv1d(512$\to$512)};
\node[bn, below=0.2cm of conv3b] (bn3b) {BatchNorm1d};
\node[activation, below=0.2cm of bn3b] (relu3b) {ReLU};
\begin{scope}[on background layer]
    \node[resblock, fit=(conv3a)(bn3a)(relu3a)(conv3b)(bn3b)(relu3b), label={[font=\tiny\bfseries, yshift=1pt]above:Conv Block 3}] (block3) {};
\end{scope}
\node[dimtext, below=0.15cm of block3] {$[B, 512, K_v/16]$};

\node[pool] (gpool) at (13.9, \centerY - 0.26) {AdaptiveAvgPool1d(1)};
\node[dimtext, below=0.15cm of gpool] {$[B, 512]$};

\node[fc] (fc1) at (16.5, \centerY + 1.97) {Linear(512$\to$512)};
\node[activation, below=0.2cm of fc1] (relu4) {ReLU};
\node[dropout, below=0.2cm of relu4] (drop1) {Dropout(0.3)};
\node[fc, below=0.2cm of drop1] (fc2) {Linear(512$\to$256)};
\node[activation, below=0.2cm of fc2] (relu5) {ReLU};
\node[dropout, below=0.2cm of relu5] (drop2) {Dropout(0.15)};
\node[fc, below=0.2cm of drop2] (fc3) {Linear(256$\to$1024)};
\node[softmax, below=0.2cm of fc3] (softmax) {Softmax};
\begin{scope}[on background layer]
    \node[resblock, fit=(fc1)(relu4)(drop1)(fc2)(relu5)(drop2)(fc3)(softmax), label={[font=\tiny\bfseries, yshift=1pt]above:Classifier}] (classifier) {};
\end{scope}

\node[output] (output) at (19.3, \centerY - 0.26) {\textbf{Location Grid}};
\node[dimtext, below=0.15cm of output] {$[B, H, W]$};

\draw[bigarrow] (input.east) -- (initblock.west);
\draw[bigarrow] (initblock.east) -- (block1.west);
\draw[bigarrow] (block1.east) -- (block2.west);
\draw[bigarrow] (block2.east) -- (block3.west);
\draw[bigarrow] (block3.east) -- (gpool.west);
\draw[bigarrow] (gpool.east) -- (classifier.west);
\draw[bigarrow] (classifier.east) -- (output.west);

\draw[arrow] (conv0) -- (bn0);
\draw[arrow] (bn0) -- (relu0);
\draw[arrow] (relu0) -- (pool0);

\draw[arrow] (conv1a) -- (bn1a);
\draw[arrow] (bn1a) -- (relu1a);
\draw[arrow] (relu1a) -- (conv1b);
\draw[arrow] (conv1b) -- (bn1b);
\draw[arrow] (bn1b) -- (relu1b);

\draw[arrow] (conv2a) -- (bn2a);
\draw[arrow] (bn2a) -- (relu2a);
\draw[arrow] (relu2a) -- (conv2b);
\draw[arrow] (conv2b) -- (bn2b);
\draw[arrow] (bn2b) -- (relu2b);

\draw[arrow] (conv3a) -- (bn3a);
\draw[arrow] (bn3a) -- (relu3a);
\draw[arrow] (relu3a) -- (conv3b);
\draw[arrow] (conv3b) -- (bn3b);
\draw[arrow] (bn3b) -- (relu3b);

\draw[arrow] (fc1) -- (relu4);
\draw[arrow] (relu4) -- (drop1);
\draw[arrow] (drop1) -- (fc2);
\draw[arrow] (fc2) -- (relu5);
\draw[arrow] (relu5) -- (drop2);
\draw[arrow] (drop2) -- (fc3);
\draw[arrow] (fc3) -- (softmax);

\node[above=1.9cm of input, xshift=-1.2cm] (legendtitle) {\scriptsize\bfseries Legend:};
\node[conv, right=0.15cm of legendtitle, minimum width=0.8cm, minimum height=0.3cm] (legconv) {\tiny Conv};
\node[bn, right=0.08cm of legconv, minimum width=0.6cm, minimum height=0.3cm] (legbn) {\tiny BN};
\node[activation, right=0.08cm of legbn, minimum width=0.6cm, minimum height=0.3cm] (legrelu) {\tiny ReLU};
\node[pool, right=0.08cm of legrelu, minimum width=0.6cm, minimum height=0.3cm] (legpool) {\tiny Pool};
\node[fc, right=0.08cm of legpool, minimum width=0.5cm, minimum height=0.3cm] (legfc) {\tiny FC};
\node[dropout, right=0.08cm of legfc, minimum width=0.7cm, minimum height=0.3cm] (legdrop) {\tiny Dropout};

\end{tikzpicture}

%% file: figures_tex/cusense_dataset.tex
\begin{tikzpicture}
\definecolor{lightgray204}{RGB}{204,204,204}
\definecolor{avgcolor}{RGB}{230,159,0}
\definecolor{kalmancolor}{RGB}{86,180,233}
\definecolor{unseengray}{RGB}{220,220,220}

\node[font=\footnotesize, anchor=east] at (0.55,1.4) {Training runs};

\draw[draw=black,fill=white] (0.6,1.2) rectangle (2.1,1.6);

\draw[draw=black,fill=avgcolor,postaction={pattern=north west lines,pattern color=black}] (2.1,1.2) rectangle (2.65,1.6);
\node[font=\tiny\bfseries,white] at (2.375,1.4) {10\%};

\draw[draw=black,fill=white] (2.65,1.2) rectangle (3.85,1.6);

\draw[draw=black,fill=kalmancolor,postaction={pattern=north east lines,pattern color=black}] (3.85,1.2) rectangle (4.4,1.6);
\node[font=\tiny\bfseries,white] at (4.125,1.4) {10\%};

\draw[draw=black,fill=white] (4.4,1.2) rectangle (5.2,1.6);

\node[font=\footnotesize, anchor=east] at (0.55,0.7) {Unseen runs};

\draw[draw=black,fill=unseengray,postaction={pattern=vertical lines,pattern color=black}] (0.6,0.5) rectangle (5.2,0.9);

\draw[->] (0.6,0.2) -- (5.2,0.2);
\node[font=\footnotesize] at (2.9,0.0) {Samples};

\node[
    draw=lightgray204,
    fill=white,
    fill opacity=0.8,
    font=\footnotesize,
    inner sep=3pt,
    anchor=west,
] at (5.4,0.9) {
    \begin{tabular}{@{}cl@{}}
        \tikz{\draw[draw=black,fill=white] (0,0) rectangle (0.25,0.15);} & Train \\[2pt]
        \tikz{\draw[draw=black,fill=avgcolor,postaction={pattern=north west lines,pattern color=black}] (0,0) rectangle (0.25,0.15);} & Val \\[2pt]
        \tikz{\draw[draw=black,fill=kalmancolor,postaction={pattern=north east lines,pattern color=black}] (0,0) rectangle (0.25,0.15);} & Test \\[2pt]
        \tikz{\draw[draw=black,fill=unseengray,postaction={pattern=vertical lines,pattern color=black}] (0,0) rectangle (0.25,0.15);} & Unseen \\
    \end{tabular}
};

\end{tikzpicture}

%% file: figures_tex/cusense_cdf.tex
\begin{tikzpicture}
\definecolor{cs0}{HTML}{1F77B4}
\definecolor{cs1}{HTML}{FF7F0E}
\definecolor{cs2}{HTML}{2CA02C}
\begin{axis}[
name=bars, at={(0,0)}, anchor=south west, width=0.30\linewidth, height=0.34\linewidth, scale only axis,
xmin=-0.5, xmax=2.5, ymin=0, ymax=100, xtick={0,1,2}, ytick={0,20,40,60,80,100},
xticklabels={{$\leq$0.5~m\\Cat.~4},{$\leq$1~m\\Cat.~3},{$\leq$2~m\\Cat.~2}},
xticklabel style={font=\scriptsize, align=center}, yticklabel style={font=\scriptsize},
ylabel={Within threshold [\%]}, ylabel style={font=\footnotesize}, ymajorgrids, grid style={black!12},
axis x line*=bottom, axis y line*=left, xtick style={draw=none}, clip=false,
]
\draw[draw=black, line width=0.2pt, fill=cs0, postaction={pattern=north east lines, pattern color=black!70}] (axis cs:-0.390,0) rectangle (axis cs:-0.130,33.02);
\node[font=\fontsize{6.5pt}{7.5pt}\selectfont, rotate=90, anchor=west, inner sep=1pt] at (axis cs:-0.260,33.02) {33.0};
\draw[draw=black, line width=0.2pt, fill=cs0, postaction={pattern=north east lines, pattern color=black!70}] (axis cs:0.610,0) rectangle (axis cs:0.870,65.05);
\node[font=\fontsize{6.5pt}{7.5pt}\selectfont, rotate=90, anchor=west, inner sep=1pt] at (axis cs:0.740,65.05) {65.1};
\draw[draw=black, line width=0.2pt, fill=cs0, postaction={pattern=north east lines, pattern color=black!70}] (axis cs:1.610,0) rectangle (axis cs:1.870,87.93);
\node[font=\fontsize{6.5pt}{7.5pt}\selectfont, rotate=90, anchor=west, inner sep=1pt] at (axis cs:1.740,87.93) {87.9};
\draw[draw=black, line width=0.2pt, fill=cs1, postaction={pattern=north west lines, pattern color=black!70}] (axis cs:-0.130,0) rectangle (axis cs:0.130,36.81);
\node[font=\fontsize{6.5pt}{7.5pt}\selectfont, rotate=90, anchor=west, inner sep=1pt] at (axis cs:0.000,36.81) {36.8};
\draw[draw=black, line width=0.2pt, fill=cs1, postaction={pattern=north west lines, pattern color=black!70}] (axis cs:0.870,0) rectangle (axis cs:1.130,69.73);
\node[font=\fontsize{6.5pt}{7.5pt}\selectfont, rotate=90, anchor=west, inner sep=1pt] at (axis cs:1.000,69.73) {69.7};
\draw[draw=black, line width=0.2pt, fill=cs1, postaction={pattern=north west lines, pattern color=black!70}] (axis cs:1.870,0) rectangle (axis cs:2.130,90.56);
\node[font=\fontsize{6.5pt}{7.5pt}\selectfont, rotate=90, anchor=west, inner sep=1pt] at (axis cs:2.000,90.56) {90.6};
\draw[draw=black, line width=0.2pt, fill=cs2, postaction={pattern=horizontal lines, pattern color=black!70}] (axis cs:0.130,0) rectangle (axis cs:0.390,42.58);
\node[font=\fontsize{6.5pt}{7.5pt}\selectfont, rotate=90, anchor=west, inner sep=1pt] at (axis cs:0.260,42.58) {42.6};
\draw[draw=black, line width=0.2pt, fill=cs2, postaction={pattern=horizontal lines, pattern color=black!70}] (axis cs:1.130,0) rectangle (axis cs:1.390,78.64);
\node[font=\fontsize{6.5pt}{7.5pt}\selectfont, rotate=90, anchor=west, inner sep=1pt] at (axis cs:1.260,78.64) {78.6};
\draw[draw=black, line width=0.2pt, fill=cs2, postaction={pattern=horizontal lines, pattern color=black!70}] (axis cs:2.130,0) rectangle (axis cs:2.390,94.41);
\node[font=\fontsize{6.5pt}{7.5pt}\selectfont, rotate=90, anchor=west, inner sep=1pt] at (axis cs:2.260,94.41) {94.4};
\end{axis}
\node[font=\footnotesize, anchor=north] at ([yshift=-24pt]bars.south) {(a)};
\begin{axis}[
name=cdf, at={(0.44\linewidth,0)}, anchor=south west, width=0.40\linewidth, height=0.34\linewidth, scale only axis,
xmin=0, xmax=6, ymin=0, ymax=100, xtick={0,1,2,3,4,5,6}, ytick={0,20,40,60,80,100},
xticklabel style={font=\scriptsize}, yticklabel style={font=\scriptsize},
xlabel={Localization error [m]}, ylabel={CDF [\%]}, xlabel style={font=\footnotesize},
ylabel style={font=\footnotesize}, xmajorgrids, ymajorgrids, grid style={black!12}, clip=false,
]
\draw[black!55, dashed, thin] (axis cs:0.5,0) -- (axis cs:0.5,100);
\node[black!65, font=\fontsize{6.5pt}{7.5pt}\selectfont, rotate=90, anchor=north west, inner sep=1pt] at (axis cs:0.5,1) {Cat.~4};
\draw[black!55, dashed, thin] (axis cs:1.0,0) -- (axis cs:1.0,100);
\node[black!65, font=\fontsize{6.5pt}{7.5pt}\selectfont, rotate=90, anchor=north west, inner sep=1pt] at (axis cs:1.0,1) {Cat.~3};
\draw[black!55, dashed, thin] (axis cs:2.0,0) -- (axis cs:2.0,100);
\node[black!65, font=\fontsize{6.5pt}{7.5pt}\selectfont, rotate=90, anchor=north west, inner sep=1pt] at (axis cs:2.0,1) {Cat.~2};
\draw[black!60, densely dotted] (axis cs:0,95) -- (axis cs:6,95);
\addplot[const plot, cs0, thick] coordinates {(0.0000,0.000) (0.0000,2.452) (0.1736,9.970) (0.2455,16.343) (0.3471,21.212) (0.3881,29.692) (0.4909,33.020) (0.5207,35.965) (0.5489,41.711) (0.6258,46.343) (0.6943,48.213) (0.7157,51.688) (0.7364,53.394) (0.7762,56.443) (0.8679,60.047) (0.8850,62.290) (0.9347,64.138) (0.9819,65.052) (1.0121,66.646) (1.0414,67.378) (1.0558,68.792) (1.0978,69.860) (1.1114,71.144) (1.1644,72.058) (1.2150,72.579) (1.2273,73.992) (1.2516,74.917) (1.2636,75.692) (1.3219,76.330) (1.3556,77.103) (1.3886,77.441) (1.3994,78.723) (1.4313,79.247) (1.4728,79.594) (1.4830,80.076) (1.4931,80.593) (1.5525,81.052) (1.5621,81.298) (1.5718,81.843) (1.6002,82.679) (1.6375,83.071) (1.6466,83.416) (1.7095,83.734) (1.7183,83.952) (1.7357,84.452) (1.7444,84.805) (1.7701,85.130) (1.7870,85.432) (1.8121,85.671) (1.8451,86.012) (1.8694,86.277) (1.8775,86.535) (1.9093,86.641) (1.9172,86.910) (1.9406,87.386) (1.9637,87.509) (1.9790,87.929) (2.0242,88.084) (2.0316,88.291) (2.0829,88.399) (2.0901,88.793) (2.0973,88.967) (2.1116,89.204) (2.1187,89.355) (2.1470,89.545) (2.1748,89.652) (2.1955,89.797) (2.2092,89.894) (2.2228,90.049) (2.2564,90.273) (2.2631,90.568) (2.2830,90.734) (2.3157,90.872) (2.3287,90.988) (2.3352,91.122) (2.3608,91.344) (2.4113,91.482) (2.4176,91.561) (2.4300,91.607) (2.4362,91.721) (2.4547,91.943) (2.4669,92.042) (2.4852,92.255) (2.5033,92.354) (2.5272,92.470) (2.5628,92.561) (2.5803,92.735) (2.6036,92.901) (2.6094,93.015) (2.6266,93.128) (2.6438,93.179) (2.6495,93.240) (2.6551,93.369) (2.6946,93.446) (2.7001,93.498) (2.7113,93.558) (2.7168,93.647) (2.7444,93.808) (2.7771,93.836) (2.7826,93.916) (2.7988,94.058) (2.8041,94.119) (2.8255,94.310) (2.8468,94.358) (2.8626,94.440) (2.8731,94.485) (2.8888,94.526) (2.9096,94.586) (2.9456,94.621) (2.9507,94.694) (2.9558,94.810) (2.9660,94.868) (2.9711,94.916) (2.9862,94.965) (2.9963,95.036) (3.0313,95.146) (3.0363,95.183) (3.0708,95.245) (3.0757,95.303) (3.0904,95.329) (3.1049,95.377) (3.1243,95.387) (3.1291,95.536) (3.1435,95.584) (3.1674,95.663) (3.1863,95.696) (3.1911,95.760) (3.2005,95.864) (3.2286,95.889) (3.2426,95.930) (3.2611,95.956) (3.2749,95.991) (3.2933,96.023) (3.2979,96.032) (3.3024,96.055) (3.3161,96.141) (3.3342,96.176) (3.3387,96.264) (3.3522,96.299) (3.3701,96.413) (3.4101,96.478) (3.4190,96.501) (3.4234,96.542) (3.4365,96.559) (3.4453,96.598) (3.4584,96.635) (3.4714,96.659) (3.4758,96.669) (3.4887,96.691) (3.4931,96.723) (3.5103,96.771) (3.5146,96.831) (3.5402,96.898) (3.5614,96.924) (3.5741,96.941) (3.5783,97.029) (3.6118,97.053) (3.6243,97.089) (3.6450,97.096) (3.6491,97.128) (3.6615,97.152) (3.6779,97.195) (3.6820,97.223) (3.6902,97.242) (3.7105,97.292) (3.7146,97.301) (3.7267,97.318) (3.7388,97.344) (3.7469,97.400) (3.7549,97.417) (3.7909,97.438) (3.8067,97.475) (3.8107,97.499) (3.8186,97.516) (3.8225,97.548) (3.8343,97.561) (3.8422,97.594) (3.8539,97.622) (3.8812,97.656) (3.9005,97.684) (3.9159,97.723) (3.9275,97.725) (3.9351,97.740) (3.9580,97.772) (3.9618,97.790) (3.9656,97.798) (3.9921,97.811) (3.9959,97.841) (4.0072,97.867) (4.0260,97.874) (4.0372,97.895) (4.0483,97.915) (4.0521,97.945) (4.0632,97.968) (4.0669,97.981) (4.0854,97.992) (4.0964,98.001) (4.1148,98.024) (4.1257,98.052) (4.1403,98.068) (4.1657,98.070) (4.1693,98.080) (4.1729,98.102) (4.1802,98.128) (4.1945,98.139) (4.1981,98.160) (4.2017,98.162) (4.2232,98.173) (4.2267,98.184) (4.2374,98.190) (4.2552,98.203) (4.2693,98.216) (4.2869,98.240) (4.2939,98.244) (4.2974,98.257) (4.3114,98.268) (4.3393,98.287) (4.3428,98.294) (4.3497,98.307) (4.3531,98.322) (4.3704,98.324) (4.3807,98.337) (4.3910,98.339) (4.4184,98.341) (4.4252,98.399) (4.4354,98.412) (4.4456,98.417) (4.4490,98.425) (4.4625,98.429) (4.4794,98.442) (4.5028,98.466) (4.5062,98.473) (4.5129,98.490) (4.5162,98.503) (4.5262,98.522) (4.5428,98.535) (4.5560,98.544) (4.5660,98.548) (4.5824,98.574) (4.5857,98.582) (4.5955,98.585) (4.6119,98.591) (4.6217,98.600) (4.6315,98.604) (4.6574,98.613) (4.6639,98.619) (4.6703,98.630) (4.6736,98.651) (4.6864,98.656) (4.6896,98.666) (4.6993,98.671) (4.7153,98.675) (4.7217,98.692) (4.7376,98.703) (4.7408,98.720) (4.7661,98.740) (4.7756,98.744) (4.7882,98.753) (4.8007,98.761) (4.8133,98.768) (4.8227,98.772) (4.8351,98.774) (4.8631,98.787) (4.8724,98.791) (4.8878,98.802) (4.8909,98.811) (4.9001,98.817) (4.9093,98.822) (4.9124,98.830) (4.9155,98.845) (4.9338,98.852) (4.9369,98.856) (4.9399,98.862) (4.9643,98.869) (4.9703,98.873) (4.9734,98.880) (4.9975,98.884) (5.0066,98.893) (5.0336,98.899) (5.0366,98.906) (5.0455,98.921) (5.0545,98.923) (5.0604,98.936) (5.0694,98.940) (5.0812,98.942) (5.1049,98.955) (5.1255,98.962) (5.1284,98.968) (5.1519,98.970) (5.1548,98.979) (5.1606,98.987) (5.1781,98.992) (5.2014,98.998) (5.2071,99.000) (5.2100,99.018) (5.2216,99.020) (5.2331,99.024) (5.2475,99.026) (5.2790,99.039) (5.2904,99.056) (5.2989,99.059) (5.3103,99.063) (5.3244,99.069) (5.3470,99.074) (5.3583,99.076) (5.3611,99.091) (5.3807,99.093) (5.3835,99.097) (5.3919,99.102) (5.4059,99.110) (5.4225,99.115) (5.4336,99.119) (5.4475,99.134) (5.4503,99.143) (5.4888,99.145) (5.5135,99.151) (5.5162,99.158) (5.5353,99.160) (5.5380,99.162) (5.5570,99.166) (5.5894,99.173) (5.5975,99.186) (5.6029,99.196) (5.6083,99.199) (5.6217,99.201) (5.6324,99.205) (5.6511,99.207) (5.6670,99.211) (5.6750,99.218) (5.6856,99.224) (5.7094,99.229) (5.7252,99.233) (5.7305,99.235) (5.7384,99.237) (5.7462,99.240) (5.7489,99.244) (5.7698,99.248) (5.7802,99.252) (5.8010,99.255) (5.8192,99.257) (5.8218,99.263) (5.8347,99.283) (5.8553,99.287) (5.8733,99.302) (5.8912,99.304) (5.8963,99.306) (5.9014,99.308) (5.9116,99.311) (5.9371,99.313) (5.9775,99.317) (5.9851,99.330) (5.9951,99.332) (6.0000,99.332)};
\addplot[const plot, cs1, thick] coordinates {(0.0000,0.000) (0.0000,2.708) (0.1736,11.573) (0.2455,18.454) (0.3471,23.513) (0.3881,33.089) (0.4909,36.810) (0.5207,40.046) (0.5489,46.050) (0.6258,50.899) (0.6943,52.860) (0.7157,56.559) (0.7364,58.565) (0.7762,61.568) (0.8679,65.373) (0.8850,67.462) (0.9347,68.906) (0.9819,69.729) (1.0121,71.136) (1.0414,72.045) (1.0558,73.370) (1.0978,74.378) (1.1114,75.513) (1.1644,76.418) (1.2150,76.868) (1.2273,78.133) (1.2516,78.943) (1.2636,79.656) (1.3219,80.268) (1.3556,80.979) (1.3886,81.367) (1.3994,82.578) (1.4313,83.146) (1.4728,83.409) (1.4830,83.795) (1.4931,84.243) (1.5525,84.644) (1.5621,84.857) (1.5718,85.395) (1.6002,86.175) (1.6375,86.526) (1.6466,86.863) (1.7095,87.128) (1.7183,87.373) (1.7357,87.789) (1.7444,88.116) (1.7701,88.371) (1.7870,88.621) (1.8121,88.888) (1.8451,89.097) (1.8694,89.267) (1.8775,89.463) (1.9093,89.532) (1.9172,89.747) (1.9406,90.182) (1.9637,90.256) (1.9790,90.557) (2.0242,90.700) (2.0316,90.820) (2.0829,90.889) (2.0901,91.225) (2.0973,91.391) (2.1116,91.688) (2.1187,91.848) (2.1470,92.037) (2.1748,92.167) (2.1955,92.233) (2.2092,92.307) (2.2228,92.395) (2.2564,92.511) (2.2631,92.751) (2.2830,92.929) (2.3157,93.054) (2.3287,93.149) (2.3352,93.242) (2.3608,93.384) (2.4113,93.474) (2.4176,93.550) (2.4300,93.627) (2.4362,93.686) (2.4547,93.877) (2.4669,93.927) (2.4852,94.121) (2.5033,94.196) (2.5272,94.300) (2.5628,94.366) (2.5803,94.508) (2.6036,94.618) (2.6094,94.674) (2.6266,94.752) (2.6438,94.821) (2.6495,94.840) (2.6551,94.952) (2.6946,95.006) (2.7001,95.049) (2.7113,95.129) (2.7168,95.181) (2.7444,95.275) (2.7771,95.303) (2.7826,95.355) (2.7988,95.461) (2.8041,95.521) (2.8255,95.650) (2.8468,95.693) (2.8626,95.762) (2.8731,95.790) (2.8888,95.805) (2.9096,95.846) (2.9456,95.881) (2.9507,95.930) (2.9558,96.010) (2.9660,96.040) (2.9711,96.083) (2.9862,96.120) (2.9963,96.189) (3.0313,96.303) (3.0363,96.322) (3.0708,96.383) (3.0757,96.417) (3.0904,96.458) (3.1049,96.486) (3.1243,96.497) (3.1291,96.579) (3.1435,96.631) (3.1674,96.693) (3.1863,96.723) (3.1911,96.758) (3.2005,96.891) (3.2286,96.915) (3.2426,96.954) (3.2611,96.982) (3.2749,96.990) (3.2933,97.031) (3.3161,97.070) (3.3342,97.109) (3.3387,97.178) (3.3522,97.214) (3.3701,97.301) (3.4101,97.378) (3.4190,97.404) (3.4234,97.417) (3.4365,97.428) (3.4453,97.436) (3.4584,97.501) (3.4714,97.512) (3.4758,97.527) (3.4887,97.538) (3.4931,97.568) (3.5103,97.583) (3.5146,97.647) (3.5402,97.688) (3.5614,97.695) (3.5741,97.716) (3.5783,97.835) (3.6118,97.854) (3.6243,97.884) (3.6450,97.891) (3.6491,97.912) (3.6615,97.940) (3.6779,97.958) (3.6820,97.960) (3.6902,97.971) (3.7105,98.018) (3.7146,98.024) (3.7267,98.046) (3.7388,98.059) (3.7469,98.113) (3.7549,98.121) (3.7909,98.143) (3.8067,98.180) (3.8107,98.201) (3.8225,98.210) (3.8343,98.218) (3.8422,98.236) (3.8539,98.248) (3.8812,98.302) (3.9005,98.324) (3.9159,98.343) (3.9275,98.350) (3.9351,98.354) (3.9580,98.376) (3.9618,98.395) (3.9656,98.410) (3.9921,98.412) (3.9959,98.429) (4.0072,98.462) (4.0372,98.477) (4.0483,98.496) (4.0521,98.529) (4.0632,98.548) (4.0669,98.552) (4.0854,98.561) (4.0964,98.565) (4.1148,98.574) (4.1257,98.585) (4.1403,98.589) (4.1657,98.602) (4.1693,98.608) (4.1729,98.615) (4.1802,98.651) (4.1945,98.660) (4.1981,98.673) (4.2267,98.679) (4.2374,98.682) (4.2552,98.697) (4.2693,98.701) (4.2869,98.720) (4.2974,98.731) (4.3114,98.733) (4.3393,98.744) (4.3428,98.759) (4.3497,98.763) (4.3531,98.768) (4.3807,98.778) (4.4184,98.781) (4.4252,98.809) (4.4354,98.826) (4.4490,98.839) (4.4794,98.845) (4.5028,98.871) (4.5129,98.886) (4.5162,98.893) (4.5262,98.901) (4.5560,98.903) (4.5824,98.916) (4.5857,98.938) (4.6119,98.940) (4.6217,98.970) (4.6574,98.972) (4.6703,98.994) (4.6736,99.011) (4.6864,99.013) (4.6896,99.018) (4.6993,99.022) (4.7217,99.033) (4.7376,99.035) (4.7408,99.037) (4.7661,99.041) (4.7882,99.048) (4.8007,99.054) (4.8227,99.056) (4.8258,99.061) (4.8351,99.071) (4.8600,99.076) (4.8631,99.089) (4.8724,99.097) (4.8878,99.099) (4.8909,99.102) (4.9001,99.123) (4.9093,99.125) (4.9124,99.134) (4.9155,99.147) (4.9338,99.151) (4.9643,99.153) (4.9975,99.155) (5.0066,99.158) (5.0366,99.160) (5.0455,99.171) (5.0545,99.173) (5.0604,99.177) (5.0694,99.181) (5.0812,99.192) (5.1049,99.196) (5.1078,99.201) (5.1255,99.205) (5.1284,99.211) (5.1519,99.214) (5.1548,99.216) (5.1606,99.218) (5.1781,99.220) (5.2014,99.231) (5.2100,99.242) (5.2187,99.244) (5.2475,99.250) (5.2704,99.252) (5.2790,99.261) (5.2875,99.263) (5.2904,99.268) (5.2989,99.270) (5.3244,99.272) (5.3470,99.278) (5.3611,99.283) (5.3807,99.289) (5.3835,99.306) (5.3891,99.315) (5.3919,99.319) (5.4475,99.324) (5.4503,99.328) (5.4888,99.332) (5.5135,99.343) (5.5353,99.345) (5.5380,99.347) (5.5461,99.349) (5.5570,99.352) (5.5651,99.356) (5.5786,99.358) (5.5894,99.360) (5.5975,99.380) (5.6083,99.382) (5.7489,99.390) (5.7515,99.395) (5.7698,99.399) (5.8347,99.418) (5.8938,99.425) (5.9014,99.429) (5.9040,99.433) (5.9371,99.436) (5.9649,99.438) (5.9851,99.461) (5.9926,99.464) (5.9951,99.468) (6.0000,99.468)};
\addplot[const plot, cs2, thick] coordinates {(0.0000,0.000) (0.0000,2.723) (0.1736,11.972) (0.2455,20.992) (0.3471,26.900) (0.3881,38.186) (0.4909,42.581) (0.5207,46.091) (0.5489,53.028) (0.6258,58.416) (0.6943,60.374) (0.7157,64.420) (0.7364,66.349) (0.7762,69.388) (0.8679,72.898) (0.8850,75.944) (0.9347,77.844) (0.9819,78.639) (1.0121,79.854) (1.0414,80.623) (1.0558,81.780) (1.0978,82.500) (1.1114,83.273) (1.1644,83.868) (1.2150,84.628) (1.2273,85.380) (1.2516,86.027) (1.2636,86.501) (1.3219,87.117) (1.3556,87.589) (1.3886,88.134) (1.3994,88.911) (1.4313,89.282) (1.4728,89.519) (1.4830,90.012) (1.4931,90.325) (1.5525,90.641) (1.5621,91.178) (1.5718,91.553) (1.6002,91.846) (1.6375,92.018) (1.6466,92.343) (1.7095,92.524) (1.7183,92.617) (1.7357,92.789) (1.7444,93.033) (1.7701,93.102) (1.7870,93.162) (1.8121,93.330) (1.8451,93.446) (1.8694,93.578) (1.8775,93.627) (1.9093,93.634) (1.9172,93.679) (1.9406,93.918) (1.9637,94.022) (1.9790,94.405) (2.0242,94.491) (2.0316,94.702) (2.0829,94.717) (2.0901,94.991) (2.0973,95.217) (2.1116,95.243) (2.1187,95.306) (2.1470,95.355) (2.1748,95.558) (2.1955,95.644) (2.2092,95.696) (2.2228,95.840) (2.2564,95.943) (2.2631,96.094) (2.2830,96.120) (2.3157,96.191) (2.3287,96.256) (2.3352,96.335) (2.3608,96.527) (2.4113,96.583) (2.4176,96.669) (2.4300,96.689) (2.4362,96.727) (2.4547,96.857) (2.4669,96.919) (2.4852,97.141) (2.5033,97.221) (2.5272,97.268) (2.5628,97.389) (2.5803,97.520) (2.6036,97.570) (2.6266,97.594) (2.6438,97.613) (2.6495,97.710) (2.6551,97.818) (2.6946,97.958) (2.7001,97.960) (2.7113,97.977) (2.7168,98.031) (2.7444,98.119) (2.7771,98.169) (2.7826,98.218) (2.8041,98.231) (2.8255,98.490) (2.8468,98.503) (2.8626,98.669) (2.9096,98.740) (2.9456,98.748) (2.9507,98.787) (2.9558,98.890) (2.9660,98.916) (2.9711,99.043) (2.9963,99.319) (3.0313,99.455) (3.0708,99.489) (3.0757,99.537) (3.1291,99.554) (3.1674,99.733) (3.1911,99.774) (3.2005,99.838) (3.2426,99.860) (3.2933,99.890) (3.4101,99.920) (3.4584,100.000) (6.0000,100.000)};
\fill[cs0] (axis cs:2.996,95) circle (1.3pt);
\fill[cs1] (axis cs:2.695,95) circle (1.3pt);
\fill[cs2] (axis cs:2.097,95) circle (1.3pt);
\node[font=\fontsize{6.5pt}{7.5pt}\selectfont, anchor=north east, fill=white, inner sep=0.5pt] (conf) at (axis cs:5.9,92) {95\% confidence};
\node[font=\fontsize{6.5pt}{7.5pt}\selectfont, anchor=north east, align=right, fill=white, inner sep=0.5pt] at ([yshift=-1pt]conf.south east) {\textcolor{cs0}{3.00~m}\\\textcolor{cs1}{2.69~m}\\\textcolor{cs2}{2.10~m}};
\end{axis}
\node[font=\footnotesize, anchor=north] at ([yshift=-24pt]cdf.south) {(b)};
\node[font=\scriptsize, anchor=south] at ([yshift=1pt]current bounding box.north) {\tikz[baseline=-0.6ex]\draw[draw=black, line width=0.2pt, fill=cs0, postaction={pattern=north east lines, pattern color=black!70}] (0,-0.08) rectangle (0.3,0.08);~Raw\quad \tikz[baseline=-0.6ex]\draw[draw=black, line width=0.2pt, fill=cs1, postaction={pattern=north west lines, pattern color=black!70}] (0,-0.08) rectangle (0.3,0.08);~Averaged\quad \tikz[baseline=-0.6ex]\draw[draw=black, line width=0.2pt, fill=cs2, postaction={pattern=horizontal lines, pattern color=black!70}] (0,-0.08) rectangle (0.3,0.08);~Kalman};
\end{tikzpicture}

%% file: figures_tex/run12_three_align_trajectory_includable.tex

\begin{tikzpicture}
    \begin{groupplot}[
        group style={
            group size=1 by 2,
            vertical sep=0.15cm,
            xlabels at=edge bottom,
        },
        width=\linewidth,
        height=3cm,
        xmin=0, xmax=2000,
        grid=major,
        grid style={gray!30},
        tick label style={font=\scriptsize},
        label style={font=\small},
    ]
    
    \nextgroupplot[
        ylabel={X Position},
        ymin=0, ymax=43,
        xticklabels={},
        legend columns=2,
        legend style={
            fill opacity=0.8,
            draw opacity=1,
            text opacity=1,
            draw=lightgray204,
            font=\footnotesize,
            at={(0.95, 1.72)},
            anchor=north east,
            column sep=5pt,
        },
    ]
    
    \addplot[rawcolor, opacity=0.6, line width=0.8pt] 
        table[x=sample, y=raw_x, col sep=comma] {figures_tex/run12_three_align_trajectory_tikz.csv};
    \addlegendentry{Raw}
    
    \addplot[avgcolor, line width=1.2pt] 
        table[x=sample, y=avg_x, col sep=comma] {figures_tex/run12_three_align_trajectory_tikz.csv};
    \addlegendentry{Averaged}
    
    \addplot[kalmancolor, line width=1.5pt] 
        table[x=sample, y=kalman_x, col sep=comma] {figures_tex/run12_three_align_trajectory_tikz.csv};
    \addlegendentry{Kalman Filtered}
    
    \addplot[gtcolor, line width=1.5pt, densely dashed] 
        table[x=sample, y=gt_x, col sep=comma] {figures_tex/run12_three_align_trajectory_tikz.csv};
    \addlegendentry{Ground Truth}
    
    \nextgroupplot[
        ylabel={Y Position},
        ymin=0, ymax=64,
        xlabel={Sample Index},
    ]
    
    \addplot[rawcolor, opacity=0.6, line width=0.8pt] 
        table[x=sample, y=raw_y, col sep=comma] {figures_tex/run12_three_align_trajectory_tikz.csv};
    
    \addplot[avgcolor, line width=1.2pt] 
        table[x=sample, y=avg_y, col sep=comma] {figures_tex/run12_three_align_trajectory_tikz.csv};
    
    \addplot[kalmancolor, line width=1.5pt] 
        table[x=sample, y=kalman_y, col sep=comma] {figures_tex/run12_three_align_trajectory_tikz.csv};
    
    \addplot[gtcolor, line width=1.5pt, densely dashed] 
        table[x=sample, y=gt_y, col sep=comma] {figures_tex/run12_three_align_trajectory_tikz.csv};
    
    \end{groupplot}
\end{tikzpicture}

%% file: figures_tex/cusense_serving.tex
\begin{tikzpicture}
\definecolor{o0}{HTML}{AEC7E8}
\definecolor{o1}{HTML}{1F77B4}
\definecolor{o2}{HTML}{0B3C6E}
\definecolor{o3}{HTML}{9467BD}
\begin{axis}[
width=0.82\linewidth, height=0.40\linewidth, scale only axis, ymode=log,
ymin=100, ymax=20000, xmin=-0.55, xmax=4.55,
xtick={0,1,2,3,4}, xticklabels={MIG,MPS,Time-slicing,MPS,Time-slicing},
ytick={100,200,500,1000,2000,5000,10000}, yticklabels={100,200,500,1000,2000,5000,10000},
xticklabel style={font=\scriptsize}, yticklabel style={font=\scriptsize},
ylabel={Latency [$\mu$s]}, ylabel style={font=\footnotesize}, ymajorgrids, grid style={black!12},
axis x line*=bottom, axis y line*=left, xtick style={draw=none}, clip=false,
]
\draw[draw=black, line width=0.2pt, fill=o0, postaction={pattern=horizontal lines, pattern color=black!70}] (axis cs:-0.380,100) rectangle (axis cs:-0.190,2138);
\draw[black, line width=0.4pt] (axis cs:-0.285,2138) -- (axis cs:-0.285,2521) (axis cs:-0.333,2521) -- (axis cs:-0.238,2521);
\node[font=\fontsize{5.5pt}{6pt}\selectfont, rotate=90, anchor=west, inner sep=1pt] at (axis cs:-0.285,2622) {2138};
\draw[draw=black, line width=0.2pt, fill=o0, postaction={pattern=horizontal lines, pattern color=black!70}] (axis cs:0.620,100) rectangle (axis cs:0.810,2107);
\draw[black, line width=0.4pt] (axis cs:0.715,2107) -- (axis cs:0.715,2502) (axis cs:0.667,2502) -- (axis cs:0.762,2502);
\node[font=\fontsize{5.5pt}{6pt}\selectfont, rotate=90, anchor=west, inner sep=1pt] at (axis cs:0.715,2602) {2107};
\draw[draw=black, line width=0.2pt, fill=o0, postaction={pattern=horizontal lines, pattern color=black!70}] (axis cs:1.620,100) rectangle (axis cs:1.810,5156);
\draw[black, line width=0.4pt] (axis cs:1.715,5156) -- (axis cs:1.715,5256) (axis cs:1.667,5256) -- (axis cs:1.763,5256);
\node[font=\fontsize{5.5pt}{6pt}\selectfont, rotate=90, anchor=west, inner sep=1pt] at (axis cs:1.715,5466) {5156};
\draw[draw=black, line width=0.2pt, fill=o0, postaction={pattern=horizontal lines, pattern color=black!70}] (axis cs:2.620,100) rectangle (axis cs:2.810,1228);
\draw[black, line width=0.4pt] (axis cs:2.715,1228) -- (axis cs:2.715,1335) (axis cs:2.668,1335) -- (axis cs:2.763,1335);
\node[font=\fontsize{5.5pt}{6pt}\selectfont, rotate=90, anchor=west, inner sep=1pt] at (axis cs:2.715,1388) {1228};
\draw[draw=black, line width=0.2pt, fill=o0, postaction={pattern=horizontal lines, pattern color=black!70}] (axis cs:3.620,100) rectangle (axis cs:3.810,2271);
\draw[black, line width=0.4pt] (axis cs:3.715,2271) -- (axis cs:3.715,2667) (axis cs:3.668,2667) -- (axis cs:3.763,2667);
\node[font=\fontsize{5.5pt}{6pt}\selectfont, rotate=90, anchor=west, inner sep=1pt] at (axis cs:3.715,2774) {2271};
\draw[draw=black, line width=0.2pt, fill=o1, postaction={pattern=north east lines, pattern color=black!70}] (axis cs:-0.190,100) rectangle (axis cs:0.000,1460);
\draw[black, line width=0.4pt] (axis cs:-0.095,1460) -- (axis cs:-0.095,1694) (axis cs:-0.143,1694) -- (axis cs:-0.048,1694);
\node[font=\fontsize{5.5pt}{6pt}\selectfont, rotate=90, anchor=west, inner sep=1pt] at (axis cs:-0.095,1762) {1460};
\draw[draw=black, line width=0.2pt, fill=o1, postaction={pattern=north east lines, pattern color=black!70}] (axis cs:0.810,100) rectangle (axis cs:1.000,1399);
\draw[black, line width=0.4pt] (axis cs:0.905,1399) -- (axis cs:0.905,1607) (axis cs:0.858,1607) -- (axis cs:0.953,1607);
\node[font=\fontsize{5.5pt}{6pt}\selectfont, rotate=90, anchor=west, inner sep=1pt] at (axis cs:0.905,1671) {1399};
\draw[draw=black, line width=0.2pt, fill=o1, postaction={pattern=north east lines, pattern color=black!70}] (axis cs:1.810,100) rectangle (axis cs:2.000,2736);
\draw[black, line width=0.4pt] (axis cs:1.905,2736) -- (axis cs:1.905,2845) (axis cs:1.857,2845) -- (axis cs:1.953,2845);
\node[font=\fontsize{5.5pt}{6pt}\selectfont, rotate=90, anchor=west, inner sep=1pt] at (axis cs:1.905,2959) {2736};
\draw[draw=black, line width=0.2pt, fill=o1, postaction={pattern=north east lines, pattern color=black!70}] (axis cs:2.810,100) rectangle (axis cs:3.000,1021);
\draw[black, line width=0.4pt] (axis cs:2.905,1021) -- (axis cs:2.905,1153) (axis cs:2.858,1153) -- (axis cs:2.953,1153);
\node[font=\fontsize{5.5pt}{6pt}\selectfont, rotate=90, anchor=west, inner sep=1pt] at (axis cs:2.905,1199) {1021};
\draw[draw=black, line width=0.2pt, fill=o1, postaction={pattern=north east lines, pattern color=black!70}] (axis cs:3.810,100) rectangle (axis cs:4.000,2256);
\draw[black, line width=0.4pt] (axis cs:3.905,2256) -- (axis cs:3.905,2399) (axis cs:3.858,2399) -- (axis cs:3.953,2399);
\node[font=\fontsize{5.5pt}{6pt}\selectfont, rotate=90, anchor=west, inner sep=1pt] at (axis cs:3.905,2495) {2256};
\draw[draw=black, line width=0.2pt, fill=o2, postaction={pattern=north west lines, pattern color=black!70}] (axis cs:0.000,100) rectangle (axis cs:0.190,522);
\draw[black, line width=0.4pt] (axis cs:0.095,522) -- (axis cs:0.095,626) (axis cs:0.048,626) -- (axis cs:0.143,626);
\node[font=\fontsize{5.5pt}{6pt}\selectfont, rotate=90, anchor=west, inner sep=1pt] at (axis cs:0.095,651) {522};
\draw[draw=black, line width=0.2pt, fill=o2, postaction={pattern=north west lines, pattern color=black!70}] (axis cs:1.000,100) rectangle (axis cs:1.190,594);
\draw[black, line width=0.4pt] (axis cs:1.095,594) -- (axis cs:1.095,735) (axis cs:1.047,735) -- (axis cs:1.143,735);
\node[font=\fontsize{5.5pt}{6pt}\selectfont, rotate=90, anchor=west, inner sep=1pt] at (axis cs:1.095,764) {594};
\draw[draw=black, line width=0.2pt, fill=o2, postaction={pattern=north west lines, pattern color=black!70}] (axis cs:2.000,100) rectangle (axis cs:2.190,883);
\draw[black, line width=0.4pt] (axis cs:2.095,883) -- (axis cs:2.095,1236) (axis cs:2.048,1236) -- (axis cs:2.143,1236);
\node[font=\fontsize{5.5pt}{6pt}\selectfont, rotate=90, anchor=west, inner sep=1pt] at (axis cs:2.095,1285) {883};
\draw[draw=black, line width=0.2pt, fill=o2, postaction={pattern=north west lines, pattern color=black!70}] (axis cs:3.000,100) rectangle (axis cs:3.190,779);
\draw[black, line width=0.4pt] (axis cs:3.095,779) -- (axis cs:3.095,1409) (axis cs:3.048,1409) -- (axis cs:3.143,1409);
\node[font=\fontsize{5.5pt}{6pt}\selectfont, rotate=90, anchor=west, inner sep=1pt] at (axis cs:3.095,1465) {779};
\draw[draw=black, line width=0.2pt, fill=o2, postaction={pattern=north west lines, pattern color=black!70}] (axis cs:4.000,100) rectangle (axis cs:4.190,1094);
\draw[black, line width=0.4pt] (axis cs:4.095,1094) -- (axis cs:4.095,1425) (axis cs:4.047,1425) -- (axis cs:4.143,1425);
\node[font=\fontsize{5.5pt}{6pt}\selectfont, rotate=90, anchor=west, inner sep=1pt] at (axis cs:4.095,1482) {1094};
\draw[draw=black, line width=0.2pt, fill=o3, postaction={pattern=crosshatch, pattern color=black!70}] (axis cs:0.190,100) rectangle (axis cs:0.380,241);
\draw[black, line width=0.4pt] (axis cs:0.285,241) -- (axis cs:0.285,249) (axis cs:0.238,249) -- (axis cs:0.333,249);
\node[font=\fontsize{5.5pt}{6pt}\selectfont, rotate=90, anchor=west, inner sep=1pt] at (axis cs:0.285,259) {241};
\draw[draw=black, line width=0.2pt, fill=o3, postaction={pattern=crosshatch, pattern color=black!70}] (axis cs:1.190,100) rectangle (axis cs:1.380,275);
\draw[black, line width=0.4pt] (axis cs:1.285,275) -- (axis cs:1.285,292) (axis cs:1.237,292) -- (axis cs:1.333,292);
\node[font=\fontsize{5.5pt}{6pt}\selectfont, rotate=90, anchor=west, inner sep=1pt] at (axis cs:1.285,304) {275};
\draw[draw=black, line width=0.2pt, fill=o3, postaction={pattern=crosshatch, pattern color=black!70}] (axis cs:2.190,100) rectangle (axis cs:2.380,713);
\draw[black, line width=0.4pt] (axis cs:2.285,713) -- (axis cs:2.285,721) (axis cs:2.238,721) -- (axis cs:2.333,721);
\node[font=\fontsize{5.5pt}{6pt}\selectfont, rotate=90, anchor=west, inner sep=1pt] at (axis cs:2.285,750) {713};
\draw[draw=black, line width=0.2pt, fill=o3, postaction={pattern=crosshatch, pattern color=black!70}] (axis cs:3.190,100) rectangle (axis cs:3.380,399);
\draw[black, line width=0.4pt] (axis cs:3.285,399) -- (axis cs:3.285,470) (axis cs:3.238,470) -- (axis cs:3.333,470);
\node[font=\fontsize{5.5pt}{6pt}\selectfont, rotate=90, anchor=west, inner sep=1pt] at (axis cs:3.285,489) {399};
\draw[draw=black, line width=0.2pt, fill=o3, postaction={pattern=crosshatch, pattern color=black!70}] (axis cs:4.190,100) rectangle (axis cs:4.380,689);
\draw[black, line width=0.4pt] (axis cs:4.285,689) -- (axis cs:4.285,751) (axis cs:4.237,751) -- (axis cs:4.333,751);
\node[font=\fontsize{5.5pt}{6pt}\selectfont, rotate=90, anchor=west, inner sep=1pt] at (axis cs:4.285,781) {689};
\draw[black!50] (axis cs:2.5,100) -- (axis cs:2.5,20000);
\node[font=\footnotesize, anchor=north] at (axis cs:1,20000) {GH200};
\node[font=\footnotesize, anchor=north] at (axis cs:3.5,20000) {DGX Spark};
\end{axis}
\node[font=\scriptsize, anchor=south, inner sep=0pt] at ([yshift=3pt]current bounding box.north) {\begin{tabular}{@{}l@{\quad}l@{}}\tikz[baseline=-0.6ex]\draw[draw=black, line width=0.2pt, fill=o0, postaction={pattern=horizontal lines, pattern color=black!70}] (0,-0.08) rectangle (0.3,0.08);~Triton, Python + TorchScript & \tikz[baseline=-0.6ex]\draw[draw=black, line width=0.2pt, fill=o1, postaction={pattern=north east lines, pattern color=black!70}] (0,-0.08) rectangle (0.3,0.08);~Triton, Python + TensorRT\\\tikz[baseline=-0.6ex]\draw[draw=black, line width=0.2pt, fill=o2, postaction={pattern=north west lines, pattern color=black!70}] (0,-0.08) rectangle (0.3,0.08);~Triton, fused TensorRT & \tikz[baseline=-0.6ex]\draw[draw=black, line width=0.2pt, fill=o3, postaction={pattern=crosshatch, pattern color=black!70}] (0,-0.08) rectangle (0.3,0.08);~Native CUDA + TensorRT\end{tabular}};
\end{tikzpicture}